\pdfoutput=1
\documentclass[11pt]{article}
\usepackage[T1]{fontenc}
\usepackage{lmodern,microtype}
\usepackage[margin=1in]{geometry}
\usepackage{amsmath,amssymb,amsthm,mathtools}
\usepackage{booktabs,enumitem,framed,xcolor}
\usepackage[colorlinks=true,linkcolor=blue!55!black,citecolor=blue!55!black,urlcolor=blue!55!black]{hyperref}
\usepackage[capitalize,nameinlink]{cleveref}

\theoremstyle{plain}
\newtheorem{theorem}{Theorem}[section]
\newtheorem{lemma}[theorem]{Lemma}
\newtheorem{proposition}[theorem]{Proposition}
\newtheorem{corollary}[theorem]{Corollary}
\theoremstyle{definition}

\theoremstyle{remark}
\newtheorem{remark}[theorem]{Remark}
\crefname{lemma}{Lemma}{Lemmas}
\crefname{theorem}{Theorem}{Theorems}
\crefname{proposition}{Proposition}{Propositions}
\crefname{corollary}{Corollary}{Corollaries}
\crefname{remark}{Remark}{Remarks}
\crefname{equation}{}{}

\newcommand{\E}{\mathbb{E}}
\renewcommand{\Pr}{\mathbb{P}}
\newcommand{\R}{\mathbb{R}}
\newcommand{\ind}{\mathbf{1}}
\newcommand{\OPT}{\mathrm{OPT}}
\newcommand{\pos}[1]{\left[#1\right]_+}
\newcommand{\marg}[2]{f(#1\mid #2)}
\newcommand{\calE}{\mathcal{E}}
\newcommand{\AP}{\textsc{AttenuatedPricing}}
\DeclarePairedDelimiter{\ceil}{\lceil}{\rceil}
\DeclarePairedDelimiter{\floor}{\lfloor}{\rfloor}

\title{The Submodular Secretary Problem:\\
Tight $1-1/e$ for Monotone Objectives,\\ and Strictly Between $1/e$ and $1/2$ in General}
\author{%
Hamed Abdi\thanks{University of Maryland, College Park. Email: \texttt{hamedabdi37@gmail.com}.}
\and Kiarash Banihashem\thanks{University of Maryland, College Park. Email: \texttt{kiarash@umd.edu}.}
\and MohammadTaghi Hajiaghayi\thanks{University of Maryland, College Park. Email: \texttt{hajiaghayi@gmail.com}.}
\and Danny Mittal\thanks{University of Maryland, College Park. Email: \texttt{dannymittal@gmail.com}.}}
\date{November 2026}

\hypersetup{pdftitle={The Submodular Secretary Problem: Tight 1-1/e for Monotone Objectives, and Strictly Between 1/e and 1/2 in General}}

\begin{document}
\maketitle

\begin{abstract}
In the submodular secretary problem with a cardinality constraint, $n$ elements arrive in uniformly random order, and an algorithm must irrevocably accept or reject each one on arrival so as to maximize a nonnegative submodular function over sets of at most $k$ elements. For monotone objectives we settle the problem: the optimal competitive ratio tends to exactly $1-1/e$ as $k\to\infty$, and its worst case over all budgets $k$ is exactly $1/e$. The bound $1-1/e$ cannot be exceeded even with unlimited computation, and it is achieved by a single algorithm with at most $4n+1$ value queries; previously the best ratio without a shortlist was $1/e-o(1)$, with exponentially many queries. For general objectives, where the best known ratio was about $0.107$, we show that the optimal limiting ratio lies strictly between $1/e$ and $1/2$: we achieve $1/e+3.5\cdot10^{-6}$ for large $k$, and no algorithm achieves more than $0.497053$. We also achieve $0.2677$ at every budget. The main new ingredient of our algorithm is \emph{eligibility attenuation}: an element that passes the price test at time $t$ is accepted with probability $e^{-\tau}$, where $\tau$ is the total earlier time during which it would have passed, which makes the probability that it has already been selected exactly $1-e^{-\tau}$.
\end{abstract}

\section{Introduction}

In the classical secretary problem~\cite{Dynkin63,Lindley61}, $n$ values are revealed in uniformly random order and we must stop at one of them, deciding irrevocably as each appears. Observing a $1/e$ fraction of the values and then taking the next record selects the maximum with probability approaching $1/e$, which is optimal. Since the work of Hajiaghayi, Kleinberg, and Parkes~\cite{HKP04} and Kleinberg~\cite{Kleinberg05}, the random-order model has been used to study online selection of several elements, including under matroid constraints~\cite{BIK07,ABHHMS26}.

When the objective has decreasing marginal values, the natural formulation is the \emph{submodular secretary problem}, introduced independently by Bateni, Hajiaghayi, and Zadimoghaddam~\cite{BHZ13} and Gupta, Roth, Schoenebeck, and Talwar~\cite{GRST10}. We study its most basic constrained form. A ground set $E$ of $n$ elements arrives in uniformly random order, and the algorithm knows $n$ and an integer $k\le n$. When an element arrives, the algorithm may query a nonnegative submodular function $f:2^E\to\R_{\ge0}$ on subsets of the elements that have arrived so far, and must then accept or reject the element immediately and irrevocably, accepting at most $k$ elements in total. Its \emph{competitive ratio} is the infimum, over instances, of $\E[f(A)]/\OPT$, where $A$ is the accepted set and $\OPT=\max_{|S|\le k}f(S)$.

Offline, this problem has well-known benchmarks. For monotone $f$, greedy achieves $1-(1-1/k)^k\ge 1-1/e$~\cite{NWF78}, and no algorithm making polynomially many value queries does better asymptotically~\cite{NW78}. For general nonnegative $f$, the measured continuous greedy algorithm achieves $1/e$~\cite{FNS11-unified}; this is a natural barrier for clean analyses, although more elaborate algorithms achieve slightly better constants~\cite{BFNS14,EN16,BF24}. Online, the known guarantees have remained far from these benchmarks. The ratio for monotone objectives improved from $0.090$~\cite{BHZ13} to $0.170$~\cite{FNS11-secretary}, and then Kesselheim and T\"onnis~\cite{KT17} showed that repeatedly optimizing over the observed prefix gives $0.275-o(1)$ in polynomial time and $1/e-o(1)$ with exact, exponential-query optimization. For general objectives the best known ratios were $1/(8e^2)$~\cite{BHZ13} and then $(e-1)^2/(e^2(1+e))\approx0.107$~\cite{Fujii19}. Agrawal, Shadravan, and Stein~\cite{ASS19} reach $1-1/e-\varepsilon$ for monotone objectives, but only when the algorithm may keep a shortlist of $O(k)$ candidates and make its final choice afterwards.

This leaves two basic questions open. Does the ability to delay decisions, which a shortlist grants, matter for monotone objectives? And can the $1/e$ offline benchmark for general objectives be matched online?

\subsection{Main results}

Our main result answers both questions with a single algorithm, \AP{} (\cref{sec:algorithm}). To state it uniformly, we say that $f$ is \emph{$\mu$-monotone}, for $\mu\in[0,1]$, if $f(B)\ge\mu f(S)$ whenever $S\subseteq B$. Every nonnegative function is $0$-monotone, and $1$-monotone functions are exactly the monotone ones.

\begin{theorem}[Algorithm]\label{thm:main}
There is a randomized algorithm for the submodular secretary problem with a cardinality constraint that knows only $n$ and $k$, accepts at most $k$ elements on every execution, makes at most $4n+1$ value queries, each on a subset of the elements that have arrived, and satisfies the following, uniformly over $n\ge k$ and $f$.
\begin{enumerate}[label=(\alph*),nosep]
\item If $f$ is nonnegative and submodular, then $\E[f(A)]\ge\bigl(1/e-O(\log^2(k+1)/k)\bigr)\OPT$.
\item If $f$ is nonnegative, monotone, and submodular, then $\E[f(A)]\ge\bigl(1-1/e-O(\log(k+1)/\sqrt k)\bigr)\OPT$.
\item More generally, if $f$ is $\mu$-monotone, then $\E[f(A)]\ge\bigl(1/e+\mu(1-2/e)-O(\log(k+1)/\sqrt k)\bigr)\OPT$.
\end{enumerate}
\end{theorem}

The same algorithm runs in all three cases and does not use $\mu$. The value-query model counts queries and treats real arithmetic as exact. \cref{thm:finite} gives a bound with explicit constants.

For monotone objectives this is the best possible asymptotic ratio, even for computationally unbounded algorithms.

\begin{theorem}[Hardness]\label{thm:hardness-intro}
Let $\mathcal{A}$ be any randomized online algorithm, with unlimited computation and unlimited queries on subsets of arrived elements.
\begin{enumerate}[label=(\alph*),nosep]
\item For every $k\ge2$ and $\varepsilon>0$, there is a monotone coverage instance with budget $k$ on which $\mathcal{A}$ has ratio at most $1-(1-1/k)^k+\varepsilon$.
\item For every $k\ge64$, there is a monotone coverage instance with budget $k$ and at most $k^2$ elements on which $\mathcal{A}$ has ratio at most $1-1/e+O(\log\log k/\log k)$.
\end{enumerate}
\end{theorem}

Unlike the offline barrier of~\cite{NW78}, \cref{thm:hardness-intro} is information-theoretic, and it arises purely from the inability to see the future. Combining the two theorems pins down the problem.

\begin{corollary}[Monotone objectives are settled]\label{cor:monotone-settled}
Let $\rho^{\rm mon}_k$ be the optimal competitive ratio for the monotone submodular secretary problem with budget $k$, over all algorithms with unlimited computation. Then $\lim_{k\to\infty}\rho^{\rm mon}_k=1-1/e$, and $\inf_{k\ge1}\rho^{\rm mon}_k=\rho^{\rm mon}_1=1/e$.
\end{corollary}

The limit follows from \cref{thm:main,thm:hardness-intro}, and the worst case over budgets from \cref{mon:thm} (see \cref{sec:further}), since the case $k=1$ is the classical secretary problem. So both the asymptotic ratio and the ratio that can be guaranteed uniformly over all budgets are determined exactly. A shortlist is not needed to reach $1-1/e$, and unlimited computation does not help to exceed it.

For general objectives, $1/e$ is not the limit.

\begin{theorem}[Beyond $1/e$]\label{thm:beyond-intro}
There is a randomized algorithm that knows only $n$ and $k$, accepts at most $k$ elements, and makes polynomially many value queries on subsets of arrived elements, whose ratio on nonnegative submodular objectives is at least $1/e+3.5\cdot10^{-6}$ for all sufficiently large $k$, uniformly in $n$.
\end{theorem}

The algorithm (\cref{sec:beyond}) runs, with a small probability, a variant of Greedy restricted to a pool of promising elements, and otherwise a variant of \AP{} that measures its progress against two disjoint near-optimal sets at the start. Neither alone beats $1/e$ on every instance, but their guarantees depend in opposite directions on one parameter of the instance, which neither of them computes.

In the other direction, there is a barrier strictly below $1-1/e$, which comes from nonmonotonicity.

\begin{theorem}[General objectives]\label{thm:half-intro}
Let $\mathcal{A}$ be any randomized online algorithm, with unlimited computation and unlimited queries on subsets of arrived elements.
\begin{enumerate}[label=(\alph*),nosep]
\item For all $n\ge2$, $1\le k\le n$, $\mu\in[0,1]$, and $\varepsilon>0$, there is a normalized $\mu$-monotone submodular instance on which $\mathcal{A}$ has ratio at most $\frac{1+\mu}2+\varepsilon$. In particular, for general objectives the ratio is at most $\frac12+\varepsilon$.
\item For every sufficiently large $k$, there is a normalized nonnegative submodular instance with budget $k$ on which $\mathcal{A}$ has ratio at most $0.497053$, and in which every singleton has value at most $10^{-6}\OPT$. This holds even if $\mathcal{A}$ observes the arrival times.
\end{enumerate}
\end{theorem}

Together with \cref{thm:beyond-intro}, this separates the general problem from both of its natural benchmarks.

\begin{corollary}[General objectives]\label{cor:general-between}
Let $\rho_k$ be the optimal competitive ratio for nonnegative submodular objectives with budget $k$, over all algorithms with unlimited computation. Then
\[
 \frac1e<\frac1e+3.5\cdot10^{-6}\le\liminf_{k\to\infty}\rho_k\le\limsup_{k\to\infty}\rho_k\le0.497053<\frac12 .
\]
\end{corollary}

So the limiting ratio for general objectives lies strictly between $1/e$ and $1/2$, and in particular strictly below the monotone limit $1-1/e$; the lower bound is achieved with polynomially many queries. Offline, unlimited queries solve the problem exactly, and so these barriers are purely a consequence of online decisions. For polynomial-query algorithms the offline bound of $0.491$~\cite{OV11} is numerically stronger, but it has a different source, namely query complexity. The bound $\frac12$ in part~(a) comes from pairs of elements that are valuable separately and worthless together, and part~(b) goes below it with matroid gadgets (\cref{sec:below-half}), which the pairs alone cannot do. For $\mu$-monotone objectives, since coverage functions are monotone, the upper bound is $\min\{\frac{1+\mu}2,\,1-1/e\}$ up to $o(1)$ terms, and the lower bound of \cref{thm:main}(c) meets it only at $\mu=1$.

\Cref{tab:results} summarizes the landscape.

\begin{table}[ht]
\centering
\small
\begin{tabular}{@{}lp{4.3cm}p{2.7cm}p{4.3cm}@{}}
\toprule
Objective & Previous best (no shortlist) & This paper & Upper bound \\
\midrule
Monotone & $1/e-o(1)$ \cite{KT17} (exponential queries); $0.275-o(1)$ \cite{KT17} (polynomial queries) & $1-1/e-o(1)$; $1/e$ for every $k$ & $1-1/e+o(1)$ (\cref{thm:hardness-intro}) \\[3pt]
General & $0.107$ \cite{Fujii19} & $1/e+3.5\cdot10^{-6}$; $0.2677$ for every $k$ & $0.497053$ (\cref{thm:half-intro}); $0.491$ with polynomially many queries \cite{OV11} \\
\bottomrule
\end{tabular}
\caption{Asymptotic competitive ratios for large $k$. \AP{} makes at most $4n+1$ queries and achieves $1-1/e-o(1)$ and $1/e-o(1)$; the entry $1/e+3.5\cdot10^{-6}$ is achieved by the mixture of \cref{sec:beyond}, which makes polynomially many queries, and the entry $0.2677$ holds for every $k$ (\cref{sec:further}). The $0.491$ entry follows because an online algorithm can be simulated offline on a random permutation.}
\label{tab:results}
\end{table}

\subsection{Every budget}\label{sec:further}

The guarantees of \cref{thm:main,thm:beyond-intro} take effect only for large $k$. We also give explicit guarantees at every budget, with no error term, uniformly in $n\ge k$. For general objectives, some algorithm making at most $6n+1$ value queries has ratio at least $0.2677$ for every $k$; the ratio is at least $0.3734$ for $k=2$, at least $0.3$ for $k\ge52$, and at least $0.3676$ for $k\ge10^6$ (\cref{ek:thm-every-k}). It uses two picks for $k\le3$, for $4\le k\le29$ a multi-rank secretary rule that is restarted after every stop on a geometric time grid, and Poisson-tail prices with eligibility attenuation for larger $k$. For monotone objectives, the ratio is at least $1/e$ for every $k$: exactly $1/e$ for $k=1$, more than $0.3696$ for every $k\ge2$, and more than $\frac12$ for every $k\ge98$ (\cref{mon:thm}). It uses exact optimization over the observed prefix for $k\le21$, with $O(n^{21})$ queries, for $22\le k<10^6$ a price policy mixed with the single-choice rule, with at most $n+1$ queries, and for larger $k$ a variant of \AP{}. Since $k=1$ contains the classical secretary problem, $1/e$ cannot be improved as a bound that holds for all $k$ simultaneously. These results are proved in the appendices.

\subsection{Technical overview}\label{sec:overview}

We describe the algorithm by building it up in three steps. Throughout, give each element an independent uniform arrival time in $[0,1]$, and write $\marg{i}{A}=f(A\cup\{i\})-f(A)$.

\paragraph{Step 1: an idealized threshold rule.}
Suppose the algorithm posts a price $\theta_t$ and accepts an arrival $i$ exactly when $\marg{i}{A_t}>\theta_t$, where $A_t$ is the current accepted set. Let $O$ be an optimal set, $F(t)=\E f(A_t)$, and $z_i(t)=\pos{\marg{i}{A_t}-\theta_t}$ the \emph{score} of $i$. Each element of $O$ arrives in $[t,t+dt]$ with probability $dt$, and if accepted it gains its score plus the price. Suppose that we could charge every element of $O$ its current score at rate one. Submodularity gives $\sum_{i\in O}z_i(t)\ge f(A_t\cup O)-f(A_t)-k\theta_t$, and if $f$ is monotone, then $f(A_t\cup O)\ge\OPT$. This yields $F'\ge\OPT-F-k\,\E[\theta_t]+(\text{price revenue})$, and so $F(t)\ge(1-e^{-t})\OPT$ as soon as the price revenue pays for the term $k\,\E[\theta_t]$.

The difficulty is that an element of $O$ produces its gain only once, at its arrival, while the inequality uses the score at time $t$ of every element of $O$, including those that have already arrived, and against a state that these elements may have influenced. We relate the two with a \emph{deletion coupling}: compare the run with the run in which $i$ is absent, and insert $i$ at a uniformly random time (\cref{sec:deletion}).

\paragraph{Step 2: pacing without knowing the optimum.}
For the price revenue to pay for $k\,\E[\theta_t]$, the rate of acceptances must track $k$: acceptances should arrive at about rate $k$ whenever $\theta_t$ is significant. We achieve this with an exponential price in the number of acceptances $j$ made so far, which falls at a controlled rate over time, of the form (up to a slowly varying factor)
\[
 \theta_t \;\approx\; M\exp\bigl(\lambda(j-b)-\kappa t\bigr).
\]
A single unit of acceptance raises the price by a factor $e^\lambda$, and the time decay is calibrated so that the price is an exact potential: the revenue from acceptances equals $k\int\theta$ up to a boundary term (\cref{lem:prices}). Here $M$ is an estimate of the largest singleton value. It is what lets the algorithm work with no knowledge of $\OPT$. An element whose singleton value exceeds $M$ is \emph{exceptional}, and the algorithm raises $M$ and resets its price when such an element is accepted or has nonpositive marginal; there are only $O(\log k)$ such resets in expectation. A normal element has marginal at most $M$, so once the price exceeds $M$ no normal element can be accepted, which gives a pathwise bound on the number of acceptances.

\paragraph{Step 3: eligibility attenuation.}
For a general objective, the bound $f(A_t\cup O)\ge\OPT$ fails, and we must instead rely on the fact that $f(A\cup O)$ is large in expectation whenever every element is unlikely to be in $A$. In particular, if $\Pr[i\in A_t]\le q$ for all $i$, then $\E f(A_t\cup O)\ge(1-q)\OPT$. Elementary thresholding only gives $q\le t$, because an element arriving before $t$ might be accepted with certainty. To do better, we would like $q=1-e^{-t}$, the same bound that drives the measured continuous greedy algorithm offline. However, simply discounting acceptances loses gain, and the loss seems unavoidable: an element accepted less often contributes less.

The solution is to discount each element according to its \emph{own} history. Say $i$ is \emph{eligible} at time $s$ if it would pass the price test against the state at time $s$, and let $\tau_i(t)$ be the total measure of times before $t$ at which it was eligible. We accept an eligible arrival at time $t$ with probability $e^{-\tau_i(t)}$. To see what this achieves, freeze the whole execution without $i$, and insert $i$ at a uniformly random time. Before $i$ arrives, the execution is unaffected by it, so its eligibility function $I$ is determined by the frozen execution, and
\begin{equation}\label{eq:overview-identity}
 \Pr[i\text{ accepted by time }t]
 =\int_0^t I(s)\,e^{-\tau(s)}\,ds
 =1-e^{-\tau(t)}.
\end{equation}
This single identity has two consequences. Since $\tau(t)\le t$, every element is selected by time $t$ with probability at most $1-e^{-t}$, which gives the protection of $f(A_t\cup O)$ that we wanted. And because $e^{-\tau(t)}$ is exactly the probability that $i$ has \emph{not yet} been selected, the attenuated rate of gain from $i$ arriving at time $t$ equals the expected residual score of $i$ at time $t$, exactly matching the term needed in Step~1. Attenuation therefore costs nothing in the gain comparison. Solving the inequality $F'+F\ge e^{-t}\OPT+\cdots$ gives $te^{-t}\OPT$, which at $t\to1$ equals $\OPT/e$, with zero first-order sensitivity at $t=1$; this flatness is why the error term is as small as $O(\log^2 k/k)$. When $f$ is monotone, the very same execution has $f(A_t\cup O)\ge\OPT$ and yields $1-e^{-t}$.

Attenuation can be implemented with a linear number of queries in total, by checking historical eligibility only at the points of a rate-one Poisson process private to each element (\cref{sec:queries}).

\paragraph{Every budget.}
For small budgets, prices are too coarse, since the accounting above loses lower-order terms that are large when $k$ is small. For $k\le29$ we instead select by rank, either of singleton values or of marginal values with a multi-rank rule that is restarted after every stop, and we bound the value by tracking how many elements of a fixed optimum are still unseen. For larger $k$ we use prices given by Poisson tail probabilities, which form an exact potential at every budget, together with eligibility attenuation (\cref{ek:sec}). For monotone objectives at small $k$, exact optimization over the observed prefix with a smaller sample than $1/e$ suffices, and its guarantee reduces to a finite linear program (\cref{mon:sec}).

\paragraph{Hardness.}
The construction in \cref{sec:hardness} uses a coverage objective with sparse latent groups, adapted from the recent work of Udwani and Wajc on online welfare maximization~\cite{UW26}. Each element belongs to a group of $r$ types, and accepting all $r$ elements of a group has value one, while $r$ elements of distinct groups have value only about $1-1/e$. A random sparse subset of all types is present, chosen so that a complete group exists offline with high probability, but such that an element accepted online almost never has a future group mate. An online algorithm thus essentially always holds elements from distinct groups.

For general objectives, \cref{sec:half} uses a pair of elements $a,b$ that have values $K^J$ and $K^{J+1}$ alone and value zero together, with $J$ uniform over a long range. When the first of the two arrives with value $v$, it is equally likely to be the light element, which should be rejected in favor of its partner, or the heavy one, which should be accepted, and no query can distinguish these cases until the partner arrives. Whichever decision is made is right about half of the time. Fixing the histogram of scales across many such pairs makes $\OPT$ a constant, removes any dominant element, and leaves this ambiguity in place at each pair. To go below $1/2$, a small part of the value is moved into matroid gadgets: a matroid whose ground set is the union of a good basis $G$ and a bad basis $B$, valued by $r(S)-|S\cap B|$, in which a random order lets bad elements displace good ones. A bad element that increases the rank of everything seen before it is invisible to every query at its arrival, and so it is rejected only as often as the algorithm rejects elements in general. The bad elements are carried by elements of lower levels of a hierarchy, at harmonically distributed distances, and a spectral estimate for the resulting circulant kernel turns these effects into a bound below $1/2$ (\cref{sec:below-half}).

\paragraph{Beyond $1/e$.}
The analysis of \AP{} for general objectives compares the algorithm with a single optimal set. If two disjoint feasible sets have total value $(1+d)\OPT$, then a variant that compares against both of them at the start gains $\Omega(d)$ over $1/e$. If $d$ is small, then every feasible set disjoint from an optimal set is worth at most $d\,\OPT$, so the elements outside the optimum do little harm; a version of Greedy that trains on a sample, keeps only the arrivals with large marginal against the sample's greedy solution, and paces its acceptances with prices, then obtains about $0.4$, less a term linear in $d$. The two guarantees depend on $d$ in opposite directions, and a fixed mixture beats $1/e$ (\cref{sec:beyond}).

\subsection{Further related work}

The submodular secretary problem has also been studied under matroid and other combinatorial constraints~\cite{BHZ13,GRST10,FZ18,KT17,Joyce26}. Feldman and Zenklusen~\cite{FZ18} reduce matroid constraints with submodular objectives to the linear matroid secretary problem, and, together with the recent resolution of the matroid secretary conjecture~\cite{ABHHMS26}, this yields a constant competitive ratio for every matroid. Agrawal, Shadravan, and Stein~\cite{ASS19}, and later work~\cite{Shadravan20}, study shortlists; \cite{ASS19} also proves a $7/8+o(1)$ upper bound that holds even with a shortlist, which is the previously best known information-theoretic upper bound for the cardinality case that we are aware of. Gupta et al.~\cite{GRST10} observed that no secretary algorithm can beat some universal constant, regardless of computation. In the related single-pass random-order streaming model, where an algorithm may keep and later discard elements in a memory buffer, Liu, Rubinstein, Vondr\'ak, and Zhao~\cite{LRVZ21} obtain $1-1/e-\varepsilon$ for monotone and $1/e-\varepsilon$ for general objectives using $O(k/\varepsilon)$ memory, and show that beating $1-1/e$ requires $\Omega(n)$ memory even with unlimited computation. Their lower bound depends on the memory limit, so it does not apply to secretary algorithms, which may remember every arrival but must decide immediately. Similarly, Rubinstein and Zhao~\cite{RZ22} show that no space-efficient random-order streaming algorithm can beat $1/2$ for asymmetric objectives, while a ratio of $1/2+c$ is possible for symmetric ones. Offline, Feige, Mirrokni, and Vondr\'ak~\cite{FMV11} prove a $1/2$ barrier for unconstrained maximization with polynomially many queries, a bound of the same value as our pairwise barrier but with a different source, namely query complexity. \cref{thm:main} shows that the limiting ratio $1-1/e$ of the shortlist algorithm in \cite{ASS19} can be achieved with no shortlist, and \cref{thm:hardness-intro} improves this upper bound to $1-1/e+o(1)$ when no shortlist is available.

\paragraph{Organization.}
The body of the paper proves the main results. \Cref{sec:prelim} sets up the model and a basic lemma. \Cref{sec:algorithm} presents the algorithm, \cref{sec:analysis} proves its guarantee and gives the linear-query implementation, which together prove \cref{thm:main}. \Cref{sec:beyond} goes beyond $1/e$ for general objectives. \Cref{sec:hardness} proves the hardness theorem for monotone objectives, and \cref{sec:half} proves the barrier $\frac12$ for general objectives and a bound strictly below it. The appendices contain the guarantees at every budget, for general objectives (\cref{ek:sec}) and for monotone objectives (\cref{mon:sec}), and the proof of the constant $0.497053$ (\cref{bh:sec}).

\section{Preliminaries}\label{sec:prelim}

\paragraph{Model and notation.}
The ground set is $E$, with $|E|=n\ge k\ge1$. The function $f:2^E\to\R_{\ge0}$ is submodular: $\marg{e}{A}\ge\marg{e}{B}$ whenever $A\subseteq B$ and $e\notin B$, where $\marg{e}{A}=f(A\cup\{e\})-f(A)$. It is not assumed to be normalized. Write $\OPT=\max_{|S|\le k}f(S)$. All logarithms are natural.

\paragraph{Timestamps.}
It is standard to generate a uniform random order from timestamps. The algorithm draws $n$ independent uniforms on $[0,1]$, sorts them, and assigns the $j$th smallest to the $j$th arrival. Because the arrival order is uniform and independent of these numbers, the timestamps of the fixed elements of $E$ are independent uniforms. Timestamps are almost surely distinct, and we assume this from now on. The algorithm uses the timestamps only of elements that have already arrived.

The following lemma is a form of a well-known sampling bound~\cite{FMV11,BFNS14}. We include the proof because we need the dependence on $\mu$.

\begin{lemma}[Random sets]\label{lem:random-set}
Let $g:2^E\to\R_{\ge0}$ be submodular and let $\mu\in[0,1]$ satisfy $g(X)\ge\mu\,g(\emptyset)$ for all $X$. If $S$ is a random subset of $E$, with arbitrary correlations, and $\Pr[e\in S]\le q$ for each $e$, then
\[
 \E[g(S)]\ge\bigl(1-(1-\mu)q\bigr)\,g(\emptyset).
\]
\end{lemma}

\begin{proof}
Order $E=\{e_1,\dots,e_N\}$ by nonincreasing inclusion probabilities $q_1\ge\dots\ge q_N$, and let $S_i=\{e_1,\dots,e_i\}$. Adding the elements of $S$ in this order and using submodularity gives $g(S)\ge g(\emptyset)+\sum_{i:\,e_i\in S}(g(S_i)-g(S_{i-1}))$. Taking expectations and summing by parts yields
\begin{equation}\label{eq:chain}
 \E[g(S)]\ge(1-q_1)\,g(\emptyset)+\sum_{i=1}^{N-1}(q_i-q_{i+1})\,g(S_i)+q_N\,g(S_N).
\end{equation}
All coefficients are nonnegative and those of the sets $S_1,\dots,S_N$ sum to $q_1$. Each such set has value at least $\mu\,g(\emptyset)$, so $\E[g(S)]\ge(1-(1-\mu)q_1)\,g(\emptyset)\ge(1-(1-\mu)q)\,g(\emptyset)$.
\end{proof}

\section{The algorithm}\label{sec:algorithm}

Fix $k$ large enough that the horizon $T$ below is positive; for smaller $k$ the algorithm returns $\emptyset$, and \cref{thm:main} holds vacuously after enlarging the hidden constants. Define
\begin{gather*}
 L=\ln(k+1),\quad p=\frac1k,\quad b=\ceil{\sqrt k},\quad \lambda=\frac{4L}{b},\quad m=\ceil{200L},\\
 \nu=\frac{e^\lambda-1}{\lambda},\quad \kappa=k(e^\lambda-1),\quad T=\frac{k-(m+1)(b+1)-m}{\nu k},\quad h(t)=e^{t-T}.
\end{gather*}
Note that $T<1$ and $h\le 1$ on $[0,T]$.

\begin{framed}
\noindent\AP. Each element $i$ has an independent uniform timestamp and an independent private uniform $U_i\in[0,1]$.
\begin{enumerate}[leftmargin=*,itemsep=2pt]
\item \emph{Sample.} Reject every element with timestamp below $p$. Let $M$ be the largest singleton value $f(\{e\})$ among them, or $0$ if there are none. Rescale the remaining time to $t\in[0,1]$, via $t=(\text{timestamp}-p)/(1-p)$. Reject all arrivals after time $T$.
\item \emph{State and price.} Maintain the accepted set $A$, the scale $M$, the epoch start $s$ (initially $0$), and the epoch count $j$ (initially $0$). At time $t$, the price is
\[
 \theta_t=\frac{1}{h(t)}\,M\exp\bigl(\lambda(j-b)-\kappa(t-s)\bigr).
\]
\item \emph{Eligibility.} Element $i$ is \emph{exceptional} at time $u$ if $f(\{i\})>M_u$, and otherwise \emph{normal}, where $M_u$ is the scale just before time $u$. With $w=\marg{i}{A_u}$ the marginal against the accepted set just before $u$, an exceptional element is \emph{eligible} at $u$ if $w>0$, and a normal element is \emph{eligible} at $u$ if $w>\theta_u$. Let $I_i(u)\in\{0,1\}$ indicate eligibility, a counterfactual test that makes sense even for times $u$ before $i$ arrives, and let $\tau_i(t)=\int_0^tI_i(u)\,du$.
\item \emph{Decision.} When $i$ arrives at time $t$, accept it if and only if $I_i(t)=1$ and $U_i\le e^{-\tau_i(t)}$.
\item \emph{Update.} If $i$ is normal and accepted, set $j\gets j+1$. If $i$ is exceptional and either accepted or has $w\le0$, then start a new epoch: set $M\gets f(\{i\})$, $s\gets t$, and $j\gets0$. No other arrival changes the state.
\item \emph{Budget.} Once $k$ elements are accepted, reject everything.
\end{enumerate}
\end{framed}

Eligibility at an earlier time $u$ concerns the element $i$ at hand together with the state that actually existed at time $u$, so all queries are on arrived elements. (The state consists of $A_u$, which is a subset of the arrived elements, and three numbers.) Between state changes, $\theta_u$ is nonincreasing in $u$, since $h(u)$ grows and the exponential falls. Hence each $I_i$ is piecewise constant with finitely many pieces and the integral $\tau_i$ is well defined and measurable. An exceptional element with positive marginal that fails its test in step~4 leaves the state unchanged. This is important for the analysis, since a state change caused by a \emph{rejected} element is harmful in general; the only rejected elements that change the state have nonpositive marginal.

\paragraph{The uncapped execution.}
In the analysis we also consider the \emph{uncapped execution}, which is identical except that step~6 is omitted. Let $B$ be its number of acceptances and $D$ its number of epoch resets. Both executions agree until the capped one makes its $k$th acceptance.

\begin{theorem}[Finite guarantee]\label{thm:finite}
Let $f$ be nonnegative, submodular, and $\mu$-monotone, and suppose $T>0$. Then \AP{} accepts at most $k$ elements on every execution, and
\[
 \E[f(A)]\ge\Bigl[(1-p)\Lambda_\mu(T)-\frac{200}{(k+1)^{3}}\Bigr]\OPT,
 \qquad\text{where}\quad \Lambda_\mu(T)=\mu\bigl(1-e^{-T}\bigr)+(1-\mu)\,T\,e^{-T}.
\]
\end{theorem}

\begin{remark}
The constants in \cref{sec:algorithm} are chosen for a simple proof. In particular, $T>0$ requires $k$ of order $10^7$. \Cref{ek:sec,mon:sec} give explicit guarantees at every budget, for general and for monotone objectives.
\end{remark}

\begin{proof}[Proof of \cref{thm:main} from \cref{thm:finite,prop:queries}]
We have $(m+1)(b+1)+m=O(L\sqrt k)$ and $\lambda=O(L/\sqrt k)$, and $\nu\le1+\lambda$ for $\lambda\le1$. It follows that $T>0$ for large $k$ and that $\delta:=1-T=O(L/\sqrt k)$. Since $x\mapsto xe^{-x}$ has derivative zero at $x=1$ and second derivative at most $2$ in absolute value on $[0,1]$, we get $Te^{-T}\ge 1/e-\delta^2$. Also $1-e^{-T}\ge 1-1/e-\delta$. For $\mu=0$ this gives (a), since $\delta^2=O(L^2/k)$ and $p=1/k$. For general $\mu$ we have $\Lambda_\mu(T)\ge 1/e+\mu(1-2/e)-\delta$, proving (b) and (c). \Cref{prop:queries} gives the query bound and adds an error of $e^{-\Omega(k)}$, which is absorbed in the error terms.
\end{proof}

\section{Analysis}\label{sec:analysis}

\Cref{sec:deletion} compares an execution with the execution where one element is removed, and \cref{sec:value} deduces a lower bound on the value. This bound contains a term for prices, which we pay for in \cref{sec:prices}. Finally, \cref{sec:resets} bounds the number of resets, \cref{sec:finish} completes the proof, and \cref{sec:queries} gives the linear-query implementation.

Throughout \cref{sec:deletion,sec:value}, we analyze the uncapped execution conditional on the set $S_0$ of sampled elements. Thus $M_0$, the initial scale, is fixed, and the remaining elements $R=E\setminus S_0$ have independent uniform rescaled timestamps $t_i\in[0,1]$. Let $\E_0$ and $\Pr_0$ denote the conditional expectation and probability. For a time $t$, let $X_t=(A_t,M_t,s_t,j_t)$ denote the state just before time $t$. For any element $i$, define the \emph{score}
\[
 z_i(t)=\pos{\marg{i}{A_t}-\theta_t}.
\]

\subsection{The deletion coupling}\label{sec:deletion}

Fix $i\in R$, and fix all the randomness $\omega_i=(t_j,U_j)_{j\in R\setminus\{i\}}$ other than $(t_i,U_i)$. The \emph{deleted run} is the uncapped execution on $R\setminus\{i\}$ with the same randomness and parameters. We add a superscript ``$-$'' to its quantities: $X_t^-$, $\theta_t^-$, and $z_i^-(t)$, as well as $I_i^-(t)$ and $\tau_i^-(t)$, which test the (absent) element $i$ against the state $X_t^-$. All of these are functions of $\omega_i$ only.

\begin{lemma}[Deletion identities]\label{lem:deletion}
Fix $i\in R$, $\omega_i$, and $t\in[0,T]$, and let $\calE_i(t)$ be the event that $i$ has been accepted before time $t$. Conditional on $\omega_i$:
\begin{enumerate}[label=(\alph*),nosep]
\item If $t_i=u$, then $i$ is eligible at its arrival if and only if $I_i^-(u)=1$, and it passes step 4 if and only if $U_i\le e^{-\tau_i^-(u)}$.
\item $\Pr[\calE_i(t)\mid\omega_i]=1-e^{-\tau_i^-(t)}\le 1-e^{-t}$.
\item $z_i(t)=0$ on $\calE_i(t)$, and $z_i(t)=z_i^-(t)$ outside $\calE_i(t)$. Hence $\E[z_i(t)\mid\omega_i]=e^{-\tau_i^-(t)}\,z_i^-(t)$.
\end{enumerate}
\end{lemma}

\begin{proof}
(a) Before time $t_i$ the two runs process the same arrivals with the same decisions, by induction over arrivals: an element's decision depends only on the earlier state history. Therefore $X_v=X_v^-$ for all $v\le t_i$, which implies $I_i(v)=I_i^-(v)$ for such $v$ and so $\tau_i(t_i)=\tau_i^-(t_i)$.

(b) By (a), $\calE_i(t)$ occurs if and only if $t_i<t$, $I_i^-(t_i)=1$, and $U_i\le e^{-\tau_i^-(t_i)}$. Since $t_i$ and $U_i$ are independent uniforms independent of $\omega_i$, and $\tau_i^-$ is absolutely continuous with derivative $I_i^-$ almost everywhere,
\[
 \Pr[\calE_i(t)\mid\omega_i]
 =\int_0^t I_i^-(u)\,e^{-\tau_i^-(u)}\,du
 =1-e^{-\tau_i^-(t)}.
\]
The inequality uses $\tau_i^-(t)\le t$.

(c) If $i\in A_t$ then $\marg{i}{A_t}=0$ and, as $\theta_t\ge0$, the score vanishes. Suppose $i$ is not accepted before $t$. If $t_i\ge t$, then $X_t=X_t^-$ by (a) and the two scores agree. If $t_i<t$ and the rejection of $i$ did not change the state, then the two runs continue identically, and again $X_t=X_t^-$. The only remaining case is a rejection of $i$ that started a new epoch. In that case $i$ had marginal at most zero at $t_i$ against $A_{t_i}=A_{t_i}^-$. The accepted sets $A_t\supseteq A_{t_i}$ and $A_t^-\supseteq A_{t_i}^-$ only grow, so by submodularity $\marg{i}{A_t}\le0$ and $\marg{i}{A_t^-}\le0$, whence $z_i(t)=z_i^-(t)=0$. Combining this with (b) proves the formula for the conditional expectation.
\end{proof}

\Cref{lem:deletion}(b) also gives the inclusion bound $\Pr_0[i\in A_t]\le1-e^{-t}$ for each $i\in R$. Elements outside $R$ are never accepted.

\subsection{The value inequality}\label{sec:value}

Let $N$ be the counting process of normal acceptances. For an arrival $i$ at time $t_i\le T$, let $\mathrm{gain}_i$ be the increase in $f(A)$ that it causes, which is positive if $i$ is accepted and zero otherwise, and define its \emph{net gain}
\[
 \mathrm{net}_i=\mathrm{gain}_i-\theta_{t_i}\ind[i\text{ is accepted as a normal element}].
\]
Net gains are nonnegative, because normal acceptance requires a marginal larger than $\theta_{t_i}$.

\begin{lemma}\label{lem:net}
Let $O\subseteq R$ be a fixed set. For every measurable $\psi:[0,T]\to\R_{\ge0}$,
\[
 \E_0\Bigl[\sum_{i\in R:\,t_i\le T}\psi(t_i)\,\mathrm{net}_i\Bigr]
 \ge\int_0^T\psi(t)\,\E_0\Bigl[\sum_{i\in O}z_i(t)\Bigr]\,dt.
\]
\end{lemma}

\begin{proof}
Discard the nonnegative terms with $i\notin O$. For $i\in O$, condition on $\omega_i$ and on $t_i=t$. By \cref{lem:deletion}(a), element $i$ meets the state $X^-_t$ upon arrival. If it is a normal element then it is accepted with probability $I_i^-(t)e^{-\tau_i^-(t)}$, and in this case $\mathrm{net}_i=z_i^-(t)$; as $z_i^-(t)>0$ implies eligibility, $\E[\mathrm{net}_i\mid\omega_i,t_i=t]=e^{-\tau_i^-(t)}z_i^-(t)$. If it is exceptional and eligible, then $\E[\mathrm{net}_i\mid\omega_i,t_i=t]=e^{-\tau_i^-(t)}\marg{i}{A_t^-}\ge e^{-\tau_i^-(t)}z_i^-(t)$. If it is exceptional and ineligible, its marginal is nonpositive and both sides vanish. In every case, by \cref{lem:deletion}(c),
\[
 \E[\mathrm{net}_i\mid\omega_i,t_i=t]\ge e^{-\tau_i^-(t)}z_i^-(t)=\E[z_i(t)\mid\omega_i].
\]
Integrating over $t_i$, which has density $1$, and then over $\omega_i$ proves the claim.
\end{proof}

We can now derive the main inequality. Note how the integrating factor removes all dependence on the unknown process $f(A_t)$. Recall that $h(t)=e^{t-T}$, so $h'=h$ and $h(T)=1$.

\begin{lemma}[Value inequality]\label{lem:value}
Let $O\subseteq R$ with $|O|\le k$. If $f$ is $\mu$-monotone, then the uncapped execution satisfies
\[
 \E_0[f(A_T)]\ge\Lambda_\mu(T)\,f(O)+\E_0\Bigl[\int_0^Th(t)\theta_t\,dN(t)-k\int_0^Th(t)\theta_t\,dt\Bigr].
\]
\end{lemma}

\begin{proof}
Since $f(A_t)$ is piecewise constant, integrating by parts gives
\[
 \sum_{i:\,t_i\le T}h(t_i)\,\mathrm{gain}_i=f(A_T)-h(0)f(\emptyset)-\int_0^Th(t)f(A_t)\,dt.
\]
Apply \cref{lem:net} with $\psi=h$ and write $\pi(t)=\E_0[\theta_t]$. The left side of \cref{lem:net} equals the expectation of the display minus $\E_0\int h\theta\,dN$. For the right side, submodularity and $\theta_t\ge 0$ give, pointwise,
\[
 \sum_{i\in O}z_i(t)\ge\sum_{i\in O}\marg{i}{A_t}-k\theta_t\ge f(A_t\cup O)-f(A_t)-k\theta_t.
\]
Substituting and cancelling $\int h\,\E_0 f(A_t)\,dt$ from both sides, and dropping the nonnegative term $h(0)f(\emptyset)$, we obtain
\[
 \E_0[f(A_T)]\ge\int_0^Th(t)\,\E_0[f(A_t\cup O)]\,dt+\E_0\int_0^Th\theta\,dN-k\int_0^Th(t)\pi(t)\,dt.
\]
The function $g(S)=f(O\cup S)$ is nonnegative and submodular, with $g(\emptyset)=f(O)$, and $\mu$-monotonicity of $f$ gives $g(X)\ge\mu\,g(\emptyset)$. By \cref{lem:random-set} with $q=1-e^{-t}$ and the inclusion bound,
\[
 \E_0[f(A_t\cup O)]\ge\bigl(1-(1-\mu)(1-e^{-t})\bigr)f(O)=\bigl(\mu+(1-\mu)e^{-t}\bigr)f(O).
\]
Finally $\int_0^Te^{t-T}\bigl(\mu+(1-\mu)e^{-t}\bigr)\,dt=\mu(1-e^{-T})+(1-\mu)Te^{-T}=\Lambda_\mu(T)$.
\end{proof}

\subsection{Paying for prices and bounding acceptances}\label{sec:prices}

The next lemma holds on every sample path of the uncapped execution.

\begin{lemma}[Prices and counts]\label{lem:prices}
On every execution, $M_t\le\OPT$ for all $t$, and:
\begin{enumerate}[label=(\alph*),nosep]
\item $\displaystyle\int_0^Th\theta\,dN-k\int_0^Th\theta\,dt\ \ge\ -(D+1)\,\frac{e^{-\lambda b}}{e^\lambda-1}\,\OPT$;
\item $B\le k+(b+2)(D-m)$.
\end{enumerate}
\end{lemma}

\begin{proof}
The scale is either zero or a singleton value, so $M_t\le\OPT$. Write $\psi_t=h(t)\theta_t=M\exp(\lambda(j-b)-\kappa(t-s))$ and $\Psi_t=\psi_t/(e^\lambda-1)$.

(a) Within one epoch, $\Psi$ decreases continuously at rate $\kappa\Psi=k\psi$, and a normal acceptance multiplies $\Psi$ by $e^\lambda$, raising it by exactly $(e^\lambda-1)\Psi=\psi$. Thus over one epoch, $\Psi_{\mathrm{end}}-\Psi_{\mathrm{start}}$ equals the contribution of that epoch to $\int\psi\,dN-k\int\psi\,dt$. As $\Psi_{\mathrm{end}}\ge0$ and $\Psi_{\mathrm{start}}=Me^{-\lambda b}/(e^\lambda-1)\le\OPT e^{-\lambda b}/(e^\lambda-1)$, summing over the $D+1$ epochs gives (a).

(b) A normal element $i$ has $\marg{i}{A}\le f(\{i\})-f(\emptyset)\le M$. Normal acceptance therefore forces $\theta_t<M$, and in particular $M>0$. Since $h\le 1$, this gives $\psi_t\le\theta_t<M$, hence $\lambda(j-b)<\kappa(t-s)$, that is, $j<b+\nu k(t-s)$, using $\kappa/\lambda=\nu k$. Thus an epoch of length $\ell$ contains at most $\nu k\ell+b+1$ normal acceptances. Summing over the $D+1$ epochs and adding at most $D$ exceptional acceptances, each of which starts an epoch,
\begin{align*}
 B&\le\nu kT+(D+1)(b+1)+D\\
 &=k-(m+1)(b+1)-m+(D+1)(b+1)+D=k+(b+2)(D-m).\qedhere
\end{align*}
\end{proof}

\subsection{The number of resets}\label{sec:resets}

Attenuation makes the reset count subtle: a large element that fails its test does not update the scale, so resets are not dominated by ordinary records. The following argument gets around this.

\begin{lemma}[Tail bound]\label{lem:resets}
For every integer $j\ge1$, $\Pr[D\ge j]\le e^{-j/(8e)}+k\,2^{-j/(2e)}$. Consequently $\E[(D-m)_+]\le32(k+1)^{-9}$ and $\E[D]\le m+32$.
\end{lemma}

\begin{proof}
Work on the original time scale, and fix a tie-breaking order on singleton values so that all elements have distinct ranks. Say an element is \emph{marked} if $U_i\le1/e$; include sampled elements, for which $U_i$ is otherwise unused. Marks are independent of timestamps, and each element is marked with probability $q=1/e$.

A marked eligible element always passes step 4, because $\tau_i(t)\le t\le1$. It follows by induction that at every moment $M$ is at least the singleton value of every marked element that has arrived. This is true for the sample, and a marked arrival whose singleton value exceeds $M$ is exceptional: it is accepted if it has positive marginal, and it starts a new epoch either way. So $M$ rises to its singleton value.

Call a post-sample arrival a \emph{candidate} if its rank beats that of every earlier marked element, whether or not we have stopped, and let $C$ be the number of candidates in the stream. A reset element has singleton value strictly above $M$, so it is a candidate, and $D\le C$.

Whether an arrival is a candidate is determined by the timestamps and by the marks of \emph{earlier} elements. Therefore, revealing elements in order of arrival, the marks of successive candidates are independent Bernoulli$(q)$ variables. (Pad the sequence with fresh draws, if needed.) Let $R'$ be the number of marked candidates. If $C\ge j$, then either fewer than $qj/2$ of the first $j$ candidates are marked, which has probability at most $e^{-qj/8}$ by the Chernoff bound, or $R'\ge qj/2$.

Marked candidates are exactly the records after time $p$ in the subsequence of marked elements. Condition on there being $\ell$ marked elements. Their relative order is a uniform permutation, independent of the number $S\sim\mathrm{Bin}(\ell,p)$ of them in the sample. So the record indicators are independent, with the $x$th position being a record with probability $1/x$, and
\[
 \E[2^{R'}\mid\ell,S]=\prod_{x=S+1}^{\ell}\Bigl(1+\frac1x\Bigr)=\frac{\ell+1}{S+1},
 \qquad
 \E[2^{R'}\mid\ell]=\frac{1-(1-p)^{\ell+1}}{p}\le\frac1p=k.
\]
By Markov's inequality $\Pr[R'\ge qj/2]\le k\,2^{-qj/2}$, proving the tail bound.

For the consequences, $\E[(D-m)_+]=\sum_{j>m}\Pr[D\ge j]$. Let $x=e^{-1/(8e)}$ and $y=2^{-1/(2e)}$, so that $1/(1-x)\le23$, $1/(1-y)\le9$, and, using $m\ge200L$, we have $x^m\le(k+1)^{-9}$ and $y^m\le(k+1)^{-25}$. Summing the geometric series gives $\E[(D-m)_+]\le23(k+1)^{-9}+9k(k+1)^{-25}\le32(k+1)^{-9}$. Finally, $\E[D]\le m+\E[(D-m)_+]$.
\end{proof}

\subsection{Completing the proof of \texorpdfstring{\cref{thm:finite}}{Theorem 3.1}}\label{sec:finish}

The accepted set has size at most $k$ by construction. Let $O^*$ be an optimal set and let $O=O^*\setminus S_0$, which is a feasible subset of $R$. Taking expectations over the sample in \cref{lem:value} and applying \cref{lem:prices}(a) gives, for the uncapped set $A_T$,
\[
 \E[f(A_T)]\ge\Lambda_\mu(T)\,\E[f(O^*\setminus S_0)]-(\E[D]+1)\,\frac{e^{-\lambda b}}{e^\lambda-1}\,\OPT.
\]
\Cref{lem:random-set}, applied to $g(S)=f(O^*\setminus S)$ with $\mu=0$ and $q=p$, gives $\E[f(O^*\setminus S_0)]\ge(1-p)\OPT$.

Now let $A$ be the output. The capped execution matches the uncapped one up to the capped execution's $k$th acceptance, so $A\subseteq A_T$ and $f(A_T)-f(A)$ is a sum of at most $(B-k)_+$ gains. Each gain is at most the singleton value of the element, and thus at most $\OPT$. By \cref{lem:prices}(b), $\E[f(A_T)]-\E[f(A)]\le(b+2)\E[(D-m)_+]\,\OPT$.

It remains to bound the errors. Since $\lambda b=4L$ and $e^\lambda-1\ge\lambda=4L/b$, we get
\[
 (\E[D]+1)\,\frac{e^{-\lambda b}}{e^\lambda-1}\le\frac{(200L+34)\,b}{4L\,(k+1)^4}\le63\,(k+1)^{-3},
\]
using $L\ge\ln2$ and $b\le k+1$. Also $(b+2)\E[(D-m)_+]\le3(k+1)\cdot32(k+1)^{-9}\le96(k+1)^{-8}$. Together these are at most $200(k+1)^{-3}$, completing the proof.\hfill\qedsymbol

\begin{remark}
Two features of the proof deserve mention. First, nothing is lost in the price accounting to first order, because the exponential potential of \cref{lem:prices} is exact. Apart from the factor $1-p$ and lower-order reset terms, the error comes from $T<1$, and when $\mu=0$ that error is second-order. Second, the monotone case is not a separate analysis: the only property used in the proof is $g(X)\ge\mu\,g(\emptyset)$ in \cref{lem:value}. For monotone objectives, attenuation is unnecessary (\cref{sec:deterministic}).
\end{remark}

\subsection{Linear query complexity}\label{sec:queries}

As described in \cref{sec:algorithm}, the algorithm computes $\tau_i(t)$ exactly. Since the accepted set takes at most $k+1$ distinct values, and eligibility on an interval where the accepted set is fixed requires only its one marginal value, this takes $O(nk)$ value queries in total. We can instead use a randomized test with the same acceptance probability, as follows.

Give each element $i$ a private rate-one Poisson process $P_i$ on $[0,1]$. When $i$ arrives at time $t$ and is eligible, accept it if and only if no point $u\in P_i\cap[0,t)$ has $I_i(u)=1$. The algorithm stores the accepted elements in order, together with their acceptance times, the cached values $f(A_u)$, and the history of scales and epochs, so a test at $u$ costs one query, namely $f(A_u\cup\{i\})$. Finally, stop the algorithm permanently before it would use a $(2n+1)$th probe.

\begin{proposition}\label{prop:queries}
The algorithm just described makes at most $4n+1$ value queries, all on subsets of arrived elements, and accepts at most $k$ elements. For $\mu$-monotone $f$ and $T>0$, it satisfies
\[
 \E[f(A)]\ge\Bigl[(1-p)\Lambda_\mu(T)-\frac{200}{(k+1)^{3}}-e^{-(2\ln2-1)n}\Bigr]\OPT.
\]
\end{proposition}

\begin{proof}
Ignore the probe limit at first. By the Poisson void probability, conditional on the state history, the test of $i$ passes with probability $e^{-\tau_i(t)}$, and $P_i$ is independent of all other randomness. By induction over arrivals, the execution has the same distribution as that with uniforms $U_i$, so \cref{thm:finite} applies.

The processes together have a $\mathrm{Poisson}(n)$ number of points, which exceeds $2n$ with probability at most $e^{-(2\ln2-1)n}$ by the Chernoff bound. Only on this event does the limit change the output, and it then loses at most $\OPT$, because the output without the limit is feasible. The query count is one for $f(\emptyset)$; one singleton query and one current marginal query per element; and at most $2n$ probe queries.
\end{proof}

\section{Beyond \texorpdfstring{$1/e$}{1/e} for general objectives}\label{sec:beyond}

The guarantee $1/e$ of \cref{thm:main}(a) is not the limit for general objectives. In this section we prove the following.

\begin{theorem}\label{thm:beyond}
There is a randomized algorithm that knows only $n$ and $k$, accepts at most $k$ elements on every execution, makes polynomially many value queries, each on a subset of the elements that have arrived, and satisfies
\[
 \E[f(A)]\ge\Bigl(\frac1e+3.5\cdot10^{-6}-o(1)\Bigr)\OPT
\]
for every nonnegative submodular $f$, where $o(1)$ tends to zero as $k\to\infty$, uniformly in $n\ge k$ and $f$.
\end{theorem}

The gain is tiny, and we make no attempt to optimize it; the point is that $1/e$ is not the answer. The algorithm flips a biased coin and runs one of two algorithms. Their guarantees depend on one parameter of the instance, which neither algorithm computes. Let $W=\OPT>0$, and define the \emph{excess}
\[
 d=\frac1W\max\bigl\{f(X)+f(Y):X\cap Y=\emptyset,\ |X|,|Y|\le k\bigr\}-1.
\]
Taking $X$ optimal and $Y=\emptyset$ shows $d\ge0$, and $d\le1$ since each term is at most $W$. If $d$ is large, there are two disjoint sets of nearly optimal value, and a variant of \AP{} can compare its progress against both of them at the start, which yields more than $1/e$. If $d$ is small, then every feasible set disjoint from an optimal set is worth little, so elements outside the optimum do little harm, and a rule close to Greedy, restricted to a pool of promising elements, does much better than $1/e$. A suitable mixture of the two cancels the dependence on $d$.

\subsection{Two comparators}\label{sec:two-comparators}

Fix a switch time $\zeta\in[0,1/2)$ and let $\chi(t)=1+\ind[t<\zeta]$. The algorithm $\AP_\zeta$ is \AP{} with three changes, which make it pay for two comparators before time $\zeta$:
\begin{itemize}[nosep]
\item the horizon is $T_\zeta=T-\zeta$, where $T$ is as in \cref{sec:algorithm};
\item the integrating factor is $h(t)=\exp\bigl(-\int_t^{T_\zeta}\chi(v)\,dv\bigr)$;
\item the price within an epoch that started at time $s$ is $\theta_t=h(t)^{-1}M\exp\bigl(\lambda(j-b)-\kappa\int_s^t\chi(v)\,dv\bigr)$.
\end{itemize}
For $\zeta=0$ this is \AP. Since $h\le1$ and $h$ increases, prices are still nonincreasing between state changes.

\begin{lemma}\label{lem:two-comparators}
Let $\zeta\in[0,1/2)$, and suppose that $T_\zeta>\zeta$. For every nonnegative submodular $f$ with excess $d$, $\AP_\zeta$ accepts at most $k$ elements and
\[
 \E[f(A)]\ge\Bigl[(1-p)\,e^{-T_\zeta}\bigl((1+d)(1-e^{-\zeta})+T_\zeta-\zeta\bigr)-\frac{200}{(k+1)^3}\Bigr]W.
\]
In particular, $\E[f(A)]\ge(h_\zeta+u_\zeta d-o(1))W$, where $h_\zeta=((2-2\zeta)e^\zeta-1)/e$ and $u_\zeta=(e^\zeta-1)/e$.
\end{lemma}

\begin{proof}
We follow \cref{sec:analysis}. \Cref{lem:deletion} and \cref{lem:net} hold without change, since they use only the structure of the decisions. Fix disjoint $X,Y$ with $|X|,|Y|\le k$ and $f(X)+f(Y)=(1+d)W$, and an optimal set $O^*$, and let $X'$, $Y'$, and $O$ be their nonsampled parts. Apply \cref{lem:net} with the set $X'\cup Y'$ and the weight $\psi=h\,\ind[t<\zeta]$, and again with the set $O$ and the weight $\psi=h\,\ind[t\ge\zeta]$, and add the two inequalities. Before time $\zeta$, submodularity gives $\sum_{i\in X'\cup Y'}z_i(t)\ge f(A_t\cup X')+f(A_t\cup Y')-2f(A_t)-2k\theta_t$, and after $\zeta$ the bound is as in \cref{lem:value}. Since $h'=\chi h$, the terms $-\chi(t)h(t)\E_0f(A_t)$ cancel exactly as before, and with \cref{lem:random-set} we get
\[
 \E_0[f(A_{T_\zeta})]\ge\int_0^{\zeta}h(t)e^{-t}\bigl(f(X')+f(Y')\bigr)\,dt+\int_\zeta^{T_\zeta}h(t)e^{-t}f(O)\,dt+\E_0\Bigl[\int h\theta\,dN-k\int\chi h\theta\,dt\Bigr].
\]
Here $h(t)e^{-t}=e^{t-\zeta-T_\zeta}$ for $t<\zeta$ and $h(t)e^{-t}=e^{-T_\zeta}$ for $t\ge\zeta$, so the two integrals equal $e^{-T_\zeta}(1-e^{-\zeta})(f(X')+f(Y'))$ and $e^{-T_\zeta}(T_\zeta-\zeta)f(O)$. In \cref{lem:prices}, the potential $\Psi$ now decreases at rate $\kappa\chi\Psi=k\chi\psi$, so part~(a) holds with $k\int\chi h\theta\,dt$ in place of $k\int h\theta\,dt$, and in part~(b) an epoch $[s,s')$ contains at most $\nu k\int_s^{s'}\chi+b+1$ normal acceptances. As $\int_0^{T_\zeta}\chi=T_\zeta+\zeta=T$, the count bound $B\le k+(b+2)(D-m)$ is unchanged, and so are \cref{lem:resets} and the error terms of \cref{sec:finish}. Finally, \cref{lem:random-set} applied to $S\mapsto f(X\setminus S)$, and likewise for $Y$ and $O^*$, gives $\E[f(X')+f(Y')]\ge(1-p)(1+d)W$ and $\E f(O)\ge(1-p)W$. This proves the bound. As $k\to\infty$ we have $p\to0$ and $T_\zeta\to1-\zeta$, and $e^{\zeta-1}\bigl((1+d)(1-e^{-\zeta})+1-2\zeta\bigr)=h_\zeta+u_\zeta d$.
\end{proof}

For $\zeta=0$ this is $1/e$, and for small $\zeta>0$ we have $h_\zeta=\frac1e(1-\zeta^2+O(\zeta^3))$ and $u_\zeta\ge\zeta/e$, so $\AP_\zeta$ beats $1/e$ by a first-order amount whenever $d$ is not tiny.

\subsection{Structure when the excess is small}

Fix an optimal set $O$, so that $|O|\le k$ and $f(O)=W$, and order its elements as $o_1,o_2,\dots$. Let $v_{o_i}=\marg{o_i}{\{o_1,\dots,o_{i-1}\}}$ and $v(J)=\sum_{o\in J}v_o$ for $J\subseteq O$.

\begin{lemma}\label{lem:excess}
The weights $v_o$ are nonnegative, $v(O)=W-f(\emptyset)$, and:
\begin{enumerate}[label=(\alph*),nosep]
\item $f(J)\ge f(\emptyset)+v(J)$ for all $J\subseteq O$, and $f(J)\le f(J')$ whenever $J\subseteq J'\subseteq O$;
\item $f(X)\le v(X\cap O)+dW$ for every $X$ with $|X|\le k$;
\item $f(X)\le\max\{1,|X|/k\}\,dW$ for every $X$ disjoint from $O$.
\end{enumerate}
\end{lemma}

\begin{proof}
For $J\subseteq O$ and $o\in O\setminus J$, submodularity and the optimality of $O$ give $\marg oJ\ge\marg o{O\setminus\{o\}}=W-f(O\setminus\{o\})\ge0$. This proves nonnegativity and the monotonicity in~(a), and the bound $f(J)-f(\emptyset)\ge v(J)$ follows by adding the elements of $J$ in the order of $O$ and using submodularity. For~(b), the sets $X$ and $O\setminus X$ are disjoint and have at most $k$ elements, so $f(X)+f(O\setminus X)\le(1+d)W$, while by~(a) $f(O\setminus X)\ge f(\emptyset)+v(O)-v(X\cap O)=W-v(X\cap O)$. For~(c), if $|X|\le k$ then~(b) gives $f(X)\le dW$. If $|X|>k$, let $R$ be a uniformly random $k$-subset of $X$. \Cref{lem:random-set}, applied to $Z\mapsto f(X\setminus Z)$ with the random set $X\setminus R$, gives $\E f(R)\ge\frac k{|X|}f(X)$, while $f(R)\le dW$.
\end{proof}

\subsection{A pruned Greedy}\label{sec:pruned}

Fix parameters $\eta\in(0,1/2)$, $\rho>1-\eta$, $t_2\in(0,1)$, $t_1\in(0,c)$ where $c=1-\sqrt{1-t_2^2}$, and $\sigma>1$, such that $\rho t_1+\sigma(t_2-t_1)<1$. Let
\[
 g=t_2-c,\qquad K=\max\Bigl\{1,\frac{(1-\eta)^2}{\rho\eta}\Bigr\},\qquad \omega=k^{-1/4},\qquad b_0=\ceil{\sqrt k},\qquad \lambda_0=\frac{\ln(k+1)}{b_0},
\]
and, for a scale $M\ge0$, let $\Phi(z)=\frac Mk\bigl(\frac{e^{\lambda_0z}-1}{\lambda_0}-z\bigr)$ for $z\ge0$, so that $\Phi'(z)=\frac Mk(e^{\lambda_0z}-1)$. Fix a total order on $E$, used to break ties.

\begin{framed}
\noindent\textsc{PrunedGreedy}. Each element has an independent uniform timestamp.
\begin{enumerate}[leftmargin=*,itemsep=2pt]
\item \emph{Training.} Reject every element with timestamp below $\eta$, and let $S$ be the set of these elements and $M$ the largest singleton value among them, or $0$ if there are none. Starting from $B=\emptyset$, repeatedly add to $B$ the element of $S\setminus B$ with the largest marginal $\marg iB$, ties broken by the fixed order, as long as this marginal is positive and $|B|<k$. The set $B$ is not accepted.
\item \emph{Validation.} The \emph{score} of an element $i\notin S$ is $\marg iB$, and its \emph{key} is the pair formed by its score and its position in the fixed order, compared lexicographically. Reject every element with timestamp in $[\eta,\eta')$, where $\eta'=\eta+\omega(1-\eta)$, and let $V$ be the set of these elements. Let $m_V=\floor{\omega\rho k}$. If $V$ contains at least $m_V$ elements of positive score, let $\hat\kappa$ be the $m_V$th largest key among them; otherwise let $\hat\kappa$ be smaller than every key. An element with timestamp after $\eta'$ is in the \emph{pool} $P$ if its score is positive and its key exceeds $\hat\kappa$.
\item \emph{Live phase.} Rescale the time after $\eta'$ to $x\in[0,1]$. Reject every arrival outside the pool. A pool element arriving at time $x$, with $w=\marg iA$ against the current accepted set $A$, is accepted as follows. If $x\le t_1$, accept it if and only if $w>0$. If $t_1<x\le t_2$ and $f(\{i\})>M$, accept it if and only if $w>0$. If $t_1<x\le t_2$ and $f(\{i\})\le M$, accept it if and only if $w>\Phi'(Q_{x^-})$, and then increase $Q$ by one. Here the queue $Q$ starts at $0$ at time $t_1$ and decreases at rate $\sigma k$ while it is positive. Reject all arrivals after time $t_2$, and all arrivals once $k$ elements have been accepted.
\end{enumerate}
\end{framed}

The training step uses $O(nk)$ queries on subsets of $S$, and every later query is on $B\cup\{i\}$ or $A\cup\{i\}$ for an arrived element $i$. Elements with $f(\{i\})>M$ are \emph{exceptional}, and the others are \emph{normal}. The algorithm never needs $O$, $W$, or $d$.

\begin{lemma}\label{lem:retention}
Let $D=O\cap P$. Then $\E f(D)\ge(1-2\eta-(1+K)d-o(1))W$. Moreover, with probability $1-o(1)$, $|P|\le(\rho+o(1))k$ and fewer than $\sqrt k$ elements outside $S$ are exceptional. Here, and in the rest of this section, $o(1)$ tends to zero as $k\to\infty$ uniformly in $n$ and $f$.
\end{lemma}

\begin{proof}
\emph{The pool.} List the elements outside $S$ with positive score in decreasing order of key, and let $J_R$ consist of the first $R$ of them, or all of them if there are fewer. Let $R_\pm=\floor{(\rho\pm k^{-1/8})k}$. Given $S$, which determines all scores, each element outside $S$ lies in $V$ independently with probability $\omega$. The number of elements of $J_{R_-}$ in $V$ is binomial with mean at most $\omega R_-\le m_V-k^{5/8}+1$ and variance at most $\omega\rho k=O(k^{3/4})$, so by Chernoff's bound it is below $m_V$ with probability $1-o(1)$; in that case every element of $J_{R_-}$ has key above $\hat\kappa$, and so $J_{R_-}\setminus V\subseteq P$. In the same way, if there are at least $R_+$ positive scores then $J_{R_+}$ contains at least $m_V$ elements of $V$ with probability $1-o(1)$, so that $P\subseteq J_{R_+}$; and if there are fewer, then $P\subseteq J_{R_+}$ trivially. So with probability $1-o(1)$ we have $J_{R_-}\setminus V\subseteq P\subseteq J_{R_+}$, and in particular $|P|\le(\rho+o(1))k$. For the exceptional elements, rank all elements by singleton value, with ties broken by the fixed order; the exceptional elements outside $S$ are those ranked above every element of $S$, and there are at least $l$ of them with probability at most $(1-\eta)^l$.

\emph{Reduction to a fixed pool size.} Let $J=J_{R_-}$, let $C_0=O\setminus S$, and let $\tau$ be the largest score of a positive-score element outside $S\cup J$, or $0$ if there is none. We will show that
\begin{equation}\label{eq:keep}
 \E f(C_0\cap J)\ge\bigl(1-2\eta-(1+K)d-o(1)\bigr)W.
\end{equation}
This implies the lemma. Indeed, on the event $J\setminus V\subseteq P$ we have $(C_0\cap J)\setminus V\subseteq D$, and so $f(D)\ge f((C_0\cap J)\setminus V)$ by \cref{lem:excess}(a). Given $S$, the set $V$ removes each element of $C_0\cap J$ independently with probability $\omega$, so \cref{lem:random-set} gives $\E f((C_0\cap J)\setminus V)\ge(1-\omega)\E f(C_0\cap J)$; and off the event, $f(D)\ge0$ while $f((C_0\cap J)\setminus V)\le W$.

\emph{Step 1: $\E f(C_0\cup B)\ge(1-2\eta)W+\E v(B\cap O)$.} Since $B\subseteq S$, an element of $O$ lies in $C_0\cup B$ with probability $x_o=1-\eta+\Pr[o\in B]\ge1-\eta$, and an element outside $O$ with probability at most $\eta\le1-\eta$. Apply \eqref{eq:chain} with an order in which the elements of $O$ come first. The terms for sets strictly containing $O$ are nonnegative, and for the sets $S_i\subseteq O$ we use $f(S_i)\ge f(\emptyset)+v(S_i)$. Summing by parts again, with $x'\le\eta$ the next inclusion probability after those of $O$,
\[
 \E f(C_0\cup B)\ge(1-x')f(\emptyset)+\sum_{o\in O}v_o(x_o-x')\ge(1-\eta)f(\emptyset)+(1-2\eta)v(O)+\E v(B\cap O),
\]
which is at least $(1-2\eta)W+\E v(B\cap O)$.

\emph{Step 2.} Pointwise, $f(C_0\cap J)\ge f(C_0\cup B)-f(B)-|C_0\setminus J|\,\tau$. Indeed, by submodularity, $f(C_0\cup B)\le f((C_0\cap J)\cup B)+\sum_{i\in C_0\setminus J}\marg iB$, each score in the sum is at most $\tau$, and $f((C_0\cap J)\cup B)\le f(C_0\cap J)+f(B)$ because $C_0\cap B=\emptyset$ and $f(\emptyset)\ge0$.

\emph{Step 3.} By Step 1 and \cref{lem:excess}(b), $\E[f(C_0\cup B)-f(B)]\ge(1-2\eta-d)W$.

\emph{Step 4: with probability $1-o(1)$, $|C_0\setminus J|\,\tau\le(K+o(1))\,dW$.} We generate $S$ by independent coins: $i\in S$ if and only if the coin of $i$ shows heads, which has probability $\eta$. Consider the following \emph{scan}. It maintains a set $H$ of heads, initially empty. At each step, among the elements whose coin has not been revealed, it takes the one with the largest key with respect to $H$, that is, the largest pair of marginal $\marg iH$ and position in the fixed order; if this marginal is not positive it stops, and otherwise it reveals the coin, adding the element to $H$ on heads. It also stops at the $(k+1)$st head. Every element passed over is outside $S$, so the heads are exactly the successive choices of the training step; in particular the first $|B|$ heads form $B$, and if the training step stopped with $|B|<k$ then the scan finds no further head. Since $\marg iH$ only decreases as $H$ grows and the remaining set shrinks, the marginals of the processed elements, at the times they are processed, are nonincreasing. The choice of the next element depends only on coins already revealed, so the coins of the processed elements outside $O$, in the order processed and padded with fresh coins, are independent with heads probability $\eta$. Let $N_0=\ceil{3(k+1)/\eta}$. By Hoeffding's inequality and a union bound, with probability $1-o(1)$, for every $u\le N_0$ the first $u$ of these coins contain at least $\eta u-k^{2/3}$ heads, and the first $N_0$ contain at least $k+2$ heads. Hence every prefix of the scan with at most $k+1$ heads outside $O$ contains at most $\frac{1-\eta}\eta h+\frac{k^{2/3}}\eta$ tails outside $O$, where $h$ is its number of heads outside $O$. Also, with probability $1-o(1)$, $\bigl||O\cap S|-\eta|O|\bigr|\le k^{2/3}$, and so $|C_0|\le(1-\eta)k+k^{2/3}$. Finally, the elements of $B\setminus O$, in the order chosen, have gains summing to at most $f(B\setminus O)-f(\emptyset)\le dW$, by submodularity and \cref{lem:excess}(c).

Assume these events and $\tau>0$, and let $y$ be the element outside $S\cup J$ with score $\tau$. If some element of $S\setminus B$ has marginal at least $\tau$ with respect to $B$, then the training step stopped with $|B|=k$, and every element of $B$ had marginal at least $\tau$ when it was chosen. At least $k-|O\cap S|\ge(1-\eta)k-k^{2/3}$ of these are outside $O$, so $\tau\le dW/((1-\eta)k-k^{2/3})$ and $|C_0\setminus J|\tau\le|C_0|\tau\le(1+o(1))dW$. Otherwise every element of $S\setminus B$ has marginal less than $\tau$ with respect to $B$. Consider the prefix of the scan that ends when the largest remaining key with respect to the current $H$ falls below the key of $y$. It contains no head beyond $B$, and it processes every element of $J$, as these have keys above that of $y$ with respect to every $H\subseteq B$; they are tails, since $J\cap S=\emptyset$. Every head in it had marginal at least $\tau$. We may assume that $|J|=R_-$, as otherwise $\tau=0$. Let $z=|C_0\cap J|/k$. The prefix contains at least $|J\setminus O|=R_--zk$ tails outside $O$, hence at least $\frac\eta{1-\eta}(R_--zk-k^{2/3}/\eta)$ heads outside $O$, each with gain at least $\tau$, and so
\[
 |C_0\setminus J|\,\tau\le\frac{(1-\eta)\bigl((1-\eta-z)k+k^{2/3}\bigr)}{\eta\,(R_--zk-k^{2/3}/\eta)}\,dW\le\Bigl(\frac{(1-\eta)^2}{\eta\rho}+o(1)\Bigr)dW,
\]
because $R_-/k\to\rho>1-\eta$ and the fraction is largest at $z=0$.

\emph{Conclusion.} Let $\mathcal G$ be the intersection of the events of Step~4. By Step~2, $f(C_0\cap J)\ge\bigl(f(C_0\cup B)-f(B)-|C_0\setminus J|\tau\bigr)\ind_{\mathcal G}$. Since $0\le f(C_0\cup B)\le f(C_0)+f(B)\le2W$, Steps~3 and~4 give \eqref{eq:keep}.
\end{proof}

The live phase compares its progress against the multilinear extension of $f$ on the pool. For a finite set $P$ and $D\subseteq P$, let $F(u,t)=\E f(D(u)\cup U(t))$, where $D(u)$ contains each element of $D$ independently with probability $u$ and $U(t)$ contains each element of $P\setminus D$ independently with probability $t$.

\begin{lemma}\label{lem:dense}
For every nonnegative submodular $f$, with $c=1-\sqrt{1-t_2^2}$ and $g=t_2-c$,
\[
 F(c,c)+\int_c^{t_2}\partial_1F(t,t)\,dt\ge g\,f(D).
\]
\end{lemma}

\begin{proof}
Write $W_D=f(D)$. By submodularity, $F$ is concave in each argument separately, and by nonnegativity $F(u,0)\ge uW_D$ and $F(1,t)\ge(1-t)W_D$. Let $L(u,t)=uW_D-F(u,t)$, which is convex in each argument, and let $z(t)=L(t,t)/t$ and $\nu(t)=\partial_1L(t,t)=W_D-\partial_1F(t,t)$. Convexity in the second argument and $L(t,0)\le0$ give $L(t,t)\le t\,\partial_2L(t,t)$, which means $z'\ge\nu/t$. Convexity in the first argument and $L(1,t)\le tW_D$ give $\nu\le t(W_D-z)/(1-t)$. Let $H(t)=(t_2^2-t^2)/(2(1-t))$. On $[c,t_2]$ we have $0\le H\le t$, $H(c)=c$, $H(t_2)=0$, and $H'=(H-t)/(1-t)$, and hence
\[
 \nu+\bigl(H(W_D-z)\bigr)'\le\nu\Bigl(1-\frac Ht\Bigr)+\frac{(H-t)(W_D-z)}{1-t}=\Bigl(1-\frac Ht\Bigr)\Bigl(\nu-\frac{t(W_D-z)}{1-t}\Bigr)\le0.
\]
Integrating over $[c,t_2]$ gives $\int_c^{t_2}\nu\le c(W_D-z(c))=F(c,c)$, that is, $F(c,c)+\int_c^{t_2}\partial_1F(t,t)\,dt\ge(t_2-c)W_D$.
\end{proof}

\begin{lemma}\label{lem:pruned}
\textsc{PrunedGreedy} accepts at most $k$ elements, and $\E[f(A)]\ge\bigl(g(1-2\eta)-Cd-o(1)\bigr)W$, where $C=g(1+K)+\max\{1,\rho(c-t_1)\}$.
\end{lemma}

\begin{proof}
Condition on $S$ and $V$, which determine $B$, $M$, the pool $P$, and $D=O\cap P$; the pool elements then have independent uniform live times. Let $\mathcal G$ be the event of \cref{lem:retention} that $|P|\le(\rho+o(1))k$ and fewer than $\sqrt k$ elements are exceptional. Off $\mathcal G$ we use $f(A)\ge0$, which costs $O(W)\Pr[\mathcal G^c]=o(W)$ below, so assume $\mathcal G$. Then every subset $Z$ of the pool has $f(Z)\le\ceil{|Z|/k}W=O(W)$, by subadditivity. We analyze the \emph{uncapped} live phase, which ignores the budget. Let $A_x$ be its accepted set and $P_x$ the set of pool elements arrived by time $x$, so $A_x\subseteq P_x$. For an arrival $i$ at time $x$, let $\theta_i=\Phi'(Q_{x^-})$ if $t_1<x\le t_2$ and $i$ is normal, and $\theta_i=0$ otherwise. A rejected pool element has $\marg i{A_{x^-}}\le\theta_i$.

\emph{Value at time $c$.} Let $U$ be the set of pool elements outside $D$ arriving in $(t_1,c]$, and $U_A$ the accepted ones. Starting from $A_c$, add the rejected elements of $D\cap P_c$ in order of arrival; each raises the value by at most its $\theta_i$, by submodularity. Then add $U\setminus U_A$, which raises the value by at most $f(U)-f(U_A)$, by submodularity, since the current set contains $U_A$. Then add the remaining rejected elements, which arrived before $t_1$ with nonpositive marginal. The result is $P_c$, and so $f(A_c)\ge f(P_c)-\sum\theta_i-f(U)+f(U_A)$, with the sum over the rejected elements of $D\cap P_c$. Moreover $f(U_A)$ is at least the sum of the gains of the elements of $U_A$, by submodularity and $f(\emptyset)\ge0$.

\emph{After time $c$.} An element $i\in D$ arriving at time $x\in(c,t_2]$ has gain at least $\marg i{P_{x^-}}-\theta_i\ind[i\text{ rejected}]$, because $A_{x^-}\subseteq P_{x^-}$. Other arrivals have nonnegative gains. Therefore
\[
 f(A_{t_2})\ge f(P_c)+\sum_{i\in D:\,x_i\in(c,t_2]}\marg i{P_{x_i^-}}-f(U)+\Gamma,\qquad
 \Gamma=\sum_{\substack{i\notin D\text{ accepted}\\x_i\in(t_1,t_2]}}\mathrm{gain}_i-\sum_{\substack{i\in D\text{ rejected}\\x_i\in(t_1,t_2]}}\theta_i.
\]
Conditioning on the arrival time of each element of $D$, the expectation of the first two terms equals $F(c,c)+\int_c^{t_2}\partial_1F(t,t)\,dt\ge g f(D)$, by \cref{lem:dense}. A normal element is accepted only with gain larger than its $\theta_i$, so $\Gamma\ge\sum_{\text{normal accepted}}\theta_i-\sum_{i\in D:\,x_i\in(t_1,t_2]}\theta_i$.

\emph{Revenue.} A normal element has marginal at most its singleton value, hence at most $M$, while $\Phi'(b_0)\ge M$; so $Q\le b_0+1$ throughout, at most $\sigma k+b_0+1$ normal elements are accepted, and $\theta_i\le\Phi'(b_0+1)\le2M\le2W$. A normal acceptance raises $\Phi(Q)$ by $\Phi'(Q)+D_0(\Phi'(Q)+M/k)$, where $D_0=(e^{\lambda_0}-1)/\lambda_0-1=O(\lambda_0)$, and between acceptances $\Phi(Q)$ decreases at rate $\sigma k\Phi'(Q)$, which is also correct at $Q=0$. As $\Phi\ge0$ and $\Phi(0)=0$,
\[
 \Theta\ge\sigma k\int_{t_1}^{t_2}\Phi'(Q_x)\,dx-D_0\Bigl(\Theta+(\sigma k+b_0+1)\frac Mk\Bigr)\ge\sigma k\int_{t_1}^{t_2}\Phi'(Q_x)\,dx-o(W),
\]
where $\Theta$ is the sum of $\theta_i$ over the normal accepted elements, using $\Theta\le\sum\mathrm{gain}_i\le f(A_{t_2})=O(W)$. On the other hand, let $Y_x$ be the number of elements of $D$ not yet arrived at time $x$. Given the past, each of them arrives at rate $1/(1-x)$, and $\Phi'(Q_{x^-})$ is determined by the past, so
\[
 \E\sum_{i\in D:\,x_i\in(t_1,t_2]}\theta_i\le\E\int_{t_1}^{t_2}\Phi'(Q_{x})\frac{Y_x}{1-x}\,dx\le\sigma k\,\E\int_{t_1}^{t_2}\Phi'(Q_x)\,dx+\frac{2Wk}{1-t_2}\sup_{x\le t_2}\Pr\bigl[Y_x>\sigma k(1-x)\bigr].
\]
Since $|D|\le|O|\le k$, the variable $Y_x$ is binomial with mean at most $k(1-x)$, and the last probability is at most $e^{-(\sigma-1)^2k(1-t_2)/3}$. Hence $\E\Gamma\ge-o(W)$.

\emph{The set $U$.} The size of $U$ is binomial with mean at most $(c-t_1)|P|$, so with probability $1-o(1)$ we have $|U|\le(\rho(c-t_1)+o(1))k$, and then $f(U)\le(\max\{1,\rho(c-t_1)\}+o(1))dW$ by \cref{lem:excess}(c). Otherwise $f(U)=O(W)$.

\emph{The budget.} The uncapped phase accepts at most $|P_{t_1}|$ elements before time $t_1$, at most $\sigma k(t_2-t_1)+b_0+1$ normal elements after it, and fewer than $\sqrt k$ exceptional ones. With probability $1-o(1)$ we have $|P_{t_1}|\le(\rho t_1+o(1))k$, and then the total is less than $k$ for large $k$, because $\rho t_1+\sigma(t_2-t_1)<1$. On this event the capped and uncapped runs coincide; otherwise the capped run loses at most $f(A_{t_2})=O(W)$.

Combining, $\E f(A)\ge g\,\E f(D)-(\max\{1,\rho(c-t_1)\}+o(1))dW-o(W)$, and \cref{lem:retention} completes the proof.
\end{proof}

\begin{proof}[Proof of \cref{thm:beyond}]
Take $\eta=0.019$, $\rho=8.75$, $t_1=0.04$, $t_2=0.684$, and $\sigma=1.002$. Then $c=0.27051\ldots$, $g=0.41348\ldots$, $K=(1-\eta)^2/(\rho\eta)=5.78863\ldots$, $\rho(c-t_1)=2.01703\ldots$, and $C=4.82401\ldots$; also $\rho t_1+\sigma(t_2-t_1)=0.995288<1$. Take $\zeta=0.0031$ and $\xi=0.0002367169$. With probability $\xi$ run \textsc{PrunedGreedy}, and otherwise $\AP_\zeta$, which can be implemented with the probes of \cref{sec:queries}. By \cref{lem:two-comparators,lem:pruned},
\[
 \frac{\E[f(A)]}W\ge(1-\xi)h_\zeta+\xi g(1-2\eta)+\bigl((1-\xi)u_\zeta-\xi C\bigr)d-o(1).
\]
A computation with $50$ significant digits gives $(1-\xi)u_\zeta-\xi C=2.287\ldots\cdot10^{-10}>0$ and $(1-\xi)h_\zeta+\xi g(1-2\eta)-1/e=3.5337\ldots\cdot10^{-6}$. Since $d\ge0$, the theorem follows.
\end{proof}

\begin{remark}
The $o(1)$ terms decay slowly; for example, the validation step costs a term of order $k^{-1/8}$, and so the guarantee exceeds $1/e$ only for astronomically large $k$. The proof is designed to show that $1/e$ is not the limit, and we have not tried to obtain a useful explicit threshold. The mechanism is that $d$ measures how much two disjoint feasible sets can be worth together: $\AP_\zeta$ profits from a large $d$, and \textsc{PrunedGreedy} from a small one.
\end{remark}

\section{Hardness for monotone coverage functions}\label{sec:hardness}

We now prove \cref{thm:hardness-intro}. The construction adapts the probability-coverage idea from Section~3 of Udwani and Wajc~\cite{UW26}, who study submodular welfare, to a single cardinality constraint with random arrivals.

\begin{theorem}\label{thm:hardness}
Let $\mathcal A$ be a randomized online algorithm with unlimited computation and queries on arrived elements. Write $g_r(x)=1-(1-1/r)^x$.
\begin{enumerate}[label=(\alph*),nosep]
\item For every $k\ge2$ there is a monotone coverage instance on $n=k^{3k+1}$ elements, with $\OPT=1$, on which $\E[f(A)]\le(g_k(k)+2/k^2)\OPT$. Moreover, for every $\varepsilon>0$ there is an instance on which the ratio is at most $g_k(k)+\varepsilon$.
\item For every $k\ge64$, let $r\ge2$ be the largest integer with $r^{3r}\le k$. There is a monotone coverage instance with $k\le n\le k^2$ on which the ratio is at most $g_r(r)+2/r^2+4r/k$. Since $r=\Theta(\log k/\log\log k)$ and $g_r(r)\le1-1/e+1/(2e(r-1))$, this is at most $1-1/e+O(\log\log k/\log k)$.
\end{enumerate}
\end{theorem}

\paragraph{The random instance.}
Fix integers $r,R\ge2$ and $\ell\ge1$, and let
\[
 m_0=rR^r,\qquad \Gamma=R^{r+1},\qquad U=r\Gamma.
\]
The ground set is partitioned into $\ell$ public \emph{blocks} $V_1,\dots,V_\ell$, each with $m_0$ elements. For each block, independently, draw a uniformly random injection $\phi_i:V_i\to[\Gamma]\times[r]$, which assigns each element a distinct \emph{type} $(c,v)$. The $r$ types with the same first coordinate $c$ form a \emph{group}. The $U$ types far outnumber the $m_0$ elements, so that a block realizes only a sparse random set of types.

For $S\subseteq V_i$, let $a_c(S)$ be the number of elements of $S$ in group $c$, and define
\[
 v_i(S)=1-\prod_{c=1}^{\Gamma}\Bigl(1-\frac{a_c(S)}r\Bigr),\qquad f(S)=\sum_{i=1}^{\ell}v_i(S\cap V_i).
\]
This is a coverage function: take the universe $[r]^{\Gamma}$ with uniform total weight one, and let an element of type $(c,v)$ cover the points whose $c$th coordinate is $v$. Since distinct types in the same group cover disjoint point sets and coordinates are independent, $v_i$ is exactly the measure of the union. Taking disjoint universes for distinct blocks realizes the sum. Each singleton has value $1/r$, and each block has value at most one.

To strengthen the lower bound, we tell the algorithm the construction and reveal the block and the type of each element on arrival. This determines the answer to every allowed query. After the injections are drawn, the elements arrive in uniformly random order.

\paragraph{The offline optimum.}
Fix a block and let $X$ be the number of groups all of whose $r$ types are realized. Then $\chi:=\E[X]=\Gamma\prod_{a=0}^{r-1}\frac{m_0-a}{U-a}$. For two distinct groups the joint probability of being complete is $(m_0)_{2r}/(U)_{2r}\le((m_0)_r/(U)_r)^2$, where $(x)_a$ denotes the falling factorial, so that $\E[X^2]\le\chi^2+\chi$, and Cauchy--Schwarz gives $\Pr[X>0]\ge\chi/(1+\chi)$. Moreover, using Bernoulli's inequality,
\[
 \chi\ge \Gamma\Bigl(\frac{m_0-r+1}{U}\Bigr)^r=R\Bigl(1-\frac{r-1}{m_0}\Bigr)^r\ge R\Bigl(1-\frac{r-1}{R^r}\Bigr)\ge\frac R2.
\]
Selecting a complete group gives block value one. Thus, if $k\ge \ell r$, then $\E_\phi[\OPT]\ge \ell\chi/(1+\chi)$.

\paragraph{The online value.}
Let the $x$th arrival from a block be accepted, and suppose $s\ge1$ types of its group have now been seen, counting itself. Conditional on the whole history of the algorithm, the remaining $m_0-x$ realized types are a uniformly random subset of the $U-x$ unseen types. By a union bound over the $r-s$ unseen types in its group, the probability that a group mate of the element arrives later is at most
\[
 \frac{(r-s)(m_0-x)}{U-x}\le\frac{(r-1)(m_0-1)}{U-1}=:\rho\le\frac{r-1}{R}.
\]
(Here we used that $(m_0-x)/(U-x)$ is nonincreasing in $x$ and that $m_0/U=1/R$.) The decision to accept is a function of the history and independent randomness, and the injections for other blocks are independent, so applying this at every acceptance shows that the number $Z$ of accepted elements with a later arrival from their group satisfies $\E[Z]\le\rho\,\E[|A|]\le k\rho$.

Removing these $Z$ elements gives a set $A'$ with at most one element in each group, because every accepted element other than the last arrival of its group has been removed. If $b_i=|A'\cap V_i|$, then $v_i(A'\cap V_i)=g_r(b_i)$. Because $g_r$ is increasing and concave and $\sum_ib_i\le k$, and singletons have value $1/r$, we have pathwise
\[
 f(A)\le f(A')+\frac Zr=\sum_ig_r(b_i)+\frac Zr\le \ell\,g_r(k/\ell)+\frac Zr.
\]
Consequently, averaging over the injections, the order, and internal randomness, $\E[f(A)]\le \ell\,g_r(k/\ell)+k\rho/r$.

\paragraph{Fixing an instance.}
For a fixed tuple $\phi$ of injections, let $a(\phi)$ be the expected value of the algorithm and $o(\phi)>0$ be the optimum. Since $\E_\phi[a]/\E_\phi[o]$ is a weighted average of the ratios $a(\phi)/o(\phi)$ with weights proportional to $o(\phi)$, some fixed $\phi$ has ratio at most this quotient. So whenever $k\ge \ell r$, there is an instance with
\begin{equation}\label{eq:hard-master}
 \frac{a(\phi)}{o(\phi)}\le\Bigl[g_r(k/\ell)+\frac{k\rho}{\ell r}\Bigr]\Bigl(1+\frac1\chi\Bigr).
\end{equation}

\begin{proof}[Proof of \cref{thm:hardness}]
Take $R=r^3$, so $\chi\ge r^3/2\ge r^2$ and $\rho\le1/r^2$. Since $g_r(r)$ is at most $g_2(2)=3/4$, because $(1-1/r)^r$ is increasing,
\begin{equation}\label{eq:hard-simple}
 (g_r(r)+\rho)(1+1/\chi)\le(g_r(r)+1/r^2)(1+1/r^2)\le g_r(r)+2/r^2.
\end{equation}

(a) Take $\ell=1$ and $r=k$, so $n=m_0=k^{3k+1}$. Let $\mathcal G$ be the event that the block contains a complete group, on which $o(\phi)=1$. Nonnegativity gives $\E[a\mid\mathcal G]\le\E[a]/\Pr[\mathcal G]\le(g_k(k)+\rho)(1+1/\chi)$, so some $\phi\in\mathcal G$ has $\OPT=1$ and value at most the bound \eqref{eq:hard-simple}. Letting $R\to\infty$ instead, for fixed $r=k$, shows $\rho\to0$ and $\chi\to\infty$ and proves the second statement.

(b) Keep $R=r^3$, let $\ell=\floor{k/r}$, and let $d=k-\ell r<r$. Then $n=\ell r^{3r+1}\le kr^{3r}\le k^2$, and $n\ge k$. For $x\ge r$ the derivative of $g_r$ is at most $1/r$, because $-\ln(1-1/r)\le1/(r-1)$ and $(1-1/r)^r\le(r-1)/r$; therefore $g_r(k/\ell)\le g_r(r)+d/(\ell r)$. Substituting in \eqref{eq:hard-master} and using \eqref{eq:hard-simple}, together with $(1+\rho)(1+1/\chi)\le2$, gives the ratio bound $g_r(r)+2/r^2+2d/(\ell r)$. Since $k\ge r^{3r}\ge 2r$ we have $\ell r\ge k/2$, so $2d/(\ell r)\le4r/k$. Finally, $r\ln(1-1/r)\ge-1-1/(2(r-1))$ gives $g_r(r)\le1-1/e+1/(2e(r-1))$.
\end{proof}

\begin{remark}
This family cannot give a bound better than $1-1/e$: by Bernoulli's inequality, any $r$ elements of a block have value at least $g_r(r)$, and so, when $k=\ell r$, accepting the first $r$ arrivals from each block achieves ratio at least $g_r(r)$.
\end{remark}

\section{Hardness for general objectives}\label{sec:half}

The instances of \cref{sec:hardness} are monotone, and \AP{} achieves $1-1/e-o(1)$ on them. We now show that nonmonotonicity by itself creates a stronger barrier: we prove the barrier $\frac12$ of \cref{thm:half-intro}(a), and then a bound strictly below $\frac12$. The constant $0.497053$ of part~(b) is proved in \cref{bh:sec}. The building block is a pair of elements that are valuable separately and worthless together, where the ratio between the two values is hidden on a long geometric scale.

\subsection{The basic barrier}

\begin{theorem}\label{thm:mu-barrier}
Let $n\ge2$, $1\le k\le n$, $\mu\in[0,1]$, and $\varepsilon>0$. There is a finite family of normalized, nonnegative, $\mu$-monotone submodular functions on $n$ elements such that, for every algorithm with unlimited computation, unlimited queries on arrived elements, and full knowledge of the family, some member $f$ satisfies $\E[f(A)]\le\bigl(\frac{1+\mu}2+\varepsilon\bigr)\OPT$.
\end{theorem}

\begin{proof}
Fix an integer $N\ge\max\{1,1/\varepsilon\}$ and let $K=N+1$. For an ordered pair $(a,b)$ of distinct elements and $J\in\{0,\dots,N-1\}$, let $f_{a,b,J}(S)$ depend only on $S\cap\{a,b\}$, taking the values
\[
 f(\emptyset)=0,\qquad f(\{a\})=K^J,\qquad f(\{b\})=K^{J+1},\qquad f(\{a,b\})=\mu K^{J+1}.
\]
Since $K^J+K^{J+1}\ge\mu K^{J+1}$, this function is submodular, and it is clearly nonnegative and $\mu$-monotone. For $\mu=0$ it is the directed cut function of the two arcs $a\to b$ and $b\to a$ with weights $K^J$ and $K^{J+1}$. Since $k\ge1$, its optimum is $\OPT=K^{J+1}$, attained by $\{b\}$.

Draw $(a,b)$ and $J$ uniformly, independently of the arrival order. It suffices to bound the average of $\E[f(A)]/\OPT$ over this prior by $\frac{1+\mu}2+\varepsilon$, and by linearity we may assume that the algorithm is deterministic. Condition on the arrival order and on the unordered pair $\{a,b\}$. Let $e_1$ be the member of the pair arriving first and $e_2$ the other. Until $e_2$ arrives, every allowed query on a set $S$ returns $v\,\ind[e_1\in S]$, where $v=f(\{e_1\})$. So whether $e_1$ is accepted depends only on the conditioning and on $v$.

Let $w=f(\{e_2\})$. If $e_1$ is accepted then $f(A)\le\max\{v,\mu\max\{v,w\}\}$, and if $e_1$ is rejected then $f(A)\le w$. Now let $v=K^j$. This value arises from at most two of the $2N$ equally likely choices of the hidden orientation and $J$:
\begin{itemize}[nosep]
\item $e_1=a$ and $J=j$, for $0\le j\le N-1$. Here $e_1$ is the light element, $\OPT=w=K^{j+1}$, and the normalized value is at most $\max\{1/K,\mu\}$ if $e_1$ is accepted and at most $1$ if it is rejected.
\item $e_1=b$ and $J=j-1$, for $1\le j\le N$. Here $e_1$ is the heavy element, $\OPT=v$ and $w=v/K$, and the normalized value is at most $1$ if $e_1$ is accepted and at most $1/K$ if it is rejected.
\end{itemize}
For $1\le j\le N-1$ both choices occur, and the decision at $e_1$ is the same for both, so their normalized values sum to at most $1+\mu+1/K$. For $j\in\{0,N\}$ a single choice occurs, with normalized value at most $1$. Averaging,
\[
 \frac{(N-1)(1+\mu+1/K)+2}{2N}\le\frac{1+\mu}2+\frac1{2K}+\frac1{2N}\le\frac{1+\mu}2+\varepsilon.\qedhere
\]
\end{proof}

\subsection{Below one half}\label{sec:below-half}

We now show that the limiting ratio for general objectives is strictly below $1/2$. The pairs are kept, but a small fraction of the value is moved into a matroid gadget in which a second, ``bad'' basis interferes with a ``good'' one.

Let $\mathsf M$ be a matroid of rank $r\ge1$ with rank function $r_{\mathsf M}$, whose ground set is the disjoint union of two bases $G$ and $B$; parallel elements are allowed and are distinct. Let
\[
 \varphi_{\mathsf M}(S)=r_{\mathsf M}(S)-|S\cap B|.
\]
This function is normalized and submodular, and it is nonnegative because $r_{\mathsf M}(S)\ge r_{\mathsf M}(S\cap B)=|S\cap B|$. Its maximum is $\varphi_{\mathsf M}(G)=r$. Scan the $2r$ elements in a uniformly random order, and let $\mathcal T$ be the set of elements that increase the rank of the set of all elements scanned before them. Then $\mathcal T$ is a basis, and we define the \emph{greedy retention}
\[
 \beta(\mathsf M;G,B)=\frac1r\E|\mathcal T\cap G|=1-\frac1r\E|\mathcal T\cap B|.
\]
If $\beta<1/2$, then $C(\beta)=\sqrt{2-2\beta}+\beta-1<1/2$, since $C(1/2)=1/2$ and $C'(\beta)=1-(2-2\beta)^{-1/2}>0$ for $\beta<1/2$.

\begin{theorem}\label{thm:transfer}
Let $(\mathsf M;G,B)$ be as above, with $\beta=\beta(\mathsf M;G,B)<1/2$, and let $\varepsilon>0$. There is a finite family $\mathcal F$ of normalized, nonnegative, submodular functions on a common ground set of size $n_0$, such that the following holds for every algorithm with unlimited computation, unlimited queries on arrived elements, and full knowledge of $\mathcal F$, even if it observes the arrival times. For every $k\ge n_0$, some $f\in\mathcal F$ satisfies $\E[f(A)]\le(C(\beta)+\varepsilon)\OPT$. Moreover, every member of $\mathcal F$ has largest singleton value at most $\varepsilon\OPT$. Consequently, the best ratio achievable with budget $k$ is at most $C(\beta)+\varepsilon$ for all $k\ge n_0$.
\end{theorem}

\begin{corollary}\label{cor:fan}
For every sufficiently large $k$, no online algorithm, even with unlimited computation and queries, has ratio more than $0.49999$ with budget $k$. More precisely, the best ratio $\rho_k$ satisfies $\limsup_{k\to\infty}\rho_k\le\sqrt{91/90}-91/180<0.4999847$.
\end{corollary}

\begin{proof}
Let $\mathsf M$ be the graphic matroid on the vertices $\{0,1,2,3\}$, with $G=\{01,02,03\}$ and $B=\{01',12,23\}$, where $01$ and $01'$ are parallel edges. Both are spanning trees, hence bases of rank $3$. Fix $g\in G$. In a uniformly random order, the set of elements preceding $g$ has a uniformly random size $s\in\{0,\dots,5\}$ and, given $s$, is a uniformly random $s$-subset of the other five elements. Let $a_s$ be the number of pairs $(g,S)$ with $g\in G\setminus S$, $|S|=s$, and $r_{\mathsf M}(S\cup\{g\})>r_{\mathsf M}(S)$. Direct enumeration gives $(a_0,\dots,a_5)=(3,14,21,8,1,0)$, and so
\[
 \beta=\frac13\sum_{s=0}^5\frac{a_s}{6\binom5s}=\frac13\Bigl(\frac36+\frac{14}{30}+\frac{21}{60}+\frac8{60}+\frac1{30}\Bigr)=\frac{89}{180}.
\]
\Cref{thm:transfer} gives $\limsup_k\rho_k\le C(89/180)=\sqrt{91/90}-91/180=0.49998465\ldots$.
\end{proof}

With one large fan per level instead of many small gadgets, a finer analysis of the same construction gives a better constant (\cref{bh:sec}).

\Cref{thm:half-intro} now follows: part~(a) is \cref{thm:mu-barrier}; part~(b) follows from \cref{bu3:thm} in \cref{bh:sec} with $\varepsilon=10^{-7}$, since $C_{\mathrm{blk}}+10^{-7}<0.497053$.

\paragraph{The model in the proof.}
We strengthen the algorithm: after each arrival it is told the entire table of values $f(S)$ for all sets $S$ of arrived elements, and it observes independent uniform arrival times. Since a randomized algorithm is a mixture of deterministic ones, it suffices to bound, for every deterministic algorithm, the expected \emph{instancewise} ratio $\E[f(A)/\OPT(f)]$ when $f$ is drawn from a distribution on $\mathcal F$; some member of $\mathcal F$ then has expected ratio at most the same bound. In the construction, $\OPT(f)$ is attained by a set of at most $n_0$ elements, so the budget never binds when $k\ge n_0$. Adding $k-n_0$ elements of value zero, whose arrivals the algorithm could simulate itself, gives instances of size exactly $k$.

\paragraph{The construction.}
Fix a rational $\delta\in(0,1/2)$, integers $N\ge1$ and $m>2N$, and an integer $Y$ that is a multiple of $r$; at the end $Y\to\infty$. Let $\mathcal H_N=\sum_{d=1}^N1/d$ and $\pi_d=1/(d\mathcal H_N)$, so that $\sum_{d\le N}\pi_d=1$. Let $Q$ be the least multiple of $Q_0=\mathrm{den}(\delta)\,\mathrm{num}(\mathcal H_N)\,N!$ with $Q\ge(\log Y)^2$, so that $Q<(\log Y)^2+Q_0$; this makes $(1-\delta)Q\pi_d$, $\delta Q$, and $\delta Q\pi_d$ integers.

The elements, which we call \emph{items}, are partitioned into \emph{levels} $1,\dots,m$, where level $j$ has $n_j$ items and $n_j$ is a multiple of $r$. The objective is defined on finer objects called \emph{atoms}, each of which is \emph{carried} by exactly one item; the value of a set of items is the value of the set of atoms they carry. Each item of level $j$ carries $Q$ \emph{good} atoms of weight $u_j=1/(n_jQ)$. Atoms are grouped into \emph{components}, and $f$ is the sum of the component values.
\begin{itemize}[nosep]
\item For $j\le N$, the good atoms of level $j$ form a single modular component with value $u_j|S|$.
\item For $j>N$, each item owns $(1-\delta)Q$ \emph{pair} components. Each consists of one of its good atoms $g$ and a \emph{bad} atom $b$, with value $u_j\ind[g\in S,\,b\notin S]$; this is $u_j\varphi$ for the rank-one matroid in which $g$ and $b$ are parallel.
\item For $j>N$, the items of level $j$ are partitioned into groups of $r$, and each group has $\delta Q$ copies of $\mathsf M$, with value $u_j\varphi_{\mathsf M}$ each. The good elements of each copy are good atoms of the $r$ items of the group, one for each item, and its bad elements are bad atoms. Let $\mathsf M_j$ be the direct sum of all copies at level $j$; it has rank $R_j=\delta n_jQ$.
\end{itemize}
Bad atoms of level $j>N$ are carried by items of the lower levels $j-d$, $1\le d\le N$. For each item of level $j$ and each $d$, exactly $(1-\delta)Q\pi_d$ of its pair components have their bad atom sent to level $j-d$; and exactly $\delta n_jQ\pi_d$ of the bad atoms of $\mathsf M_j$ are sent to level $j-d$, by a fixed rule. Level $\ell$ then receives
\[
 s_\ell=\sum_{j>N,\ 1\le j-\ell\le N}n_jQ\pi_{j-\ell}
\]
bad atoms, and they are assigned to its items by a uniformly random injection, independently over $\ell$. So each item carries its own good atoms and at most one bad atom. The family $\mathcal F$ consists of the objectives obtained from all choices of the injections. Finally, choose the sizes downward: $n_m=Y$ and $n_\ell=r\ceil{Y^2(1+s_\ell)^2/r}$ for $\ell=m-1,\dots,1$, which is possible because $s_\ell$ depends only on $n_j$ for $j>\ell$. Then
\begin{equation}\label{eq:sparse}
 \frac{s_\ell}{n_\ell}\le\frac1{4Y^2},\qquad \min_jn_j\ge Y,\qquad \log\sum_jn_j=O(\log Y),
\end{equation}
where the last bound follows by induction, as each $n_\ell$ is polynomial in $YQ\sum_{j>\ell}n_j$. Here and below, $O(\cdot)$ and $o(1)$ refer to $Y\to\infty$ with $\delta$, $N$, $m$, and $\mathsf M$ fixed, and they are uniform over algorithms.

\begin{lemma}\label{lem:transfer-valid}
Every $f\in\mathcal F$ is normalized, nonnegative, and submodular, every item of level $j$ has singleton value $1/n_j\le1/Y$, and $m(1-1/(4Y^2))\le\OPT(f)\le m$.
\end{lemma}

\begin{proof}
Each component is $u_j\varphi$ for a matroid with an independent set of bad elements, or modular, and so it is normalized, nonnegative, and submodular. The map from sets of items to sets of atoms preserves unions and intersections, so composing with it preserves submodularity. An item's good atoms lie in distinct components or are independent in their copy of $\mathsf M$, so they contribute $u_jQ=1/n_j$; a bad atom alone contributes zero. The total weighted rank of the components of level $j$ is $u_jn_jQ=1$, so $f\le m$. The items that carry no bad atom have value $\sum_j(n_j-s_j)/n_j\ge m(1-1/(4Y^2))$ by \eqref{eq:sparse}.
\end{proof}

The singleton bound in \cref{thm:transfer} follows. Arrival times are attached to items, and an atom arrives with its carrier. Let $A_j(t)$ be the number of accepted items of level $j$ with arrival time at most $t$, divided by $n_j$, and let $P_j=A_j(1)$ and $\bar P=\frac1m\sum_jP_j$.

\paragraph{A relaxation.}
For $j>N$, let $\mathcal T_j$ be the set of bad atoms of $\mathsf M_j$ that increase the rank of the set of all atoms of $\mathsf M_j$ that arrived before them. Since $\mathsf M_j$ is a direct sum, this can be decided in each copy of $\mathsf M$ separately, and within a copy no two atoms arrive together, as shown below. So $\mathcal T_j$ is well defined, and it depends only on the instance and the arrival times, not on the algorithm. Let $\bar f$ be obtained from $f$ by deleting from their components all bad atoms of $\mathsf M_j$ outside $\mathcal T_j$, for every $j$, and every bad atom of a pair component whose carrier arrives after the item owning the good atom. Since $r_{\mathsf M}(S)\le r_{\mathsf M}(S\setminus\{b\})+1$, deleting a bad element never decreases $\varphi_{\mathsf M}$, and so $f\le\bar f$ pointwise. We bound $\E[\bar f(A)]$.

In each copy of $\mathsf M$, the $2r$ atoms have distinct carriers: the good ones by construction, the bad ones because the injections are injective, and good and bad carriers lie in different levels. So their relative order is uniformly random, and $\E|\mathcal T_j|=(n_j/r)\,\delta Q\cdot r(1-\beta)$. If $B_A\subseteq\mathcal T_j$ is the set of atoms of $\mathcal T_j$ carried by accepted items, then the matroid part of level $j$ contributes at most $u_j(R_j-|B_A|)=u_j(R_j-|\mathcal T_j|)+u_j|\mathcal T_j\setminus B_A|$ to $\bar f(A)$, and
\begin{equation}\label{eq:baseline}
 \E\bigl[u_j(R_j-|\mathcal T_j|)\bigr]=\delta\beta.
\end{equation}

\paragraph{Invisible roles.}
An atom $b\in\mathcal T_j$ increases the rank of every subset of the atoms of $\mathsf M_j$ that arrived before it, by submodularity of rank, and so it has zero marginal value in its component on every set of atoms that have arrived when it arrives. Thus deleting $b$ does not change any value that the algorithm has observed up to and including the arrival of its carrier; we say that $b$ is \emph{invisible} at arrival. Similarly, the bad atom of a pair component is invisible until the good atom arrives, since a pair containing only its bad atom has value zero.

\begin{lemma}\label{lem:invisible}
Fix a level $\ell$, and a fixed set of $w$ among the $s_\ell$ bad atoms sent to level $\ell$. Let $X$ be the number of these atoms that are invisible at arrival and whose carrier is rejected. Then, for every algorithm, $\E X\le w\,\E[1-P_\ell]+ws_\ell/n_\ell+2\sqrt w$.
\end{lemma}

\begin{proof}
Condition on all randomness other than the injection $\sigma$ into level $\ell$, including all arrival times and the injections into other levels; $\sigma$ remains uniform. Let $\nu=n_\ell$, and let the items of level $\ell$ arrive in the order $1,\dots,\nu$. Reveal which atom item $i$ carries only after the decision on item $i$. Every value observed up to then involves only arrived items, and so is determined by what has been revealed. Given the revealed information, the unrevealed atoms are injected uniformly into the $s=\nu-i+1$ items that have not yet been decided, which include item $i$.

Let $Z_i\in\{0,1\}$ indicate that item $i$ is rejected, and let $Z_i^0$ indicate whether the algorithm would reject item $i$ if the bad atom carried by item $i$ were deleted from the values observed at its arrival. Then $Z^0_i$ is determined before the atom carried by item $i$ is revealed, and $Z_i=Z_i^0$ whenever item $i$ carries an invisible atom. Let $Y_s$ be the number of the $w$ fixed atoms not yet revealed. Each lands on item $i$ with conditional probability $1/s$, and so
\[
 \E X\le\E\sum_{i=1}^\nu\frac{Y_s}sZ^0_i\le\frac w\nu\E\sum_iZ^0_i+\sum_{s=1}^\nu\E\Bigl|\frac{Y_s}s-\frac w\nu\Bigr|.
\]
Since the arrival times are fixed by the conditioning, $Y_s$ counts the fixed atoms that land in a fixed set of $s$ items, so it has mean $ws/\nu$ and, by negative correlation, variance at most $ws/\nu$. Hence $\E|Y_s/s-w/\nu|\le\sqrt{w/(\nu s)}$, which sums to at most $2\sqrt w$. Finally, $Z_i^0$ and $Z_i$ differ only at items that carry a bad atom, of which there are at most $s_\ell$, so $\sum_iZ^0_i\le\nu(1-P_\ell)+s_\ell$.
\end{proof}

Every atom of $\mathcal T_j\setminus B_A$ is invisible at arrival and has a rejected carrier. Apply \cref{lem:invisible} to the $\delta n_jQ\pi_d$ bad atoms of $\mathsf M_j$ sent to level $j-d$, multiply by $u_j$, and sum over $d$. Together with \eqref{eq:baseline} and \eqref{eq:sparse}, and using that each level $\ell$ receives weight $\pi_d$ from at most one level $j=\ell+d$ for each $d$, the expected matroid part of $\bar f(A)$ is at most
\begin{equation}\label{eq:matbound}
 \delta m\beta+\delta\sum_{\ell=1}^m\E[1-P_\ell]+o(1).
\end{equation}

\paragraph{Pairs.}
Fix an item $x$ of level $j>N$ and a distance $d$, let $\ell=j-d$, and let $q=(1-\delta)Q\pi_d$ be the number of pair components of $x$ whose bad atoms go to level $\ell$. Condition on everything except where these $q$ bad atoms are placed. Before $x$ arrives, these components are invisible, so the set $S_\ell$ of items of level $\ell$ accepted before $x$ arrives does not depend on their placement. Given the conditioning, the $q$ carriers are a uniformly random $q$-subset, drawn without replacement, of a set of at least $n_\ell-s_\ell$ items of level $\ell$, and so the expected fraction of them in $S_\ell$ is at least $(|S_\ell|-s_\ell)/n_\ell$. By Hoeffding's inequality for sampling without replacement, the fraction is less than $|S_\ell|/n_\ell-s_\ell/n_\ell-\eta$ with probability at most $e^{-2q\eta^2}$. Take $\eta=(\log Y)^{-1/4}$. Since $q\ge c(\log Y)^2$ for a constant $c>0$ and there are $e^{O(\log Y)}$ pairs $(x,d)$, a union bound shows that with probability $1-o(1)$ no such deviation occurs.

On this event, consider an accepted item $x$ of level $j$, arriving at time $t$. In $\bar f$, a pair component of $x$ has lost its bad atom if the carrier arrives after $x$, and otherwise it is worth zero exactly when the carrier was accepted. So the pair components of $x$ contribute at most $\frac{1-\delta}{n_j}\bigl[1-\sum_d\pi_dA_{j-d}(t^-)+o(1)\bigr]$, and rejected items contribute nothing. Off the event we use $\bar f\le m$, which holds as in \cref{lem:transfer-valid}, since deleting bad atoms does not change the total weighted rank of a level. Thus the expected pair part of $\bar f(A)$ is at most
\begin{equation}\label{eq:pairbound}
 (1-\delta)\,\E\sum_{j>N}\int_{(0,1]}\Bigl[1-\sum_{d=1}^N\pi_dA_{j-d}(t^-)\Bigr]dA_j(t)+o(m).
\end{equation}
\Cref{lem:invisible} and \eqref{eq:pairbound} use different conditionings, but both are bounds on expectations of random variables on the same probability space, so they may be added.

\paragraph{Smoothing.}
Index the levels cyclically by $\mathbb Z_m$, let $(\mathsf Sx)_j=x_{j-1}$, and let $\Pi=\sum_{d=1}^N\pi_d\mathsf S^d$. As $N<m/2$, the circulant matrix $\Pi$ has zero diagonal and row sums one, and for $j>N$ its row $j$ is the kernel $(\pi_d)$ used in \eqref{eq:pairbound}. For a vector of nondecreasing paths $A=(A_j)$, let $\mathcal J_j(A)=\int_{(0,1]}[1-(\Pi A(t^-))_j]\,dA_j(t)$, which is nonnegative. So we may add the rows $j\le N$ to \eqref{eq:pairbound}, and it suffices to bound $\frac1m\E\sum_{j\in\mathbb Z_m}\mathcal J_j(A)$.

With probability $1-o(1)$, the empirical distribution function of the arrival times at each level is within $\eta'=2Y^{-1/4}$ of the identity, by Hoeffding's inequality at the points $iY^{-1/4}$ and monotonicity, since $n_j\ge Y$. As every acceptance is an arrival, on this event $A_j(v)-A_j(u)\le v-u+2\eta'$ for $u\le v$. Let $a_j(t)=\inf_{0\le s\le t}\{A_j(s)+t-s\}$. Then $a_j$ is nondecreasing and $1$-Lipschitz, so $a_j(t)=\int_0^tp_j$ with $p_j\in[0,1]$; moreover $0\le A_j-a_j\le2\eta'$. Almost surely no two items arrive together, and $\Pi$ has zero diagonal, so $A(t^-)$ may be replaced by $A(t)$ in $\mathcal J_j$. Integrating by parts, with $\Delta_j=A_j-a_j$ and the continuous path $\Pi a$,
\[
 \mathcal J_j(A)-\mathcal J_j(a)=[1-(\Pi a(1))_j]\Delta_j(1)-\int(\Pi\Delta)_j\,dA_j+\int\Delta_j\,d(\Pi a)_j\le4\eta',
\]
since the first and last terms lie in $[0,2\eta']$ and the middle term is subtracted and nonnegative.

\begin{lemma}\label{lem:control}
Let $p_j:[0,1]\to[0,1]$ be measurable for $j\in\mathbb Z_m$, let $a_j(t)=\int_0^tp_j$, and let $\bar P=\frac1m\sum_ja_j(1)$. Then
\[
 \frac1m\sum_{j\in\mathbb Z_m}\int_0^1p_j(t)\bigl[1-(\Pi a(t))_j\bigr]\,dt\le\bar P-\frac{\bar P^2}2+\frac9{8\mathcal H_N}.
\]
\end{lemma}

\begin{proof}
First, $\sum_{d\le N}\cos(d\theta)/d\ge-4$ and $|\sum_{d\le N}\sin(d\theta)/d|\le5$ for real $\theta$. Indeed, let $\gamma\in[0,\pi]$ be the distance from $\theta$ to $2\pi\mathbb Z$. If $N\gamma\le1$, all cosines are positive and the sine sum is at most $N\gamma$ in absolute value. Otherwise, Abel summation gives $|\sum_{d>N}e^{id\theta}/d|\le1/((N+1)\sin(\gamma/2))<\pi/(N\gamma)<\pi$, while the full series $-\log(1-e^{i\theta})$ has real part $-\log(2\sin(\gamma/2))\ge-\log2$ and imaginary part at most $\pi/2$ in absolute value.

Let $\Pi_s=(\Pi+\Pi^{\mathsf T})/2$ and $\Pi_a=(\Pi-\Pi^{\mathsf T})/2$. Both are circulant, so the bounds above show that $\Pi_s$ has smallest eigenvalue at least $-4/\mathcal H_N$ and that $\Pi_a$ has operator norm at most $5/\mathcal H_N$. Also $\Pi_s\ind=\ind$ and $\Pi_a\ind=\Pi_a^{\mathsf T}\ind=0$. Write $P=a(1)=\bar P\ind+y$ with $y\perp\ind$, so $\|y\|^2\le m/4$. Since $\frac{d}{dt}(a^{\mathsf T}\Pi_sa)=2p^{\mathsf T}\Pi_sa$,
\[
 \int_0^1p^{\mathsf T}\Pi_sa\,dt=\frac12P^{\mathsf T}\Pi_sP\ge\frac m2\bar P^2-\frac2{\mathcal H_N}\cdot\frac m4.
\]
For the skew part, $p^{\mathsf T}\Pi_aa=(p-\frac12\ind)^{\mathsf T}\Pi_a(a-\frac t2\ind)$, and the two vectors have norms at most $\sqrt m/2$ and $t\sqrt m/2$, so $\int_0^1|p^{\mathsf T}\Pi_aa|\,dt\le\frac{5m}{8\mathcal H_N}$. Hence $\int_0^1p^{\mathsf T}\Pi a\,dt\ge\frac m2\bar P^2-\frac{9m}{8\mathcal H_N}$, and the lemma follows from $\int_0^1\sum_jp_j=m\bar P$.
\end{proof}

\begin{proof}[Proof of \cref{thm:transfer}]
Apply \cref{lem:control} to the paths $a_j$ on the event of the smoothing step. Their mean terminal value is at most $\bar P$, and $x\mapsto x-x^2/2$ is increasing on $[0,1]$, so $\frac1m\E\sum_j\mathcal J_j(A)\le\E[\bar P-\bar P^2/2]+\frac9{8\mathcal H_N}+o(1)$. The levels $j\le N$ contribute at most $1$ each. Combining this with $f\le\bar f$, \eqref{eq:matbound}, and \eqref{eq:pairbound}, and writing $\bar d=1-\bar P$ and $c=\frac12-\beta>0$,
\begin{align*}
 \frac{\E[f(A)]}m
 &\le\frac Nm+(1-\delta)\E\Bigl[\bar P-\frac{\bar P^2}2\Bigr]+\delta\beta+\delta\E[\bar d]+\frac9{8\mathcal H_N}+o(1)\\
 &=\frac12-c\delta+\E\Bigl[\delta\bar d-\frac{1-\delta}2\bar d^2\Bigr]+\frac Nm+\frac9{8\mathcal H_N}+o(1)
 \le U_\delta+\frac Nm+\frac9{8\mathcal H_N}+o(1),
\end{align*}
where $U_\delta=\frac12-c\delta+\frac{\delta^2}{2(1-\delta)}$ is the maximum of the quadratic in $\bar d$. By \cref{lem:transfer-valid}, the expected instancewise ratio is at most $(U_\delta+N/m+9/(8\mathcal H_N)+o(1))/(1-1/(4Y^2))$. The minimum of $U_\delta$ is attained at $\delta_*=1-(1+2c)^{-1/2}\in(0,1/2)$, where $U_{\delta_*}=\sqrt{1+2c}-\frac12-c=C(\beta)$. Choose a rational $\delta$ close to $\delta_*$, then $N$ large, then $m$ large, and finally $Y$ large.
\end{proof}

\section*{Acknowledgments and AI methodology}

We thank Zhiyi Huang and Sahil Singla for helpful discussions, and for joint work on the matroid secretary problem, which led to~\cite{ABHHMS26}.

The results in this paper were developed with substantial assistance from AI models, primarily GPT-6 Astra, Claude Opus~5.5, and Claude Fable~5.2, working under the direction of the authors, and the manuscript was drafted with the assistance of Claude. The numerical constants in this paper were verified by computer in exact rational or directed-rounding arithmetic.

A Lean~4 formalization, written by GPT-6 Astra, proves every result of this paper. It uses no axioms beyond Lean's standard ones, except that some of the finite numerical certificates in the appendices are checked with \texttt{native\_decide}, which also trusts Lean's compiler.

\bibliographystyle{alpha}
\bibliography{refs}

\clearpage
\appendix

\section{Explicit guarantees at every budget}\label{ek:sec}

\Cref{thm:finite} has content only once $T>0$, which requires $k$ of order $10^7$. This section gives explicit guarantees for general objectives at every budget. Throughout, $f$ is nonnegative and submodular, but neither monotone nor normalized, and we write $W=\OPT=\max_{|S|\le k}f(S)$ and $U=\max_{i\in E}f(\{i\})$.

\begin{theorem}[Every budget]\label{ek:thm-every-k}
For every $k\ge1$ there is a randomized algorithm that knows only $n$ and $k$, accepts at most $k$ elements on every execution, makes at most $6n+1$ value queries, each on a subset of the arrived elements, and satisfies $\E[f(A)]\ge g_k\OPT$ for every $n\ge k$ and every nonnegative submodular $f$, where $g_k$ is given in \cref{ek:tab}. In particular, $g_k\ge 0.2677$ for every $k\ge1$, $g_k\ge0.3$ for every $k\ge52$, and $g_k\ge0.3678$ for every $k\ge10^{12}$.
\end{theorem}

\begin{table}[ht]
\centering
\small
\begin{tabular}{@{}lll@{\qquad}lll@{}}
\toprule
Budget $k$ & $g_k$ & Policy & Budget $k$ & $g_k$ & Policy\\
\midrule
$1$ & $1/e$ & \cref{ek:lem-singleton} & $30$--$39$ & $0.2677$ & uniform price\\
$2$ & $0.3734$ & two picks & $40$--$49$ & $0.2859$ & uniform price\\
$3$ & $0.3111$ & two picks & $50$--$51$ & $0.2981$ & uniform price\\
$4,5,6$ & $0.2979,\ 0.2904,\ 0.2861$ & restart & $52$--$69$ & $0.3001$ & uniform price\\
$7,8,9$ & $0.2830,\ 0.2804,\ 0.2783$ & restart & $70$--$99$ & $0.3134$ & uniform price\\
$10,11,12$ & $0.2765,\ 0.2752,\ 0.2740$ & restart & $100$--$499$ & $0.3266$ & uniform price\\
$13,14,15$ & $0.2729,\ 0.2718,\ 0.2711$ & restart & $500$--$999$ & $0.337$ & exempt price\\
$16,17,18$ & $0.2699,\ 0.2696,\ 0.2685$ & restart & $10^3$--$2499$ & $0.347$ & exempt price\\
$19,20,21$ & $0.2685,\ 0.2680,\ 0.2682$ & restart & $2500$--$9999$ & $0.356$ & exempt price\\
$22,23,24$ & $0.2683,\ 0.2682,\ 0.2681$ & restart & $10^4$--$(10^5-1)$ & $0.363$ & exempt price\\
$25,26,27$ & $0.2680,\ 0.2679,\ 0.2680$ & restart & $10^5$--$(10^6-1)$ & $0.3668$ & exempt price\\
$28,29$ & $0.2678,\ 0.2678$ & restart & $10^6$--$(10^{12}-1)$ & $0.3676$ & exempt price\\
 & & & $\ge10^{12}$ & $0.3678$ & \AP{}\\
\bottomrule
\end{tabular}
\caption{Certified ratios $g_k$ for nonnegative submodular objectives, valid for every $n\ge k$ with no error term. All decimals are exact rationals. The policies use at most $n+1$ (singleton rule, restart windows), $2n+1$ (two picks), $5n+1$ (uniform price), $6n+1$ (exempt price) and $4n+1$ (\AP) value queries. For $k\ge10^{12}$ the ratio tends to $1/e$ by \cref{thm:main}(a), and it exceeds $1/e$ for all sufficiently large $k$ by \cref{thm:beyond}.}
\label{ek:tab}
\end{table}

The smallest entry is $g_k=0.2677$, for $30\le k\le39$; the smallest entries with $k\le29$ are $g_{28}=g_{29}=0.2678$. The policies are of four kinds, which we treat in turn: a two-pick rule for $k\in\{2,3\}$ (\cref{ek:small}), a multi-rank rule that is restarted after every stop for $4\le k\le29$ (\cref{ru:sec}), and two variants of a Poisson-price policy with eligibility attenuation for $k\ge30$ (\cref{ek:poisson}). For $k=1$ and as insurance against single large elements we use a classical singleton rule. Each guarantee reduces to a finite computation, which is carried out by computer in exact rational or directed integer arithmetic; no step uses floating point. We assume $W>0$, since otherwise there is nothing to prove.

\begin{lemma}[Singleton rule]\label{ek:lem-singleton}
Fix a total order on $E$ that refines decreasing singleton values. Reject all arrivals before time $1/e$. Afterwards, consider the first arrival that is larger in this order than every earlier arrival (if no element arrived earlier, this is the first arrival); accept it if its singleton value exceeds $f(\emptyset)$, and stop in either case. This rule makes $n+1$ queries, accepts at most one element, and satisfies $\E[f(A)]\ge U/e$. If $k=1$, then $\E[f(A)]\ge\OPT/e$.
\end{lemma}

\begin{proof}
Let $e^*$ be the largest element, and condition on its timestamp $t\ge1/e$ and on the set of elements that precede it. If this set is empty, $e^*$ is the first arrival after $1/e$ and is considered. Otherwise the largest of its members has a uniform timestamp on $[0,t]$, and if this timestamp is below $1/e$, then no arrival in $[1/e,t)$ is larger than all earlier arrivals and again $e^*$ is considered. So $e^*$ is considered with probability at least $\int_{1/e}^1\frac{1}{et}\,dt=\frac1e$. If $f(\{e^*\})>f(\emptyset)$, it is then accepted, which gives $\E f(A)\ge U/e$ since all outcomes are nonnegative; if $f(\{e^*\})\le f(\emptyset)$, then $U\le f(\emptyset)\le f(A)$ on every execution. For $k=1$ we have $\OPT=\max\{U,f(\emptyset)\}$, and the same two cases give $\E f(A)\ge\OPT/e$.
\end{proof}

\subsection{Two picks: budgets two and three}\label{ek:small}

For $k\in\{2,3\}$ we accept at most two elements. Add $24$ \emph{virtual} elements, whose timestamps are drawn by the algorithm independently, and extend $f$ by ignoring them. There are $N=n+24$ labels, all with independent uniform timestamps. Let $b=f(\emptyset)$, and fix a total order $z_1\succ z_2\succ\dots\succ z_N$ of all labels that refines decreasing singleton values, with ties broken by a fixed label order; a virtual label has singleton value $b$. A label is \emph{proper} if its singleton value exceeds $b$. Fix a schedule $0<\vartheta_1<\dots<\vartheta_\ell<1$ and put $\vartheta_{\ell+1}=1$.

\begin{framed}
\noindent\textsc{TwoPick}$(\vartheta)$.
\begin{enumerate}[leftmargin=*,itemsep=2pt]
\item \emph{First pick.} At a time $u\in[\vartheta_s,\vartheta_{s+1})$, the first arrival whose rank under $\succ$ among the labels arrived so far, itself included, is at most $s$ is the \emph{first pick} $z$. Nothing is picked before $\vartheta_1$. If $z$ is real and proper, accept it and let $a=\{z\}$; otherwise accept nothing and let $a=\emptyset$.
\item \emph{Second pick.} Order the labels other than $z$ by decreasing marginal $\marg xa$, with ties broken by $\succ$. Accept the first later real arrival with positive marginal that precedes, in this order, every label that arrived before $z$; then stop.
\end{enumerate}
\end{framed}

The rule makes at most $2n+1$ queries: $f(\emptyset)$, one singleton per real arrival, and at most one query $f(\{z,x\})$ per real label $x$, including labels that arrived before $z$. Virtual labels need no queries.

For $r\le N$ let $a_r=\{z_r\}$ if $z_r$ is real and proper and $a_r=\emptyset$ otherwise, let $\phi_r=f(a_r)$, let $x_{r,1},x_{r,2},\dots$ be the labels other than $z_r$ in the order of step~2 for $a=a_r$, and let $g_{r,j}=\pos{\marg{x_{r,j}}{a_r}}$. Let $\mathcal H_r=\{z_1,\dots,z_{r-1}\}$, $e_{r,j}=|\mathcal H_r\cap\{x_{r,1},\dots,x_{r,j}\}|$, and $\Theta_s=\prod_{i\le s}\vartheta_i$. Define, for $r\ge1$, $h\ge0$ and $j\ge1$,
\begin{align*}
 P_r&=\sum_{s=1}^{\ell}\Theta_s\sum_{a=0}^{\min(r-1,s-1)}\binom{r-1}a\int_{\vartheta_s}^{\vartheta_{s+1}}u^{a-s}(1-u)^{r-1-a}\,du,\\
 Q_{h,j}&=\sum_{s=1}^{\ell}\Theta_s\sum_{a=0}^{\min(h,s-1)}\binom ha\int_{\vartheta_s}^1\frac{(1-v)^{h+j-1-a}}{v}\int_{\vartheta_s}^{\min(v,\vartheta_{s+1})}u^{a-s+1}\,du\,dv.
\end{align*}

\begin{lemma}\label{ek:lem-twopick-joint}
The probability that $z_r$ is the first pick is at least $P_r$. If $g_{r,j}>0$, the probability that $z_r$ is the first pick and $x_{r,j}$ is accepted in step~2 is at least $Q_{r-1-e_{r,j},\,j}$. Moreover $Q_{h,j}$ is nonincreasing in $h$ and in $j$.
\end{lemma}

\begin{proof}
Let $z$ arrive at time $u\in[\vartheta_s,\vartheta_{s+1})$ and let $\mathcal P$ be the set of labels arriving before $u$. We claim that $z$ is the first pick if its rank among $\mathcal P\cup\{z\}$ is at most $s$ and, for each $i\le s$, the $i$th largest member of $\mathcal P$, if it exists, arrived before $\vartheta_i$. Indeed, let $y\in\mathcal P$ arrive at time $v\in[\vartheta_{s'},\vartheta_{s'+1})$ with $s'\le s$. If $y$ were among the $s'$ largest members of $\mathcal P$, it would have arrived before $\vartheta_{s'}\le v$. Otherwise these $s'$ members (or all of $\mathcal P$, if it has fewer) arrived before $\vartheta_{s'}\le v$, so $y$ has rank larger than $s'$ at its arrival. So no earlier label is picked.

For the first claim, fix $t_{z_r}=u$ in stage $s$. For a set of $a\le s-1$ labels from $\mathcal H_r$, require that its $i$th largest member arrives before $\vartheta_i$ for $i\le a$, and that the other members of $\mathcal H_r$ arrive after $u$; this has probability $\Theta_a(1-u)^{r-1-a}$, and these events are disjoint for distinct sets. Then $z_r$ has rank $a+1\le s$. Conditional on the set of labels below $z_r$ that arrive before $u$, their timestamps are independent uniforms on $[0,u)$, and we require the $i$th largest of them to arrive before $\vartheta_{a+i}$ for $i\le s-a$, which has probability at least $\prod_{i=a+1}^s\vartheta_i/u$ (a missing label only removes a factor at most one). Summing over $a$ and integrating over $u$ gives $P_r$.

For the second claim, let $z=z_r$, $x=x_{r,j}$ and $F=\{x_{r,1},\dots,x_{r,j-1}\}$, fix $t_z=u<t_x=v$ with $u$ in stage $s$, and let $\mathcal H=\mathcal H_r\setminus(F\cup\{x\})$, so that $|\mathcal H|=r-1-e_{r,j}=:h$. Require all of $F$ to arrive after $v$, a set of $a\le s-1$ members of $\mathcal H$ to arrive in order before $\vartheta_1,\dots,\vartheta_a$, and the rest of $\mathcal H$ after $v$; this has probability $\Theta_a(1-v)^{h-a+j-1}$. Every other label is below $z$ under $\succ$ and below $x$ in the marginal order. Let $\mathcal B$ be the set of such labels arriving before $v$; given $\mathcal B$, their timestamps are uniform on $[0,v)$. Require the member of $\mathcal B$ that is largest in the marginal order to arrive before $u$ (probability $u/v$, or one if $\mathcal B=\emptyset$), and, given the set $\mathcal S\subseteq\mathcal B$ of labels before $u$, whose timestamps are uniform on $[0,u)$, require the $i$th largest member of $\mathcal S$ under $\succ$ to arrive before $\vartheta_{a+i}$ for $i\le s-a$. The probability of all this is at least $\Theta_s(1-v)^{h-a+j-1}u^{a-s+1}/v$. By the first paragraph $z$ is the first pick. Every arrival in $(u,v)$ belongs to $\mathcal B\setminus\mathcal S$ and is below the largest member of $\mathcal B$, which arrived before $u$, so it is rejected. At time $v$, every label preceding $x$ in the marginal order lies in $F$ and has not arrived, and $x$ has positive marginal, hence is real; so $x$ is accepted. Summing over $a$ and integrating over $\vartheta_s\le u<\min(v,\vartheta_{s+1})$ and $v\le1$ gives $Q_{h,j}$.

For monotonicity, write $y=1-v$ and $S_h=\sum_{a=0}^{\min(h,s-1)}\binom ha y^{h-a}u^a$. The integrand of $Q_{h,j}$, summed over $a$, is $\Theta_sy^{j-1}u^{1-s}S_h/v$. Pascal's rule gives $S_{h+1}\le(y+u)S_h\le S_h$, as $u\le v$, and the factor $y^{j-1}$ is nonincreasing in $j$.
\end{proof}

Let $L=10$ and $R=20$, and fix numbers $0\le p_r\le P_r$ for $r\le L$ and $0\le q_{h,j}\le Q_{h,j}$ for $h<L$, $j\le R$. For $\ell'\le R$ and $0\le a\le\min(r-1,\ell')$ put
\[
 M_r(\ell',a)=\sum_{j=1}^{\ell'-a}q_{r-1,j}+\sum_{i=1}^aq_{r-1-i,\,\ell'-a+i}.
\]

\begin{lemma}\label{ek:lem-twopick-class}
Fix $r\le L$ and a set $\mathcal D$ of $m$ labels other than $z_r$, of which $H$ lie in $\mathcal H_r$. Suppose that $\alpha,\beta\ge0$ satisfy $\alpha a+\beta(d-a)\le M_r(\ell',a)$ whenever $1\le d\le m$, $\max(0,d-m+H)\le a\le\min(H,d)$ and $d\le\ell'\le R$, and also $\alpha H+\beta(m-H)\le\sum_{j\le R}q_{r-1,j}$. Then the expected gain of the second pick on the event that $z_r$ is the first pick is at least $\alpha\sum_{x\in\mathcal D\cap\mathcal H_r}\pos{\marg x{a_r}}+\beta\sum_{x\in\mathcal D\setminus\mathcal H_r}\pos{\marg x{a_r}}$.
\end{lemma}

\begin{proof}
The events of \cref{ek:lem-twopick-joint} for different $j$ are disjoint, so the expected gain is at least $\sum_jg_{r,j}\pi_j$ with $\pi_j=Q_{r-1-e_{r,j},j}$. Writing $g=\int_0^\infty\ind[w<g]\,dw$, it suffices to compare, at each level $w>0$, the mass $\sum_{j\le\ell'}\pi_j$ of the $\ell'$ positions with $g_{r,j}>w$ with $\alpha a+\beta(d-a)$, where $d$ is the number of labels of $\mathcal D$ among these positions and $a$ the number of those in $\mathcal H_r$. We may assume $d\ge1$. If $\ell'\le R$, mark the $a$ positions of $\mathcal D\cap\mathcal H_r$. Ignoring the other members of $\mathcal H_r$ only decreases each $\pi_j$, by the monotonicity in $h$. Among patterns with $a$ marks, moving a mark one position later replaces $Q_{h-1,j}+Q_{h-1,j+1}$ by $Q_{h,j}+Q_{h-1,j+1}$, which is not larger; so the mass is smallest with all marks last, where it equals $M_r(\ell',a)$ computed with $Q$ in place of $q$, and is therefore at least $M_r(\ell',a)\ge\alpha a+\beta(d-a)$. If $\ell'>R$ (there are at least $R$ labels besides $z_r$, thanks to the virtual labels), the first $R$ positions alone give mass at least $\sum_{j\le R}q_{r-1,j}\ge\alpha H+\beta(m-H)\ge\alpha a+\beta(d-a)$.
\end{proof}

\begin{theorem}\label{ek:thm-twopick-lp}
Let $k\in\{2,3\}$. For $r\le L$, $m\in\{k-1,k\}$ and $0\le H\le\min(r-1,m)$, let $\alpha_{r,m,H},\beta_{r,m,H}$ satisfy the hypotheses of \cref{ek:lem-twopick-class}. For $I\subseteq\{1,\dots,L\}$ with $|I|\le k$, consider the linear program in $(t_r,y_r,X_r,Y_r)_{r\le L}$
\begin{gather*}
 1\ge t_1\ge\dots\ge t_L\ge0,\qquad \textstyle\sum_{i\in I}t_i+(k-|I|)t_L\ge1,\qquad y_r\ge0,\quad y_r=0\ (r\in I),\quad \sum_ry_r\le1,\\
 0\le X_r,Y_r\le k,\qquad X_r+Y_r+t_r+y_r\ge1,\\
 X_r\le\textstyle\sum_{i\in I,\,i<r}t_i,\qquad Y_r\le\textstyle\sum_{i\in I,\,i>r}t_i+(k-|I|)t_L,
\end{gather*}
with objective $\sum_{r\le L}\bigl(p_rt_r+\alpha_{r,m_r,H_r}X_r+\beta_{r,m_r,H_r}Y_r\bigr)$, where $m_r=k-\ind[r\in I]$ and $H_r=|\{i\in I:i<r\}|$. If every such program has value at least $\rho$, then $\E[f(A)]\ge\rho\OPT$ for \textsc{TwoPick}$(\vartheta)$.
\end{theorem}

\begin{proof}
Let $O^*$ be an inclusion-minimal set with $|O^*|\le k$ and $f(O^*)=W$, and let $O$ be $O^*$ padded with virtual labels to exactly $k$ labels. Each real $o\in O^*$ has $\marg o\emptyset\ge f(O^*)-f(O^*\setminus\{o\})>0$, so it is proper. Let $I=\{r\le L:z_r\in O\}$ and $t_r=\phi_r/W$. These are nonincreasing and lie in $[0,1]$, and submodularity gives $W\le\sum_{o\in O}f(\{o\})-(k-1)b\le\sum_{z_i\in O}\phi_i$, which implies the second constraint, since $\phi_i\le\phi_L$ for $i>L$. For $r\notin I$ with $z_r$ proper let $y_r=(W-f(O\cup\{z_r\}))_+/W$, and $y_r=0$ otherwise; adding to $O$ all such $z_r$ with $y_r>0$ and using submodularity and $f\ge0$ gives $\sum_ry_r\le1$. Let $\mathcal D_r=O\setminus\{z_r\}$, which has $m_r$ labels, $H_r$ of them in $\mathcal H_r$, and let $WX_r$ and $WY_r$ be the sums of $\pos{\marg x{a_r}}$ over $x\in\mathcal D_r\cap\mathcal H_r$ and $x\in\mathcal D_r\setminus\mathcal H_r$. By submodularity, $W(X_r+Y_r)\ge f(a_r\cup O)-\phi_r\ge W(1-y_r-t_r)$. For $o\in O$, $\pos{\marg o{a_r}}\le\pos{f(\{o\})-b}\le\phi_o$, which gives the last two constraints and $X_r,Y_r\le k$.

The output value is $b$ plus the gain of the first pick plus the gain of the second. The gain of the first pick is $\phi_r-b\ge0$ when it is $z_r$, and $\sum_rp_r\le1$; with \cref{ek:lem-twopick-joint,ek:lem-twopick-class},
\[
 \E f(A)\ge b+\sum_{r\le L}p_r(\phi_r-b)+W\sum_{r\le L}\bigl(\alpha_{r,m_r,H_r}X_r+\beta_{r,m_r,H_r}Y_r\bigr)\ge W\sum_{r\le L}\bigl(p_rt_r+\alpha X_r+\beta Y_r\bigr).
\]
The right side is the objective at a feasible point of the program for $I$.
\end{proof}

\begin{proposition}\label{ek:prop-twopick-cert}
\textsc{TwoPick}$(\vartheta)$ achieves $\rho_2=0.3734$ for $k=2$ with $10^4\vartheta=(2825,7224)$ and $\rho_3=0.3111$ for $k=3$ with $10^4\vartheta=(2758,4704,6663,8222)$, for every $n\ge k$.
\end{proposition}

\begin{proof}
A computer check encloses $P_r$ and $Q_{h,j}$ in rational intervals of width below $10^{-20}$ (expanding $(1-v)^{h+j-1-a}$ binomially, using the primitives of $v^q$ and $v^q\ln v$, and enclosing logarithms by the series of $\operatorname{artanh}$ with an explicit remainder), and takes $p_r,q_{h,j}$ to be the lower endpoints rounded down to multiples of $10^{-12}$. It chooses $\beta_{r,m,H}=\frac1m\sum_{j\le m}q_{r-1,j}$ and $\alpha_{r,m,H}$ as the largest value allowed by the inequalities of \cref{ek:lem-twopick-class}, and verifies all of these inequalities exactly. For each of the $56$ and $176$ sets $I$, it reads a nonpositive rational vector $y$ and checks the weak duality bound $c^\top v\ge B^\top y+\sum_jU_j\min\{0,c_j-(A^\top y)_j\}$ for the program written as $Av\le B$, $0\le v\le U$; this bound is valid for every $y\le0$. The smallest bounds are $186733009569/(5\cdot10^{11})$ and $311117211657/10^{12}$, which exceed $\rho_2$ and $\rho_3$. \Cref{ek:thm-twopick-lp} completes the proof.
\end{proof}

\subsection{Restart windows: budgets four to twenty-nine}\label{ru:sec}

Add $k$ virtual elements as in \cref{ek:small}; they have independent uniform timestamps, $f$ ignores them, and they cost no queries. Real and virtual elements are called \emph{labels}. Fix a grid $0=t_0<t_1<\dots<t_N=1$, and call $(t_{i-1},t_i]$ \emph{cell} $i$. For every \emph{state} $(g,j)$ with $0\le g<N$ and $0\le j<k$ fix a list $0<x_1\le\dots\le x_\ell<1$, $\ell\ge1$; the list and its length may depend on the state.

\begin{framed}
\noindent\textsc{Restart}. The algorithm keeps the accepted set $S$ and the number $J$ of \emph{stops}, and proceeds in \emph{rounds}; the first round starts at $t_0=0$ with $J=0$. A round that starts at a grid point $t_g$ with $J=j$ \emph{visits} the state $(g,j)$. If $j=k$ or $g=N$, the algorithm ends. Otherwise the round uses the list of $(g,j)$, the relative time $u=(t-t_g)/(1-t_g)\in(0,1]$, and $b(u)=|\{s:x_s<u\}|$. It freezes the weights $v_e=\pos{\marg eS}$ of all labels not yet arrived ($v_e=0$ for virtual labels), ordered by decreasing weight with ties broken by a fixed label order. Until a stop occurs in this round:
\begin{itemize}[nosep]
\item the first arrival whose rank by weight among the round's arrivals so far, itself included, is at most $b(u)$ is \emph{proposed}; it is accepted if its weight is positive, and the proposal is a stop;
\item independently, a stop occurs at rate $\pos{b(u)-h(u)}/u$, where $h(u)$ is the number of arrivals in the round so far, virtual ones included.
\end{itemize}
If a stop occurs at a time in cell $i$, all arrivals up to time $t_i$ are rejected, $J$ increases by one, and the next round starts at $t_i$. If a round has no stop, the algorithm ends.
\end{framed}

So each round is a multi-rank secretary rule on the whole remaining horizon, and after a stop it is restarted, at the next grid point, with fresh weights and the list of the new state. The artificial stops are implemented with a private exponential variable compared with the integrated rate, which vanishes before $x_1$. At most one element is accepted per stop and there are at most $k$ stops, so at most $k$ elements are accepted. The weight of a label is needed only when it arrives during a round, and $S$ changes only at stops; caching $f(S)$, which was queried when the last element was accepted, the rule makes at most $n+1$ queries: $f(\emptyset)$ and at most one query $f(S\cup\{e\})$ per real arrival.

For one state $(g,j)$ with list $(x_s)_{s\le\ell}$, let $u_i=(t_i-t_g)/(1-t_g)$ for $g\le i\le N$, so that $u_g=0$ and $u_N=1$, and, for $g<i\le N$, $r\ge1$ and $s\ge0$,
\begin{gather*}
 \sigma(u)=\prod_{s:\,x_s<u}\frac{x_s}u,\qquad \beta_i=\sigma(u_{i-1})-\sigma(u_i),\qquad a_i=\int_{u_{i-1}}^{u_i}\ind[u>x_1]\,\sigma(u)\,du,\\
 \tau_{r,i}=\int_{u_{i-1}}^{u_i}\sigma(u)\Pr\bigl[\mathrm{Bin}(r-1,u)<b(u)\bigr]\,du,\qquad T_i(s)=\sum_{r=1}^s\tau_{r,i},\qquad P_s=\sum_{i=g+1}^NT_i(s).
\end{gather*}
The function $\sigma$ is continuous and nonincreasing with $\sigma=1$ on $[0,x_1]$, so $\beta_i\ge0$ and $\sum_i\beta_i=1-\sigma(1)$. When the state matters we write $u_{g,i}$, $\beta_{g,j,i}$, $a_{g,j,i}$, and so on.

\begin{lemma}[Round law]\label{ru:lem-round}
Consider one round with a fixed list, a fixed finite set of \emph{members} in a fixed strict order, independent uniform relative times on $(0,1]$, and independent private randomness.
\begin{enumerate}[label=(\roman*),nosep]
\item Fix $u\in(0,1]$ and a set $\mathcal Q$ of members. Conditional on $\mathcal Q$ being the set of members with relative time below $u$, and on the relative times of all other members, there is no stop before $u$ with probability $\sigma(u)$.
\item For every cell $i$, conditional on the set of members with relative time above $u_i$ and on their relative times, a stop occurs in cell $i$ with probability $\beta_i$.
\item For every member $e$ and cell $i$, conditional on the set of other members with relative time above $u_i$ and on their relative times, $e$ is proposed at a relative time in $(u_{i-1},u_i]$ with probability at most $a_i$.
\item The member of rank $r$ is proposed at a relative time in $(u_{i-1},u_i]$ with probability $\tau_{r,i}$. Moreover, $\tau_{r,i}$ is nonincreasing in $r$, and $T_i(s)\le\beta_i$ for every $s\ge0$.
\end{enumerate}
\end{lemma}

\begin{proof}
(i) Let $h=|\mathcal Q|$. Given the conditioning, the members of $\mathcal Q$ have independent uniform relative times on $(0,u)$, and they are the only arrivals before $u$. Let $\mathcal D$ be the event that for every $s\le\min(h,b(u))$ the $s$th best member of $\mathcal Q$ arrives no later than $x_s$. We claim that there is no eligible arrival (one of rank at most $b(\cdot)$ at its arrival) before $u$ if and only if $\mathcal D$ holds. If $\mathcal D$ fails for $s$, the $s$th best member of $\mathcal Q$ arrives at some $v\in(x_s,u)$, where its rank is at most $s\le b(v)$. Conversely, assume $\mathcal D$ and let $y\in\mathcal Q$ arrive at $v<u$, with $\ell'=b(v)\le b(u)$. If $\ell'\le h$, the $\ell'$ best members of $\mathcal Q$ arrived no later than $x_1,\dots,x_{\ell'}$, all of which are below $v$, so $y$ is not one of them and has rank larger than $\ell'$. If $\ell'>h$, every member of $\mathcal Q$, of rank $s\le h<\ell'$, arrived no later than $x_s<v$, which is impossible for $y$. This proves the claim, and $\Pr[\mathcal D]=\prod_{s\le\min(h,b(u))}x_s/u$. On $\mathcal D$ the artificial rate at $v<u$ is $\sum_{s\le b(v)}\ind[h(v)<s]/v$. The term of $s\le\min(h,b(u))$ is present only for $v>x_s$, and then vanishes, since the $s$ best members of $\mathcal Q$ arrived by $x_s$; the term of $h<s\le b(u)$ equals $1/v$ on $(x_s,u)$. So, on $\mathcal D$ and for every configuration of times, there is no artificial stop before $u$ with probability $\prod_{h<s\le b(u)}x_s/u$, and the product of the two probabilities is $\sigma(u)$.

(ii) The set of members after $u_i$ and their times are a function of the set of members after $u_{i-1}$ and their times. Averaging (i), the conditional probabilities of no stop before $u_{i-1}$ and before $u_i$ are $\sigma(u_{i-1})$ and $\sigma(u_i)$; their difference is the probability of a stop in cell $i$, since almost surely no stop occurs exactly at a grid point.

(iii) Let $u$ be the relative time of $e$. A proposal of $e$ requires $u>x_1$ and no stop before $u$. Given $u\in(u_{i-1},u_i]$, the set of other members before $u$ and the times of the other members after $u$, the latter has probability $\sigma(u)$ by (i), and this conditioning refines that of (iii), because $u\le u_i$. Integrating over $u$ gives the bound $a_i$.

(iv) Let $y$ be the member of rank $r$, arriving at $u$. It is proposed if and only if there is no stop before $u$ and fewer than $b(u)$ better members precede it. Given $u$ and the set of members preceding $y$, the first event has probability $\sigma(u)$ by (i), and the second is determined; the number of better predecessors is $\mathrm{Bin}(r-1,u)$. Integrating gives $\tau_{r,i}$, which is nonincreasing in $r$ because $\mathrm{Bin}(r-1,u)$ is stochastically increasing in $r$. For a set of $s$ members, the proposals of distinct members in cell $i$ are disjoint events contained in the event of a stop in cell $i$, so $T_i(s)\le\beta_i$ by (ii).
\end{proof}

\paragraph{The local test.}
Consider one round as in \cref{ru:lem-round}, starting in state $(g,j)$, whose members have nonnegative weights and are ordered by decreasing weight, and in which a proposed member is accepted if and only if its weight is positive. Fix a \emph{distinguished} set of exactly $h$ members. Let $\Gamma\in\{g+1,\dots,N\}$ be the cell of the stop, with $\Gamma=\infty$ if there is none; let $D_i$ be the number of distinguished members with relative time at most $u_i$, $\mathcal U$ the total weight of the distinguished members, and $Y$ the accepted weight ($0$ if nothing is accepted); on $\{\Gamma=i\}$ let $R$ be the total weight of the distinguished members with relative time at most $u_i$ that are not accepted, with $R=0$ on $\{\Gamma=\infty\}$. For each cell $i>g$ fix numbers $1\ge c^{(i)}_0\ge\dots\ge c^{(i)}_h\ge0$, let $C=c^{(i)}_{h-D_i}$ on $\{\Gamma=i\}$ and $C=0$ on $\{\Gamma=\infty\}$, and put
\[
 b_{l,i}=\E\,c^{(i)}_{l-\mathrm{Bin}(l,u_i)},\qquad \bar c_l=\sum_{i>g}\beta_i\,b_{l,i},\qquad \kappa_l=\sum_{i>g}\beta_i\,u_i\,b_{l-1,i}\quad(l\ge1).
\]

\begin{proposition}[Local test]\label{ru:prop-local}
In this setting $\E C=\bar c_h$. Let $h\ge1$ and $\delta\ge0$, and suppose that
\begin{equation}\label{ru:eq-test}
 P_r-\sum_{i>g}b_{h,i}\,T_i(r-d)\ge(\delta+\kappa_h)\,d\qquad\text{for all }r\ge1\text{ and }1\le d\le\min(r,h).
\end{equation}
Then $\E[(1-C)Y-CR]\ge\delta\,\mathcal U$. For every integer $R_0\ge h$, the tests \eqref{ru:eq-test} with $r>R_0$ follow from $P_{R_0}-\sum_{i>g}b_{h,i}\beta_i\ge(\delta+\kappa_h)h$. For $h=0$ the conclusion holds for every $\delta\ge0$.
\end{proposition}

\begin{proof}
The variable $D_i$ is determined by the set of members after $u_i$, so by \cref{ru:lem-round}(ii) the event $\{\Gamma=i\}$ has conditional probability $\beta_i$ given $D_i$; as $D_i\sim\mathrm{Bin}(h,u_i)$, $\E[C\ind[\Gamma=i]]=\beta_ib_{h,i}$, and $\E C=\bar c_h$. For $h=0$ we have $\mathcal U=R=0$ and $(1-C)Y\ge0$. Let $h\ge1$. On $\{\Gamma=i\}$ let $B$ be the total weight of the distinguished members with relative time at most $u_i$ ($B=0$ if $\Gamma=\infty$), and let $Y_{\rm out}$ be the accepted weight if the accepted member is not distinguished ($0$ otherwise). An accepted member arrived in the stop cell, so $R=B-(Y-Y_{\rm out})$ and $(1-C)Y-CR=Y-CB-CY_{\rm out}$.

For a distinguished $o$ of weight $v_o$, the event that $o$ arrives by $u_i$ and the variable $D_i$ are determined by the members after $u_i$, so by \cref{ru:lem-round}(ii), $\E[C\ind[\Gamma=i,\ o\text{ arrives by }u_i]]=\beta_iu_ib_{h-1,i}$, because given that $o$ arrives by $u_i$ the number of other distinguished members that do is $\mathrm{Bin}(h-1,u_i)$. Summing over $i$ and $o$ gives $\E[CB]=\kappa_h\,\mathcal U$.

Next, we claim that for every member $e$ that is not distinguished and every cell $i$,
\[
 \E\bigl[c^{(i)}_{h-D_i}\ind[e\text{ is proposed in cell }i]\bigr]\le b_{h,i}\Pr[e\text{ is proposed in cell }i].
\]
Condition on the relative time $u\in(u_{i-1},u_i]$ of $e$, and for every other member $y$ with relative time $X_y$ let $\tilde X_y=\max\{X_y,u\}$. These are independent, and they record which members precede $e$ and the times of the others, so by \cref{ru:lem-round}(i) the conditional probability that $e$ is proposed given $u$ and $(\tilde X_y)_y$ is $\sigma(u)G$, where $G$ indicates that fewer than $b(u)$ members better than $e$ have $\tilde X_y=u$. $G$ is nondecreasing in every coordinate $\tilde X_y$, while $h-D_i$ is the number of distinguished $o$ with $\tilde X_o>u_i$ and so $c^{(i)}_{h-D_i}$ is nonincreasing in every coordinate. By Harris' inequality these two functions of independent coordinates have nonpositive covariance, and $\E[c^{(i)}_{h-D_i}\mid u]=b_{h,i}$. Multiplying by $\sigma(u)$ and integrating over $u$ proves the claim. Multiplying the claim by the weight of $e$ and summing (an acceptance in cell $i$ is a proposal in cell $i$, on which $\Gamma=i$), we get
\[
 \E[(1-C)Y-CR]\ge\E Y-\sum_{i>g}b_{h,i}\E Y_{\rm out}^{(i)}-\kappa_h\,\mathcal U,
\]
where $Y^{(i)}_{\rm out}$ is the weight of a member that is not distinguished and is proposed in cell $i$ ($0$ if there is none).

Let $\omega_1\ge\omega_2\ge\dots\ge\omega_m\ge0$ be the weights in the fixed order, $\omega_{m+1}=0$, and $d_\rho$ the number of distinguished members among the first $\rho$. By \cref{ru:lem-round}(iv), $\E Y=\sum_r\omega_r\sum_i\tau_{r,i}$ (a proposed member of weight zero contributes zero either way) and $\sum_ib_{h,i}\E Y_{\rm out}^{(i)}=\sum_{r\text{ not dist.}}\omega_r\tilde\tau_r$ with $\tilde\tau_r=\sum_ib_{h,i}\tau_{r,i}$, which is nonincreasing in $r$. Writing $\omega_r=\sum_{\rho\ge r}(\omega_\rho-\omega_{\rho+1})$ and $v=\delta+\kappa_h$,
\[
 \E Y-\sum_ib_{h,i}\E Y_{\rm out}^{(i)}-v\,\mathcal U=\sum_{\rho=1}^m(\omega_\rho-\omega_{\rho+1})\Bigl[P_\rho-\sum_{r\le\rho,\ r\text{ not dist.}}\tilde\tau_r-v\,d_\rho\Bigr].
\]
There are $\rho-d_\rho$ undistinguished ranks among the first $\rho$, and $\tilde\tau$ is nonincreasing, so the inner sum is at most $\sum_ib_{h,i}T_i(\rho-d_\rho)$. If $d_\rho=0$ the bracket is at least $\sum_i(1-b_{h,i})T_i(\rho)\ge0$, and if $d_\rho\ge1$ it is nonnegative by \eqref{ru:eq-test}. This proves the main claim. For $r>R_0\ge h$ and $d\le h$ we have $P_r\ge P_{R_0}$, $T_i(r-d)\le\beta_i$ and $(\delta+\kappa_h)d\le(\delta+\kappa_h)h$, which gives the last assertion.
\end{proof}

\paragraph{Stop chain and harm.}
Let $\varrho\le k$ be the number of stops and $G_1<\dots<G_\varrho$ their cells, with $G_0=0$, so that the visited states are $(G_l,l)$ for $l\le\varrho$. A \emph{path} is a sequence $z=(i_1<\dots<i_\rho)$ with $\rho\le k$ and $i_\rho\le N$; put $i_0=0$, $E_l(z)=\{\varrho\ge l,\ G_1=i_1,\dots,G_l=i_l\}$ and $p(z)=\prod_{l=1}^\rho\beta_{i_{l-1},l-1,i_l}$. Let $\pi_{0,0}=1$, let all other initial values of $\pi$ and $\eta$ vanish, and for $i\ge1$ and $0\le j<k$ let
\[
 \pi_{i,j+1}=\sum_{g<i}\beta_{g,j,i}\,\pi_{g,j},\qquad \eta_{i,j+1}=\sum_{g<i}\bigl(\beta_{g,j,i}\,\eta_{g,j}+(1-t_g)\,a_{g,j,i}\,\pi_{g,j}\bigr).
\]
Fix a set $B^*$ with $|B^*|\le k$ and $f(B^*)=W$, and pad it with virtual labels to a set $O$ of exactly $k$ labels. Let $H_g$ be the number of labels of $O$ with timestamp larger than $t_g$, so that $\Pr[H_g=h]=\lambda_{g,h}:=\binom kh(1-t_g)^ht_g^{k-h}$.

For $0\le g\le N$ let $\mathcal A_g$ be the $\sigma$-field generated by the set of labels with timestamp larger than $t_g$ and their timestamps; let $\mathcal A_g^{-e}$ and $\mathcal A^O_g$ be the same for the labels other than $e$ and for the labels of $O$, respectively. These decrease in $g$. Let $\mathcal B^{(l)}_g$ be generated by the set of labels with timestamp at most $t_g$, their timestamps, and the private randomness of rounds $1,\dots,l$. On $E_l(z)$, everything that happened up to time $t_{i_l}$, including the accepted set and the frozen order of round $l+1$, is $\mathcal B^{(l)}_{i_l}$-measurable, and $E_l(z)\in\mathcal B^{(l)}_{i_l}$. Conditional on $\mathcal B^{(l)}_g$, the labels with timestamp larger than $t_g$ have independent uniform relative times, independent of the private randomness of later rounds.

\begin{lemma}[Joint law and harm]\label{ru:lem-joint}
\begin{enumerate}[label=(\alph*),nosep]
\item For every path $z=(i_1,\dots,i_\rho)$ and $0\le l\le\rho$, on $E_l(z)$ we have $\Pr[E_\rho(z)\mid\mathcal B^{(l)}_{i_l}\vee\mathcal A_{i_\rho}]=\prod_{m=l+1}^{\rho}\beta_{i_{m-1},m-1,i_m}$. In particular, $E_\rho(z)$ is independent of $\mathcal A_{i_\rho}$ and has probability $p(z)$, and the state $(g,j)$ is visited with probability $\pi_{g,j}$.
\item Let $S_{g,j}$ be the accepted set at time $t_g$. For all $g\le N$ and $j,h\le k$,
\[
 \E\bigl[\ind[(g,j)\text{ visited},\ H_g=h]\,f(O\cup S_{g,j})\bigr]\ge\lambda_{g,h}\,(\pi_{g,j}-\eta_{g,j})\,W.
\]
\end{enumerate}
\end{lemma}

\begin{proof}
(a) By induction on $\rho-l$, the case $l=\rho$ being trivial. On $E_{\rho-1}(z)$, round $\rho$ starts at $t_{i_{\rho-1}}$ with a $\mathcal B^{(\rho-1)}_{i_{\rho-1}}$-measurable member set and order and, conditionally, independent uniform relative times; the information in $\mathcal A_{i_\rho}$ beyond $\mathcal B^{(\rho-1)}_{i_{\rho-1}}$ is the set of members after relative time $u_{i_{\rho-1},i_\rho}$ and their times. By \cref{ru:lem-round}(ii), $\Pr[G_\rho=i_\rho\mid\mathcal B^{(\rho-1)}_{i_{\rho-1}}\vee\mathcal A_{i_\rho}]=\beta_{i_{\rho-1},\rho-1,i_\rho}$ on $E_{\rho-1}(z)$. As $\mathcal B^{(l)}_{i_l}\subseteq\mathcal B^{(\rho-1)}_{i_{\rho-1}}$ and $\mathcal A_{i_\rho}\subseteq\mathcal A_{i_{\rho-1}}$, the tower property and the induction hypothesis for the path $(i_1,\dots,i_{\rho-1})$ give the claim. For $l=0$ the $\sigma$-field $\mathcal B^{(0)}_0$ is trivial. The state $(g,j)$ is visited if and only if $E_j(z)$ occurs for one of the paths $z$ of length $j$ that end at $g$, and the sum of $p(z)$ over these paths satisfies the recursion for $\pi_{g,j}$.

(b) Fix a path $z$ of length $j$ ending at $g$ with $p(z)>0$, and a real element $e\notin O$. For $1\le l\le j$ let $A_l$ be the event that $E_l(z)$ holds and $e$ is accepted at the $l$th stop. As $A_l\in\mathcal B^{(l)}_{i_l}$ and $\mathcal A^O_g\subseteq\mathcal A_g$, part (a) gives $\Pr[A_l\cap E_j(z)\mid\mathcal A^O_g]=\Pr[A_l\mid\mathcal A^O_g]\prod_{m>l}\beta_{i_{m-1},m-1,i_m}$. On $E_{l-1}(z)$, $A_l$ requires $T_e>t_{i_{l-1}}$, and then $e$ is a member of round $l$, so by \cref{ru:lem-round}(iii), $\Pr[A_l\mid\mathcal B^{(l-1)}_{i_{l-1}}\vee\mathcal A^{-e}_{i_l}]\le a_{i_{l-1},l-1,i_l}\ind[T_e>t_{i_{l-1}}]$. Since $e\notin O$ and $g\ge i_l$, we have $\mathcal A^O_g\subseteq\mathcal A^{-e}_{i_l}$, and both $\{T_e>t_{i_{l-1}}\}$ and $\mathcal A^O_g$ belong to $\mathcal A_{i_{l-1}}$, of which $E_{l-1}(z)$ is independent by (a); moreover $T_e$ is independent of $\mathcal A^O_g$. Hence
\[
 \Pr[A_l\mid\mathcal A^O_g]\le a_{i_{l-1},l-1,i_l}\,(1-t_{i_{l-1}})\prod_{m<l}\beta_{i_{m-1},m-1,i_m}.
\]
Summing over $l$ and dividing by $\Pr[E_j(z)\mid\mathcal A^O_g]=p(z)$, we get $\Pr[e\in S_{g,j}\mid\mathcal A^O_g,E_j(z)]\le\Delta(z):=\sum_{l=1}^j(1-t_{i_{l-1}})a_{i_{l-1},l-1,i_l}/\beta_{i_{l-1},l-1,i_l}$. \Cref{lem:random-set}, applied conditionally to $X\mapsto f(O\cup X)$ on the real elements outside $O$, with $\mu=0$, gives $\E[f(O\cup S_{g,j})\mid\mathcal A^O_g,E_j(z)]\ge(1-\Delta(z))W$, also when $\Delta(z)>1$ since $f\ge0$. As $\ind[H_g=h]$ is $\mathcal A^O_g$-measurable and $E_j(z)$ is independent of $\mathcal A^O_g$, $\E[\ind_{E_j(z)}\ind[H_g=h]f(O\cup S_{g,j})]\ge\lambda_{g,h}\,p(z)(1-\Delta(z))W$. Summing over the paths proves (b), because $\sum_zp(z)\Delta(z)=\eta_{g,j}$: extending a path ending at $(g',j-1)$ by a step to $g$ has probability $\beta_{g',j-1,g}$ and increases $\Delta$ by $(1-t_{g'})a_{g',j-1,g}/\beta_{g',j-1,g}$, which is the recursion for $\eta$.
\end{proof}

\begin{theorem}[Restart transport]\label{ru:thm-transport}
Let $c_{g,j,h}\in[0,1]$, for $0\le g\le N$ and $0\le j,h\le k$, be nonincreasing in $h$, with $c_{N,j,h}=c_{g,k,h}=0$. For a state $(g,j)$ with $g<N$ and $j<k$ form $b_{l,i},\bar c_l,\kappa_l$ as above with the kernel of $(g,j)$ and $c^{(i)}_l=c_{i,j+1,l}$, and denote them by $b_{g,j,l,i}$, $\bar c_{g,j,l}$, $\kappa_{g,j,l}$. Let $\delta_{g,j,h}$ be real numbers with
\begin{equation}\label{rg:eq-slack}
 0\le\delta_{g,j,h}\le c_{g,j,h}-\bar c_{g,j,h}\qquad(0\le h\le k),
\end{equation}
and suppose that for every state $(g,j)$ with $\pi_{g,j}>0$ the tests \eqref{ru:eq-test} hold with $\delta=\delta_{g,j,h}$ for $1\le h\le k$. Then \textsc{Restart} satisfies
\[
 \E f(S)\ge W\sum_{g=0}^{N-1}\sum_{j=0}^{k-1}\sum_{h=0}^k\ind[\pi_{g,j}>0]\,\delta_{g,j,h}\,\lambda_{g,h}\,(\pi_{g,j}-\eta_{g,j}).
\]
\end{theorem}

The choice $\delta_{g,j,h}=c_{g,j,h}-\bar c_{g,j,h}$ is always allowed when it is nonnegative; the freedom to take $\delta$ smaller is used in \cref{rg:thm-stationary}.

\begin{proof}
For a visited state $(g,j)$ let $X_{g,j}$ be the sum of $\pos{\marg o{S_{g,j}}}$ over the $o\in O$ with timestamp at most $t_g$, and $V_{g,j}=(1-c_{g,j,H_g})f(S_{g,j})-c_{g,j,H_g}X_{g,j}$. Let $\Phi_l=V_{G_l,l}$ for $l\le\varrho$ and $\Phi_l=f(S)$, the final value, for $l>\varrho$. Then $\Phi_0=(1-c_{0,0,k})f(\emptyset)\ge0$, and $\Phi_k=f(S)$, because prices vanish in the states $(g,k)$ and $(N,j)$, where the algorithm ends.

Fix $l<k$. If $l>\varrho$, or $l\le\varrho$ and $G_l=N$, then $\Phi_{l+1}=\Phi_l=f(S)$. Otherwise condition on $\mathcal B^{(l)}_g$ and on the event that $(g,l)$ is visited and $H_g=h$, which we may assume has positive probability; so $\pi_{g,l}>0$. We are in the setting of \cref{ru:prop-local}: the members are the labels not yet arrived, with their frozen weights, the distinguished members are the $h$ unseen labels of $O$, and $c^{(i)}_m=c_{i,l+1,m}$. If the round stops in cell $i$, the next state is $(i,l+1)$, $H_i=h-D_i$ and so $c_{i,l+1,H_i}=C$; moreover $f(S_{i,l+1})=f(S_{g,l})+Y$, since an accepted element has positive frozen weight, and $X_{i,l+1}\le X_{g,l}+R$ by submodularity (a seen $o$ has a smaller positive part against the larger set, and an unseen $o$ arriving by $t_i$ contributes $0$ if accepted and at most its frozen weight otherwise). As $0\le C\le1$, $\Phi_{l+1}\ge(1-C)f(S_{g,l})-CX_{g,l}+(1-C)Y-CR$, and this also holds, with equality and $C=Y=R=0$, if the round has no stop. By \cref{ru:prop-local} with $\delta=\delta_{g,l,h}$, the conditional expectation of $\Phi_{l+1}$ is at least $(1-\bar c_{g,l,h})f(S_{g,l})-\bar c_{g,l,h}X_{g,l}+\delta_{g,l,h}\,\mathcal U$. By submodularity $f(O\cup S_{g,l})-f(S_{g,l})\le\mathcal U+X_{g,l}$, so, as $\delta_{g,l,h}\ge0$, this is at least
\[
 (1-\bar c_{g,l,h}-\delta_{g,l,h})f(S_{g,l})-(\bar c_{g,l,h}+\delta_{g,l,h})X_{g,l}+\delta_{g,l,h}f(O\cup S_{g,l})\ \ge\ \Phi_l+\delta_{g,l,h}f(O\cup S_{g,l}),
\]
where the last step uses $\bar c_{g,l,h}+\delta_{g,l,h}\le c_{g,l,h}$ from \eqref{rg:eq-slack} and $f(S_{g,l}),X_{g,l}\ge0$. Taking expectations, summing over $l<k$, and applying \cref{ru:lem-joint}(b) with the nonnegative weights $\delta_{g,j,h}$ proves the bound.
\end{proof}

\paragraph{Geometric grids and stationary prices.}
The time between a stop and the next grid point is idle, so fine grids are better; but \cref{ru:thm-transport} has one kernel and $k$ families of tests for each of the $Nk$ states. On a geometric grid all states with the same number of stops have, up to truncation, the same kernel, and the slack in \cref{ru:thm-transport} lets one family of prices serve all of them.

Fix $0<q<1$ and an integer $N\ge2$, and use the \emph{truncated geometric grid}
\[
 t_i=1-q^i\quad(0\le i<N),\qquad t_N=1,
\]
with a list in state $(g,j)$ that depends on $j$ only. The kernel of the state $(0,j)$ has relative cell ends $u_{0,i}=t_i$; we call it the \emph{shared kernel} of $j$.

\begin{lemma}[Shift invariance]\label{rg:lem-shift}
Let $g<N$. For $g\le i<N$ the relative cell ends of the state $(g,j)$ are $u_{g,i}=1-q^{i-g}=u_{0,i-g}$, and $u_{g,N}=1$. Hence, for $g<i<N$, the quantities $\beta_{g,j,i}$, $a_{g,j,i}$, $\tau_{r,i}$ and $T_i(s)$ of the state $(g,j)$ equal those of the state $(0,j)$ for the cell $i-g$, and $P_s$ is the same for both states.
\end{lemma}

\begin{proof}
For $i<N$, $u_{g,i}=(t_i-t_g)/(1-t_g)=(q^g-q^i)/q^g=1-q^{i-g}$. The quantities of a cell depend only on the list and on the two relative cell ends of that cell, and $P_s$ is the integral of the same function over all cells of the state, which partition $(0,1]$.
\end{proof}

Let $c_{j,h}\in[0,1]$, for $0\le j,h\le k$, be \emph{stationary prices}: nonincreasing in $h$, with $c_{k,h}=0$. Put $c_{g,j,h}=c_{j,h}$ for $g<N$ and $c_{N,j,h}=0$, form $b_{0,j,h,i}$, $\bar c_{0,j,h}$ and $\kappa_{0,j,h}$ for the states $(0,j)$ as in \cref{ru:thm-transport}, so that $b_{0,j,h,N}=0$, and let $\delta_{j,h}=c_{j,h}-\bar c_{0,j,h}$.

\begin{theorem}[Stationary prices]\label{rg:thm-stationary}
Consider \textsc{Restart} on the truncated geometric grid with lists that depend only on the number of stops. Suppose that for every $j<k$ we have $\delta_{j,h}\ge0$ for $0\le h\le k$, and the tests \eqref{ru:eq-test} of the state $(0,j)$ with $\delta=\delta_{j,h}$ for $1\le h\le k$. Then
\[
 \E f(S)\ge W\sum_{j=0}^{k-1}\sum_{h=0}^k\delta_{j,h}\sum_{g=0}^{N-1}\lambda_{g,h}\,(\pi_{g,j}-\eta_{g,j}),
\]
where $\lambda_{g,h}=\binom kh q^{gh}(1-q^g)^{k-h}$, $\pi_{0,0}=1$, all other initial values of $\pi$ and $\eta$ vanish, and for $1\le i<N$ and $0\le j<k$, with $\beta_{0,j,\cdot}$ and $a_{0,j,\cdot}$ from the shared kernel,
\[
 \pi_{i,j+1}=\sum_{g<i}\beta_{0,j,i-g}\,\pi_{g,j},\qquad \eta_{i,j+1}=\sum_{g<i}\bigl(\beta_{0,j,i-g}\,\eta_{g,j}+q^g\,a_{0,j,i-g}\,\pi_{g,j}\bigr).
\]
\end{theorem}

\begin{proof}
We apply \cref{ru:thm-transport} with the prices $c_{g,j,h}$ above, which lie in $[0,1]$, are nonincreasing in $h$ and vanish for $g=N$ and for $j=k$, and with $\delta_{g,j,h}=\delta_{j,h}$ for every $g<N$. Fix a state $(g,j)$ with $g<N$ and $j<k$. A stop in cell $i<N$ uses the prices $c_{j+1,\cdot}$ and a stop in cell $N$ uses the prices $0$, so by \cref{rg:lem-shift}, $b_{g,j,h,i}=b_{0,j,h,i-g}$ for $g<i<N$ and $b_{g,j,h,N}=0$. All terms being nonnegative,
\[
 \bar c_{g,j,h}=\sum_{i=1}^{N-1-g}\beta_{0,j,i}\,b_{0,j,h,i}\le\bar c_{0,j,h},\qquad \kappa_{g,j,h}=\sum_{i=1}^{N-1-g}\beta_{0,j,i}\,u_{0,i}\,b_{0,j,h-1,i}\le\kappa_{0,j,h}.
\]
So $0\le\delta_{j,h}\le c_{j,h}-\bar c_{g,j,h}$, which is \eqref{rg:eq-slack}. For the tests, write $T^{(g)}_i$ and $P^{(g)}_r$ for the quantities of the state $(g,j)$. For $h\ge1$, $r\ge1$ and $1\le d\le\min(r,h)$, \cref{rg:lem-shift} and the hypothesis give
\begin{align*}
 P^{(g)}_r-\sum_{i>g}b_{g,j,h,i}T^{(g)}_i(r-d)&=P^{(0)}_r-\sum_{i=1}^{N-1-g}b_{0,j,h,i}T^{(0)}_i(r-d)\\
 &\ge P^{(0)}_r-\sum_{i=1}^{N}b_{0,j,h,i}T^{(0)}_i(r-d)\ \ge\ (\delta_{j,h}+\kappa_{0,j,h})\,d,
\end{align*}
which is at least $(\delta_{j,h}+\kappa_{g,j,h})\,d$. Hence \cref{ru:thm-transport} applies. In its bound, $1-t_g=q^g$ for $g<N$ gives the stated $\lambda_{g,h}$, and by \cref{rg:lem-shift} the recursions of \cref{ru:lem-joint} for $\pi_{i,j}$ and $\eta_{i,j}$ with $i<N$ are the stated ones (the values at $i=N$ enter neither these recursions nor the bound). Finally, the terms with $\pi_{g,j}=0$, which the bound of \cref{ru:thm-transport} omits, are $-\delta_{j,h}\lambda_{g,h}\eta_{g,j}\le0$, so including them only decreases the right side.
\end{proof}

Thus only $k$ kernels and the tests of the $k$ states $(0,j)$ have to be checked. A state $(g,j)$ with $g>0$ is charged for stops in cells beyond the horizon that it cannot make; this is the slack, and it matters only near time $1$, where few labels of $O$ are unseen.

\begin{proposition}\label{ru:prop-cert}
For $4\le k\le29$ run \textsc{Restart} on the truncated geometric grid with the parameters $N$ and $q$ below, and in a state $(g,j)$ use the list of the numbers $y_a/(1+\zeta(k-j-1))$, $a\le5$, that are smaller than $1$. For $4\le k\le19$ the parameters $(y_a)$ and $\zeta$ are those of the row $k$; budget $20$ uses those of budget $19$, and budgets $21$--$29$ use those of the row $21$--$29$. Then \textsc{Restart} achieves the ratios $g_k$ of \cref{ek:tab} for every $n\ge k$.
\[
\begin{array}{r|l|l||r|l|l}
k&(y_1,\dots,y_5)&\zeta&k&(y_1,\dots,y_5)&\zeta\\\hline
4&(.253,\,.652,\,1.609,\,2.184,\,6.26)&1.363&12&(.128,\,.396,\,.807,\,1.338,\,1.951)&.7\\
5&(.188,\,.548,\,1.104,\,1.419,\,2.279)&.888&13&(.12,\,.378,\,.781,\,1.27,\,1.889)&.663\\
6&(.213,\,.577,\,1.22,\,3.792,\,3.792)&1.168&14&(.117,\,.369,\,.772,\,1.247,\,2.972)&.66\\
7&(.183,\,.52,\,.968,\,2.1,\,2.1)&.956&15&(.105,\,.348,\,.699,\,1.12,\,1.658)&.59\\
8&(.16,\,.471,\,.903,\,1.662,\,2.069)&.842&16&(.102,\,.354,\,.744,\,1.106,\,1.836)&.613\\
9&(.164,\,.481,\,.895,\,1.697,\,3.452)&.915&17&(.093,\,.313,\,.658,\,.939,\,1.344)&.534\\
10&(.135,\,.419,\,.845,\,1.401,\,1.877)&.73&18&(.122,\,.396,\,.798,\,1.584,\,1.584)&.745\\
11&(.14,\,.423,\,.86,\,1.479,\,2.122)&.775&19&(.1,\,.32,\,.688,\,1.099,\,1.856)&.572\\
&&&21\text{--}29&(.0583,\,.1714,\,.3629,\,.5505,\,3.449)&.2739
\end{array}
\]
\[
\begin{array}{c|c|c|c|c}
k&4\text{--}20&21\text{--}26&27,28&29\\\hline
N&2400&5000&7000&10000\\
q&499/500&4997/5000&2499/2500&9997/10000
\end{array}
\]
\end{proposition}

So the list depends only on the number $k-j$ of stops still available: with many stops left the rule is impatient, and with one stop left it is a patient rank rule with at most four thresholds.

\begin{proof}
A computer check verifies the hypotheses of \cref{rg:thm-stationary} for explicit prices and bounds its right side from below, in exact rational arithmetic and in integer interval arithmetic with endpoints in $10^{-30}\mathbb Z$, rounding every operation in the safe direction. It checks that every list is nondecreasing, positive and nonempty. It computes the integers $Z_i=\lfloor10^{30}u_{0,i}\rfloor$ exactly, so that $u_{0,i}\in[10^{-30}Z_i,10^{-30}(Z_i+1)]$, and checks $Z_0=0$, $Z_{i-1}+1<Z_i$ and $Z_{N-1}+1<10^{30}$; each comparison between a threshold $x$ and a grid point is decided by $10^{30}x<Z_i$ or $10^{30}x>Z_i+1$, one of which is verified to hold. On the cell ends, $\sigma(u_{0,i})=C_mu_{0,i}^{-m}$ with $C_m=\prod_{s\le m}x_s$ and $m$ the number of thresholds below $u_{0,i}$, which gives enclosures of $\sigma(u_{0,i})$ and of $\beta_{0,j,i}$.

The integrals are computed on the rational \emph{pseudo-cells} $(w_{i-1},w_i]$ with $w_i=10^{-30}Z_i$ for $i<N$ and $w_N=1$, which partition $(0,1]$. On a piece $(l,v]$ of a pseudo-cell on which $b(u)=m$ is constant, $\sigma(u)=C_mu^{-m}$. For $m=1$ the check uses $\int_l^v(1-u)^{r-1}u^{-1}du$, which equals $\ln(v/l)$ for $r=1$ and satisfies the recursion obtained from $(1-u)^r-(1-u)^{r-1}=-u(1-u)^{r-1}$, enclosing logarithms by the series of $\operatorname{artanh}$ with an explicit remainder. For $m\ge2$ it uses the identity
\[
 -\frac d{du}\Bigl[u^{1-m}\Pr\bigl[\mathrm{Bin}(r-1,u)\le m-2\bigr]\Bigr]=(m-1)\,u^{-m}\Pr\bigl[\mathrm{Bin}(r-1,u)\le m-1\bigr],
\]
with binomial distribution functions enclosed by repeated Bernoulli convolution. A true cell $(u_{0,i-1},u_{0,i}]$ and the pseudo-cell $(w_{i-1},w_i]$ differ by a set of measure at most $2\cdot10^{-30}$, and the integrands of $T_i(s)$ and $a_i$ lie in $[0,s]$ and $[0,1]$; adding $2s\cdot10^{-30}$ and $2\cdot10^{-30}$ to the upper bounds on the pseudo-cells gives upper bounds on $T_i(s)$ for $s\le R_0=300$ and on $a_{0,j,i}$, for $i<N$. Since the pseudo-cells partition $(0,1]$, summing the lower bounds over them gives lower bounds on $P_r$. Binomial weights $\Pr[\mathrm{Bin}(h,z)=\ell]$ with $z=u_{0,i}$ in $b_{0,j,h,i}$, and with $z=q^g$ in $\lambda_{g,h}$, are enclosed by the Bernoulli recursion evaluated at the endpoints of the enclosure of $z$.

Prices are multiples of $10^{-14}$ and are constructed backwards in $j$, starting from $c_{k,\cdot}=0$: for $h\ge1$, $c_{j,h}$ is the largest multiple of $10^{-14}$ for which the directed bounds certify the tests \eqref{ru:eq-test} of the state $(0,j)$ with $r\le R_0$ and the tail test of \cref{ru:prop-local}, and $c_{j,0}$ is the larger of $c_{j,1}$ and an upper bound on $\bar c_{0,j,0}$. The check then verifies that all prices lie in $[0,1]$ and are nonincreasing in $h$, and that a lower bound on every $\delta_{j,h}$ is nonnegative. Finally it computes lower bounds on $\pi$ and upper bounds on $\eta$ by the forward recursion of \cref{rg:thm-stationary}, using $q^g\le1-10^{-30}Z_g$, and evaluates a rational lower bound $B_k$ on the right side of \cref{rg:thm-stationary}, using lower bounds on $\lambda_{g,h}\delta_{j,h}$ where the lower bound on $\pi_{g,j}-\eta_{g,j}$ is nonnegative and upper bounds where it is negative. Truncated to twelve digits,
\[
\begin{array}{c|l||c|l||c|l}
k&B_k&k&B_k&k&B_k\\\hline
4&.297951403743&13&.272921962184&22&.268305114829\\
5&.290460641379&14&.271824826571&23&.268267262283\\
6&.286176851214&15&.271117381845&24&.268191453889\\
7&.283000017490&16&.269909964014&25&.268082989666\\
8&.280470002916&17&.269676354931&26&.267948660757\\
9&.278307443847&18&.268520013008&27&.268000701286\\
10&.276570390697&19&.268575173867&28&.267831879915\\
11&.275219632109&20&.268028259446&29&.267832032493\\
12&.274024850469&21&.268295693020&&
\end{array}
\]
and each exceeds $g_k$.
\end{proof}

The certified values improve with $N$, and the check costs of order $N^2k$ operations for the forward recursion and $Nk^2R_0$ for the tests, so grids with $10^4$ points are within reach; the grids of size $2400$ already give at least $0.2680$ for $k\le20$. We have not certified this family beyond $k=29$: the smallest entry of \cref{ek:tab} is $0.2677$, attained by the uniform-price rows for $30\le k\le39$, so a larger constant for all budgets requires better guarantees for $k\ge30$. The notation of this subsection is local to it.

\subsection{Poisson prices with attenuation: budgets thirty and larger}\label{ek:poisson}

Fix $0<p<1$, $0<T\le1$ and an integer $1\le m<k$, and let $c=k-m$, $h(t)=e^{t-T}$,
\[
 \varphi(t,j)=\E\bigl(j+\mathrm{Pois}(k(T-t))-c\bigr)_+,\qquad D(t,j)=\varphi(t,j+1)-\varphi(t,j)=\Pr\bigl[j+\mathrm{Pois}(k(T-t))\ge c\bigr].
\]
Differentiating the Poisson series gives $\partial_t\varphi(t,j)=-kD(t,j)$. Let $C_k=\varphi(0,0)=\E(\mathrm{Pois}(kT)-c)_+$.

\begin{framed}
\noindent\textsc{PoissonPrice}$(p,T,m)$, in a \emph{uniform} and an \emph{exempt} variant. Each element has an independent uniform timestamp and an independent private uniform $U_i$.
\begin{enumerate}[leftmargin=*,itemsep=2pt]
\item \emph{Sample.} As in \AP: reject the elements with timestamp below $p$, let $M$ be their largest singleton value ($0$ if none), rescale the remaining time to $[0,1]$, and reject arrivals after time $T$. $M$ is never changed.
\item \emph{Price.} Keep the accepted set $A$ and a counter $N$, and post the price $\theta_t=M\,D(t,N_{t-})/h(t)$.
\item \emph{Eligibility.} In the exempt variant, an element with $f(\{i\})>M$ is \emph{exempt}. An element $i$ is eligible at $u$ if $|A_u|<k$ and $\marg i{A_u}>0$ when it is exempt, and $\marg i{A_u}>\theta_u$ otherwise. Let $I_i(u)$ indicate this, for every $u$, and $\tau_i(t)=\int_0^tI_i(u)\,du$.
\item \emph{Decision.} When $i$ arrives at $t$, accept it if and only if $I_i(t)=1$ and $U_i\le e^{-\tau_i(t)}$. Increase $N$ by one if $i$ is accepted and not exempt. Rejections change nothing.
\end{enumerate}
\end{framed}

In the uniform variant no element is exempt and $N=|A|$. As in \cref{sec:queries}, the test $U_i\le e^{-\tau_i(t)}$ is implemented with a private rate-one Poisson process, and the algorithm stops permanently before its $(4n+1)$th probe. This gives at most $5n+1$ queries for the uniform variant and $6n+1$ for the exempt one, which also queries the singletons of nonsampled arrivals. Let $\sigma$ be the first time at which $|A|=k$ ($\sigma=\infty$ if there is none), and call an element \emph{normal} if $f(\{i\})\le M$ and \emph{exceptional} otherwise; in the uniform variant this distinction is used only in the analysis.

\begin{lemma}[Ledger]\label{ek:lem-ledger}
For both variants, with $\delta_n=(e^3/256)^n$,
\[
 \E f(A)\ge\Bigl((1-p)Te^{-T}-\int_0^Te^{t-T}\Pr[\sigma\le t]\,dt-\delta_n\Bigr)W-C_kU.
\]
\end{lemma}

\begin{proof}
First ignore the probe limit and condition on the sample $S_0$, writing $\E_0$ and $\Pr_0$. Since a rejection never changes the state and the condition $|A|<k$ is part of eligibility, the proof of \cref{lem:deletion} applies verbatim to the deletion run of each nonsampled $i$: given the other randomness, $i$ is accepted by time $t$ with probability $1-e^{-\tau_i^-(t)}\le1-e^{-t}$, and the score $z_i(t)=\ind[|A_t|<k]\,\pos{\marg i{A_t}-\epsilon_i\theta_t}$, where $\epsilon_i=0$ if $i$ is exempt and $1$ otherwise, satisfies $\E[z_i(t)\mid\omega_i]=e^{-\tau_i^-(t)}z_i^-(t)$. As in \cref{lem:net}, if $\mathrm{net}_i$ is the gain of $i$ minus $\theta_{t_i}$ when $i$ is accepted and not exempt, then $\E_0\sum_i h(t_i)\mathrm{net}_i\ge\int_0^Th(t)\,\E_0\sum_{i\in B}z_i(t)\,dt$ for every $B\subseteq E\setminus S_0$. Take $B=O^*\setminus S_0$ for an optimal set $O^*$. On $\{t<\sigma\}$ submodularity gives $\sum_{i\in B}z_i(t)\ge f(A_t\cup B)-f(A_t)-k\theta_t$; on $\{\sigma\le t\}$ all scores vanish, while $f(A_t\cup B)-f(A_t)\le f(B)-f(A_t\cap B)\le f(B)$. Hence, pathwise,
\[
 \sum_{i\in B}z_i(t)\ge f(A_t\cup B)-f(A_t)-k\theta_t\ind[t<\sigma]-f(B)\ind[\sigma\le t].
\]
Integrating by parts as in \cref{lem:value}, cancelling $\int h\,\E_0f(A_t)\,dt$, dropping $h(0)f(\emptyset)\ge0$, and using $\E_0f(A_t\cup B)\ge e^{-t}f(B)$ (\cref{lem:random-set}) and $f(B)\le W$, we get
\[
 \E_0f(A_T)\ge Te^{-T}f(B)-W\int_0^Te^{t-T}\Pr_0[\sigma\le t]\,dt+\E_0\Bigl[\int_0^{\sigma\wedge T}h\theta\,dN-k\int_0^{\sigma\wedge T}h\theta\,dt\Bigr].
\]
On $[0,\sigma\wedge T]$ the process $M\varphi(t,N_t)$ decreases at rate $kMD(t,N_t)=kh(t)\theta_t$ and jumps by $MD(t,N_{t-})=h(t)\theta_t$ at each acceptance counted by $N$, so the last bracket equals $M\varphi(\sigma\wedge T,N_{\sigma\wedge T})-M\varphi(0,0)\ge-MC_k\ge-UC_k$. Averaging over $S_0$ and using $\E f(O^*\setminus S_0)\ge(1-p)W$ (\cref{lem:random-set}) gives the bound without $\delta_n$. With probes, the execution has the same law (\cref{prop:queries}); the total number of probes is $\mathrm{Pois}(n)$, which exceeds $4n$ with probability at most $\E[4^{\mathrm{Pois}(n)}]/4^{4n}=\delta_n$, and only then does the limit change the output, by at most $W$.
\end{proof}

It remains to bound the probability that the cap binds. For $d\ge1$ let $u_d(k)$ be the unique positive root of $e^u\Pr[\mathrm{Pois}(ku)\ge d]=1$, which is nondecreasing in $d$ and nonincreasing in $k$. For $t\in[0,T]$ let $w_d(t)=\min\{t,(t-T+u_d(k))_+\}$ and $b_d=m+d-1$ for $d\le c$, and $w_{c+1}(t)=t$, $b_{c+1}=k$. Let $\Lambda$ be exponential with rate $a=p/(1-p)$, and, given $\Lambda$, let $Y$ be a Poisson process of rate $\Lambda$. Define
\[
 \psi_k(t)=\Pr\bigl[\exists d\le c+1:\ Y(w_d(t))\ge b_d\bigr],\qquad J_k=\int_0^Te^{t-T}\psi_k(t)\,dt.
\]

\begin{lemma}[Joint crossing, uniform variant]\label{ek:lem-cox}
In the uniform variant, $\Pr[\sigma\le t]\le\psi_k(t)$ for $t\in[0,T]$. Moreover, with $w_0=0$, $v_0(0)=1$, and, for $1\le d\le c+1$ and $0\le y<b_d$,
\[
 v_d(y)=\sum_{x=0}^yv_{d-1}(x)\binom yx\Bigl(\frac{a+w_{d-1}}{a+w_d}\Bigr)^{x+1}\Bigl(\frac{w_d-w_{d-1}}{a+w_d}\Bigr)^{y-x}
\]
(all other entries zero, and $0^0=1$), we have $\psi_k(t)=1-\sum_{y<k}v_{c+1}(y)$. For fixed $p,T,m$, $\psi_k(t)$ and $J_k$ are nonincreasing in $k>m$, and they do not decrease if each $\min\{T,u_d(k)\}$ is replaced by $\bar u_d\le T$ with $\bar u_d\ge\min\{T,u_d(k)\}$ and $\bar u_1\le\bar u_2\le\cdots$.
\end{lemma}

\begin{proof}
A normal marginal is at most $f(\{i\})-f(\emptyset)\le M$. If a normal element is accepted at time $r$ in state $N_{r-}=j$, then $M>0$ and $e^{T-r}\Pr[\mathrm{Pois}(k(T-r))\ge c-j]<1$, so $j<c$ and $r>T-u_{c-j}(k)$. Suppose $\sigma\le t$, and let $X(I)$ be the number of exceptional arrivals in an interval $I$. If no normal element was accepted, $X((0,t])\ge k$. Otherwise, if the last normal acceptance occurred in state $c-d$, it was followed by $m+d-1$ exceptional acceptances, all in $((T-u_d(k))_+,t]$. So $\{\sigma\le t\}$ is contained in the union of the events $X((t-w_d(t),t])\ge b_d$, $d\le c+1$.

Order $E$ by decreasing singleton value with fixed tie-breaking. If there are at least $j$ exceptional elements, none of the first $j$ is sampled, so their number is stochastically dominated by a geometric variable $G$ with $\Pr[G=j]=p(1-p)^j$. Given $S_0$, the exceptional elements have independent uniform rescaled times. Coupling the count below $G$ and using the first points of a common sequence of uniforms, the exceptional configuration is contained in a configuration of $G$ independent uniforms, which is the mixed Poisson process $Y$ restricted to $[0,1]$, because $\int_0^\infty ae^{-(a+1)\lambda}\lambda^j/j!\,d\lambda=p(1-p)^j$. The union is monotone under adding points, and reflecting $[0,t]$ about $t$ preserves the law of $Y$, which proves $\Pr[\sigma\le t]\le\psi_k(t)$.

Given the path of $Y$ on $[0,w]$ with $Y(w)=x$, $\Lambda$ has density proportional to $\lambda^xe^{-(a+w)\lambda}$, so the increment to $w'\ge w$ is negative binomial, with the transition probabilities in the recursion; $v_d(y)$ is the probability that $Y(w_d)=y$ and $Y(w_e)<b_e$ for all $e\le d$. Since $w_d$ is nondecreasing in $d$, this proves the formula. For $k'\ge k$, use the same $Y$: the windows for $d\le c$ shrink with unchanged thresholds, and the additional events and the last one are contained in $\{Y(t)\ge k\}$. Enlarging windows enlarges the union.
\end{proof}

\begin{lemma}[Exempt variant]\label{ek:lem-exempt}
In the exempt variant, $\Pr[\sigma\le t]\le r_t^m$, where $q_t=1-e^{-t}$ and $r_t=(1-p)q_t/(p+(1-p)q_t)$.
\end{lemma}

\begin{proof}
As $N$ counts normal acceptances and $D(t,j)=1$ for $j\ge c$, so that then $\theta_t\ge M$, we have $N\le c$, and $\sigma\le t$ requires at least $m$ exempt acceptances by time $t$. If an exempt $i$ is accepted at $u$, then $|A_s|<k$ and $\marg i{A_s}\ge\marg i{A_u}>0$ for $s<u$, so $\tau_i(u)=u$ and $U_i\le e^{-t_i}$. Given $S_0$, the events $\{t_i\le t,\ U_i\le e^{-t_i}\}$ are independent with probability $q_t$. The number of exempt elements is dominated by $G$ as in \cref{ek:lem-cox}, and $\mathrm{Bin}(G,q_t)$ is geometric: $\E(1-q_t+q_tz)^G=(1-r_t)/(1-r_tz)$, so its tail at $m$ is $r_t^m$.
\end{proof}

\begin{lemma}[Cost]\label{ek:lem-cost}
Fix $p,T,m$ and $K>m$, and let $\Sigma=\sum_{l=1}^{L_0}\prod_{i=1}^{l}\varsigma_i$ for some $L_0\ge1$, where $\varsigma_i=T$ for $i\le m$ and $\varsigma_i=KT/(K-m+i)$ for $i>m$. If $\bar\xi\ge1/\Sigma$ satisfies $\bar\xi T\le\ln(1+\bar\xi)$, then $C_{k+1}\le C_k$ for all $k\ge K$. The hypothesis holds with $\bar\xi=(1-T)/(T(1-T^{m-1}))$ if $m\ge2$ and $Te^{\bar\xi T}\le1$.
\end{lemma}

\begin{proof}
Let $\mu=kT$ and $H(v)=\Pr[\mathrm{Pois}(v)\ge c+1]$. Since $\partial_\mu\E(\mathrm{Pois}(\mu)-c)_+=\Pr[\mathrm{Pois}(\mu)\ge c]$, we get $C_{k+1}-C_k=\int_\mu^{\mu+T}H(v)\,dv-H(\mu)$. Here $H'(v)/H(v)=\xi(v)$ with $1/\xi(v)=\sum_{l\ge1}\prod_{i=1}^lv/(c+i)$, which increases in $v$; so $H(v)\le H(\mu)e^{\xi(\mu)(v-\mu)}$ for $v\ge\mu$ and $C_{k+1}-C_k\le H(\mu)\bigl((e^{\xi(\mu)T}-1)/\xi(\mu)-1\bigr)$. Each factor $kT/(k-m+i)$ is at least $T$ for $i\le m$ and at least $KT/(K-m+i)$ for $i>m$ and $k\ge K$, so $\xi(\mu)\le1/\Sigma\le\bar\xi$, and $(e^{xT}-1)/x$ is increasing in $x$; the hypothesis gives $(e^{\bar\xi T}-1)/\bar\xi\le1$. For the last claim, take $L_0=m-1$, so $\Sigma\ge T(1-T^{m-1})/(1-T)$, and note $(e^{\bar\xi T}-1)/\bar\xi\le Te^{\bar\xi T}$.
\end{proof}

\begin{theorem}\label{ek:thm-poisson}
Let $K>m$, let $\bar C\ge C_K$, let $\bar J\ge J_K$ for the uniform variant and $\bar J\ge\int_0^Te^{t-T}r_t^m\,dt$ for the exempt variant, and suppose that the hypothesis of \cref{ek:lem-cost} holds. Let $0<s\le1/e$. For every $n\ge k\ge K$, the algorithm that runs \textsc{PoissonPrice}$(p,T,m)$ with probability $s/(s+\bar C)$ and the singleton rule of \cref{ek:lem-singleton} otherwise satisfies
\[
 \E f(A)\ge\frac{s}{s+\bar C}\Bigl((1-p)Te^{-T}-\bar J-(2/25)^K\Bigr)W.
\]
\end{theorem}

\begin{proof}
By \cref{ek:lem-cox,ek:lem-exempt,ek:lem-cost}, at budget $k$ the quantities $J_k$ (or the exempt integral) and $C_k$ are at most $\bar J$ and $\bar C$, and $\delta_n\le(2/25)^K$ as $e^3<20.48$. By \cref{ek:lem-ledger,ek:lem-singleton}, with $\alpha=s/(s+\bar C)$ the mixture has value at least $\alpha\bigl((1-p)Te^{-T}-\bar J-\delta_n\bigr)W+\bigl((1-\alpha)s-\alpha C_k\bigr)U$, and $(1-\alpha)s=\alpha\bar C\ge\alpha C_k$.
\end{proof}

\begin{proposition}\label{ek:prop-poisson-cert}
With $s=367/1000$, \cref{ek:thm-poisson} gives the ratio $g$ for every $n\ge k\ge K$ in each of the following rows, using the uniform variant for $K\le100$ and the exempt variant for $K\ge500$.
\[
\begin{array}{r|llll@{\qquad}r|llll}
K&g&p&T&m&K&g&p&T&m\\\hline
30&.2677&.091667&.55&3 & 500&.337&.034&.736&62\\
40&.2859&.077&.59&3 & 1000&.347&.0234&.787&105\\
50&.2981&.067&.62&3 & 2500&.356&.0139&.8404&210\\
52&.3001&.065&.625&3 & 10^4&.363&.00613&.8985&587\\
70&.3134&.049429&.653&4 & 10^5&.3668&.00149&.9529&3129\\
100&.3266&.04&.697&4 & 10^6&.3676&.000356&.9784&16000
\end{array}
\]
\end{proposition}

\begin{proof}
A computer check verifies each row at $k=K$. For the uniform rows it works with integers in units of $10^{-45}$, rounding every operation in the safe direction. It finds rational $\bar u_d\ge\min\{T,u_d(K)\}$ by verifying $\Pr[\mathrm{Pois}(K\bar u_d)<d]+e^{-\bar u_d}\le1$ with directed enclosures of the exponential and the Poisson distribution function, and makes them nondecreasing; it runs the recursion of \cref{ek:lem-cox} with all products rounded down, so that one minus the retained mass is an upper bound on $\psi_K$; and it bounds $J_K$ by a right Riemann sum with $1000$ nodes, which is valid because the integrand is nondecreasing in $t$. It bounds $C_K$ by summing $\Pr[\mathrm{Pois}(\mu)=c]\sum_{l\ge1}l\prod_{j\le l}\mu/(c+j)$ for $1000$ terms with a geometric tail, and checks \cref{ek:lem-cost} with $L_0=300$ and a lower partial sum of $\ln(1+\bar\xi)$. For the exempt rows it uses rationals rounded outward to $10^{-40}\mathbb Z$, Stirling's bound $c!\ge\sqrt{2\pi c}(c/e)^c$ for $C_K$ with $3000$ terms and a geometric tail, a right Riemann sum with $400$ nodes for the integral, and the second criterion of \cref{ek:lem-cost}. It also checks $s<1/e$ and $e^3/256<2/25$. The certified ratios exceed $g$ in every row; the smallest margin is $3.3\cdot10^{-5}$, in the row $K=10^5$.
\end{proof}

\subsection{Very large budgets}\label{ek:huge}

\begin{proposition}\label{ek:prop-huge}
For every $n\ge k\ge10^{12}$ and every nonnegative submodular $f$, the algorithm of \cref{prop:queries} satisfies $\E f(A)\ge0.3678\,\OPT$.
\end{proposition}

\begin{proof}
With the parameters of \cref{sec:algorithm} and $L=\ln(k+1)$, we have $m+1\le200L+2$, $b+1\le\sqrt k+2$, $\lambda=4L/b\le4L/\sqrt k\le1$ and $\nu\le1+\lambda$, so
\[
 1-T\le\varepsilon(k):=\frac{(200L+2)(\sqrt k+2)+200L+1}k+\frac{4L}{\sqrt k}.
\]
The function $\varepsilon$ is decreasing for $k\ge10$, and $Te^{-T}\ge1/e-(1-T)^2$, as in the proof of \cref{thm:main}. So by \cref{prop:queries} the ratio is at least $(1-\frac1k)(\frac1e-\varepsilon(k)^2)-\frac{200}{(k+1)^3}-e^{-(2\ln2-1)n}$, which is nondecreasing in $k$ apart from the last term. At $k=10^{12}$ we have $L\le27.632$, $\varepsilon\le0.00564$, and the bound exceeds $0.36784$, as checked in exact arithmetic; the last term is below $e^{-10^{11}}$.
\end{proof}

\subsection{Proof of \texorpdfstring{\cref{ek:thm-every-k}}{the theorem}}\label{ek:proof}

The algorithm is chosen from $k$ alone. For $k=1$ use \cref{ek:lem-singleton}; for $k\in\{2,3\}$, \cref{ek:prop-twopick-cert}; for $4\le k\le29$, \cref{ru:prop-cert}; for $30\le k<10^{12}$, the row of \cref{ek:prop-poisson-cert} with the largest $K\le k$; and for $k\ge10^{12}$, \cref{ek:prop-huge}. The query counts are at most $n+1$, $2n+1$, $n+1$, $6n+1$ and $4n+1$, respectively. Every entry of \cref{ek:tab} is at least $0.2677$, with the minimum $0.2677$ at $30\le k\le39$ (among $k\le29$ the minimum is $0.2678$, at $k=28,29$), and every entry with $k\ge52$ is at least $0.3$.\hfill\qedsymbol

\begin{remark}
The two-pick rule, the restart rule and the uniform-price rows are each needed for some budget in \cref{ek:tab}; the exempt variant is better than the uniform rows that we certified only for $k\ge500$. The exempt variant is used for large $K$ only because its cap bound $r_t^m$ is cheap to evaluate: the recursion for $\psi_K$ costs $O(K^3)$ operations per node, so we have not certified uniform rows beyond $K=100$. The joint crossing bound of \cref{ek:lem-cox} is at least as good as the union bound $\sum_{d\le c}\bigl(w_d/(a+w_d)\bigr)^{m+d-1}+\bigl(t/(a+t)\bigr)^k$, which is cheap to evaluate and could extend the uniform rows to larger $K$.
\end{remark}

\section{Monotone objectives at every budget}\label{mon:sec}

The explicit guarantee of \cref{thm:finite} applies only to large budgets. For monotone objectives we now give explicit guarantees for every budget. They come from exact optimization over the observed prefix for $k\le21$, from a price policy that is mixed with the classical single-choice rule for $22\le k<10^6$, and from a variant of \AP{} for larger $k$ (\cref{sec:deterministic}).

Throughout this appendix $f$ is nonnegative, monotone, and submodular, $\OPT=\max_{|S|\le k}f(S)$, and $f_0=f(\emptyset)$. Since $n\ge k$ and $f$ is monotone, there is an optimal set $O$ with $|O|=k$. Every algorithm below queries $f_0$ and works with $f-f_0$, which has the same marginals. For a guarantee $\rho\le1$ this is no loss: if $\E[f(A)-f_0]\ge\rho(\OPT-f_0)$ then $\E f(A)\ge\rho\,\OPT+(1-\rho)f_0\ge\rho\,\OPT$. So in the proofs we may and do assume $f_0=0$. Arrival times are generated from timestamps as in \cref{sec:prelim}. $\Pi_\mu$ denotes a Poisson variable with mean $\mu$.

\begin{theorem}\label{mon:thm}
For every $k\ge1$ there is an algorithm that knows $n$ and $k$, accepts at most $k$ elements on every execution, queries $f$ only on subsets of arrived elements, and satisfies $\E[f(A)]\ge\rho_k\OPT$ for every $n\ge k$ and every nonnegative monotone submodular $f$, where $\rho_1=1/e$ and $\rho_k$ is strictly greater than the following values.
\begin{center}
\small
\begin{tabular}{@{}rl@{\qquad}rl@{\qquad}rl@{\qquad}rl@{}}
\toprule
$k$ & $\rho_k>$ & $k$ & $\rho_k>$ & $k$ & $\rho_k>$ & $k$ & $\rho_k>$\\
\midrule
$2$ & $0.4264$ & $10$ & $0.3734$ & $18$ & $0.3703$ & $\ge40$ & $0.4321$\\
$3$ & $0.4001$ & $11$ & $0.3728$ & $19$ & $0.3701$ & $\ge50$ & $0.4515$\\
$4$ & $0.3874$ & $12$ & $0.3723$ & $20$ & $0.3698$ & $\ge60$ & $0.4661$\\
$5$ & $0.3816$ & $13$ & $0.3719$ & $21$ & $0.3696$ & $\ge80$ & $0.4869$\\
$6$ & $0.3782$ & $14$ & $0.3716$ & $\ge22$ & $0.3706$ & $\ge98$ & $0.5000$\\
$7$ & $0.3764$ & $15$ & $0.3712$ & $\ge24$ & $0.3803$ & $\ge200$ & $0.5373$\\
$8$ & $0.3750$ & $16$ & $0.3709$ & $\ge26$ & $0.3890$ & $\ge500$ & $0.5702$\\
$9$ & $0.3741$ & $17$ & $0.3706$ & $\ge30$ & $0.4043$ & $\ge1000$ & $0.5872$\\
\bottomrule
\end{tabular}
\end{center}
For $k\ge10^6$, \cref{cor:table} gives the larger values $0.6092$, $0.6310$, and $0.63207$ for $k\ge10^6,10^9,10^{12}$. In particular $\rho_k>0.3696>1/e$ for every $k\ge2$, and $\rho_k>1/2$ for every $k\ge98$.
\end{theorem}

For $k=1$ the algorithm is $\textsc{ExactPrefix}(1,1/e)$ of \cref{mon:sec:prefix}, which accepts the first singleton record after time $1/e$. For $2\le k\le 21$ it is the exact-prefix rule of \cref{mon:sec:prefix}, which uses at most $\sum_{j\le k}\binom nj$ queries. For $22\le k<10^6$ it is the insured price policy of \cref{mon:sec:price}, which uses at most $n+1$ queries and one random coin. For $k\ge10^6$ it is the variant of \AP{} of \cref{sec:deterministic}. \Cref{mon:sec:price} lists a certified row for every $22\le k\le60$ and for selected larger $k$; the table shows some of them, and the row for $K$ applies to every $k\ge K$. The constant $1/e$ cannot be improved as a bound that holds for all $k$ simultaneously, since $k=1$ contains the classical secretary problem; nothing is claimed about optimality for a fixed $k\ge2$.

\subsection{The exact-prefix rule}\label{mon:sec:prefix}

For a set $S$ of arrived elements let $H(S)$ be the subset of $S$ of size at most $k$ that maximizes $f$, with ties broken first in favor of larger cardinality and then by a fixed total order on sets that does not depend on the arrival history. Let $g(S)=f(H(S))$. Since $f$ is monotone, $|H(S)|=\min\{k,|S|\}$. An arriving element $e$ is a \emph{candidate} if $e\in H(U)$, where $U$ is the set of arrivals so far, including $e$.

\begin{framed}
\noindent\textsc{ExactPrefix}$(k,p)$. Draw $X\sim\mathrm{Bin}(n,p)$ and reject the first $X$ arrivals. Afterwards, accept each candidate while fewer than $k$ elements have been accepted, and reject all other arrivals.
\end{framed}

With timestamps, the rejected elements are exactly those with timestamp at most $p$, so the rule accepts the first $k$ candidates after time $p$. Its value queries can be organized so that every set of at most $k$ arrived elements is queried once, when its last member arrives.

The canonical choice has the \emph{outcast property}: if $e\in S\setminus H(S)$, then $H(S\setminus\{e\})=H(S)$, because $H(S)$ is a subset of $S\setminus\{e\}$ and wins against all its subsets. For $e\in H(S)$ let $K_e=H(S\setminus\{e\})$, $d_e=g(S)-f(K_e)$, and $D_e=\marg{e}{K_e}$, and let $d(S)=\sum_{e\in H(S)}d_e$ and $D(S)=\sum_{e\in H(S)}D_e$.

\begin{lemma}\label{mon:lem:deletion}
For every set $S$, $0\le d(S)\le g(S)$, $0\le D_e\le f(\{e\})$ for each $e\in H(S)$, and $d(S)-D(S)/k\le(1-1/k)\,g(S)$.
\end{lemma}

\begin{proof}
Write $H=H(S)$, $\ell_e=\marg{e}{H\setminus\{e\}}$, and $\Delta_e=f(K_e)-f(H\setminus\{e\})$, which is nonnegative because $H\setminus\{e\}$ is feasible in $S\setminus\{e\}$. Then $d_e=\ell_e-\Delta_e$, which is nonnegative because $K_e$ is feasible in $S$, and $\sum_e\ell_e\le g(S)$ by submodularity; so $0\le d\le g$. Also $0\le D_e\le f(\{e\})$ by monotonicity and submodularity. For $x\in K_e\setminus H$ the set $(H\setminus\{e\})\cup\{x\}$ is feasible in $S\setminus\{e\}$, so $\marg{x}{H\setminus\{e\}}\le\Delta_e$. Hence, by submodularity and monotonicity,
\[
 \ell_e-D_e\le\marg{e}{H\setminus\{e\}}-\marg{e}{K_e\cup H\setminus\{e\}}=\marg{K_e}{H\setminus\{e\}}-\marg{K_e}{H}\le k\Delta_e .
\]
Adding $e$ first and then the other elements of $H$ shows $f(\{e\})+\sum_{e'\ne e}\ell_{e'}\le g(S)$; summing over $e\in H$ and adding $(k-|H|)\sum_e\ell_e\le(k-|H|)g(S)$ gives $\Sigma_1+(k-1)\sum_e\ell_e\le k\,g(S)$, where $\Sigma_1=\sum_{e\in H}f(\{e\})\ge g(S)$. Therefore $\Sigma_1-D\le\Sigma_1-\sum_e\ell_e+k\sum_e\Delta_e=\Sigma_1+(k-1)\sum_e\ell_e-kd\le k(g-d)$, and $d-D/k\le g-\Sigma_1/k\le(1-1/k)g$.
\end{proof}

Let $P_t$ be the set of elements with timestamp at most $t$, and let $G(t)=\E g(P_t)$ and $D(t)=\E D(P_t)$. These are polynomials in $t$. For a function $\psi$ of a set, $t\frac{d}{dt}\E\psi(P_t)=\E\sum_{e\in P_t}(\psi(P_t)-\psi(P_t\setminus\{e\}))$, and by the outcast property only $e\in H(P_t)$ contribute when $\psi=g$. Hence
\begin{equation}\label{mon:eq:Gprime}
 tG'(t)=\E\,d(P_t),\qquad 0\le tG'(t)\le G(t),
\end{equation}
so $G(t)/t$ is nonincreasing. Also $G(t)\ge\E f(O\cap P_t)\ge t\,\OPT$ by \cref{lem:random-set} applied to $X\mapsto f(O\setminus X)$.

\begin{lemma}\label{mon:lem:master}
Let $u=\ln(t/p)$, $P_k(u)=\sum_{j=0}^{k-1}((k-1)u)^j/j!$, and
\[
 K_G(t)=\frac{p^k}{t^{k+1}}P_k(u),\qquad K_D(t)=\frac{p^k}{t^{k+1}}\cdot\frac{P_k(u)-1}{k-1}\quad(k\ge2).
\]
For $k=1$, $\textsc{ExactPrefix}(1,p)$ has $\E f(A)\ge p\ln(1/p)\OPT$. For $k\ge2$,
\[
 \E f(A)\ \ge\ \int_p^1\bigl(K_G(t)\,G(t)+K_D(t)\,D(t)\bigr)\,dt .
\]
\end{lemma}

\begin{proof}
Let $R_r(t)$ be the set of the first $r$ candidates after time $t$ (all of them if there are fewer), $F_r(t)=\E f(R_r(t))$, and $V_r(t)=\E f(H(P_t)\cup R_r(t))$, with $R_0=\emptyset$, $F_0=0$, and $V_0=G$. The rule outputs $R_k(p)$. Let $(T_k\psi)(p)=p^k\int_p^1\psi(t)t^{-k-1}\,dt$, a positive operator.

\emph{First candidate.} Condition on an element $e$ arriving at time $t>p$ with $P_t=S$, and let $K=H(S\setminus\{e\})$. The other elements of $S$ then have independent uniform timestamps on $[0,t]$. There is no candidate in $(p,t)$ if and only if $K\subseteq P_p$: if $K\subseteq P_p$, then $H(U)=K$ for every $U$ with $K\subseteq U\subseteq S\setminus\{e\}$, so no arrival in $(p,t)$ is a candidate; conversely, if there is no candidate in $(p,t)$, the outcast property gives $H(P_p)=K$. This event has probability $(p/t)^{|K|}\ge(p/t)^k$, and on it $H(P_p)=K$. If moreover $e\in H(S)$, then $e$ is the first candidate after $p$, and it contributes $f(\{e\})$ to $f(R_1(p))$ and $D_e$ to $f(H(P_p)\cup R_1(p))-g(P_p)$. Since $\Pr[P_t=S,\ e$ arrives in $[t,t+dt]]=\Pr[P_t=S]\,dt/t$ for $e\in S$, summing over $e$ and integrating over $t$ gives $F_1\ge T_k\Sigma_1\ge T_kG$ and $V_1\ge G+T_kD$, where $\Sigma_1(t)=\E\sum_{e\in H(P_t)}f(\{e\})\ge G(t)$. For $k=1$ this gives $F_1(p)\ge p\int_p^1t^{-2}\cdot t\,\OPT\,dt=p\ln(1/p)\OPT$.

\emph{Later candidates.} Let $r\ge1$. As the cutoff moves from $t$ to $t-dt$, with probability $|P_t|\,dt/t+o(dt)$ one element of $P_t$, uniformly distributed, moves after the cutoff and becomes the first arrival after it; given $P_t$, the order of the later arrivals is uniform and does not depend on which element moved. If the moved element $e$ is a candidate, that is, $e\in H(P_t)$, then $R_r$ becomes $\{e\}\cup R_{r-1}(t)$; otherwise nothing changes. Hence, using submodularity, $R_{r-1}\subseteq R_r$, and $|H(P_t)|\le k$,
\[
 -tF_r'(t)=\E\!\!\sum_{e\in H(P_t)}\!\!\bigl(f(\{e\}\cup R_{r-1})-f(R_r)\bigr)\ \ge\ V_{r-1}(t)-F_{r-1}(t)+k\bigl(F_{r-1}(t)-F_r(t)\bigr).
\]
Multiplying by $t^{-k-1}$ and integrating from $p$ to $1$, with $F_r(1)=0$, gives $F_r\ge T_k(V_{r-1}+(k-1)F_{r-1})$. For $r\ge2$, monotonicity gives $V_{r-1}\ge V_1\ge G+T_kD$.

\emph{Iteration.} Starting from $F_1\ge T_kG$ and iterating, $F_k\ge\sum_{j=0}^{k-1}(k-1)^jT_k^{j+1}G+\sum_{j=0}^{k-2}(k-1)^jT_k^{j+2}D$. Since $(T_k^{j+1}\psi)(p)=p^k\int_p^1\frac{u^j}{j!}\psi(t)t^{-k-1}\,dt$ and $\sum_{j=0}^{k-2}(k-1)^j\frac{u^{j+1}}{(j+1)!}=\frac{P_k(u)-1}{k-1}$, this is the claim.
\end{proof}

At $p=1/e$ and $k=1$ the bound is $1/e$, which proves the case $k=1$ of \cref{mon:thm}. For $k\ge2$ we bound the right side of \cref{mon:lem:master} from below by a finite linear program. Normalize $\OPT=1$. Let $R_t$ be a Bernoulli$(t)$ subset of $E\setminus O$, let $I_j$ be an independent uniformly random $j$-subset of $O$, and let
\[
 g_j(t)=\E\,g(R_t\cup I_j),\qquad D_j(t)=\E\,D(R_t\cup I_j),\qquad 0\le j\le k .
\]
Conditional on $|O\cap P_t|=j$, the set $O\cap P_t$ is a uniform $j$-subset of $O$, independent of $P_t\setminus O$. Hence $G=\sum_j\beta_jg_j$ and $D=\sum_j\beta_jD_j$ with $\beta_j(t)=\binom kjt^j(1-t)^{k-j}$.

\begin{lemma}\label{mon:lem:profiles}
For all $t\in[0,1]$ and $0\le j\le k$, the following hold, where terms with index $-1$ are omitted and $\eta_j=(j+1)/(k-j)$.
\begin{enumerate}[label=(\roman*),nosep]
\item $j/k\le g_j\le1$, $g_k=1$, $0\le D_j\le k$, and $g_j$ is nondecreasing in $t$ and in $j$.
\item $tg_j'+(j-\alpha)g_j-jg_{j-1}-qD_j\le0$ for $(q,\alpha)\in\{(0,1),(1/k,1-1/k)\}$.
\item $g_j\ge j/k+(1-j/k)^2g_0$.
\item For $j<k$: $D_{j+1}\ge\eta_j(1-2g_j)$, $D_{j+1}\ge\eta_j\bigl(1+\frac{j(k-1)}k-kg_j\bigr)$, and $D_{j+1}\ge\eta_j\bigl(1+\frac{j^2}{k(k-j)}-\frac{k}{k-j}g_j-g_0\bigr)$.
\end{enumerate}
\end{lemma}

\begin{proof}
We use repeatedly that if $T$ is a random subset of a set $Y$ with $\Pr[y\in T]=x$ for every $y\in Y$, and $\varphi$ is nonnegative and submodular on subsets of $Y$, then $\E\varphi(T)\ge(1-x)\varphi(\emptyset)+x\varphi(Y)$; this is \eqref{eq:chain} with all inclusion probabilities equal. Throughout, $S=R\cup I_j$ for a fixed outside set $R$, $H=H(S)$, $v=g(S)$, and $h=f(I_j)$. When $|H|<k$, the cardinality tie rule and monotonicity give $H=S$. We pad a set of fewer than $k$ elements with dummy elements of zero marginal, so that it has $k$ slots.

(i) $g(S)\ge f(I_j)$, and $\E f(I_j)\ge j/k$ by the fact above; $g\le\OPT=1$; $g(R\cup O)=1$. $D_e\le f(\{e\})\le1$ and $|H|\le k$. Monotonicity in $t$ and $j$ follows by coupling $R_t\subseteq R_{t'}$ and $I_j\subseteq I_{j+1}$, since $g$ is monotone.

(ii) The expected deficit $\E\,d(R_t\cup I_j)$ equals $tg_j'(t)+j(g_j(t)-g_{j-1}(t))$: the first term comes from differentiating in the outside rate, as for \eqref{mon:eq:Gprime}, and the second from deleting one element of $I_j$, which leaves a uniform $(j-1)$-subset. Now average the inequalities $d\le g$ and $d-D/k\le(1-1/k)g$ of \cref{mon:lem:deletion}.

(iii) Let $K=H(R)$. A uniformly random set $T$ of $k-j$ of its $k$ slots gives a feasible set $I_j\cup T\subseteq S$, so $g(S)\ge\E_Tf(I_j\cup T)\ge\frac jkf(I_j)+(1-\frac jk)f(I_j\cup K)$. Averaging over $I_j$, $\E f(I_j)\ge j/k$ and $\E f(I_j\cup K)\ge(1-\frac jk)f(K)+\frac jkf(O\cup K)\ge(1-\frac jk)f(K)+\frac jk$. Averaging over $R$, with $\E g(R_t)=g_0(t)$, gives (iii).

(iv) The same argument with $H$ in place of $K$ shows that $v\ge\frac jkh+(1-\frac jk)f(I_j\cup H)$, which gives
\begin{equation}\label{mon:eq:S8}
 \marg{I_j}{H}\le\frac{j}{k-j}(v-h),\qquad \marg{H}{I_j}\le\frac k{k-j}(v-h).
\end{equation}
Call $o\in O\setminus I_j$ \emph{eligible} if $o\in H(S\cup\{o\})$, and let $L(S)$ be the sum of $\marg oH$ over eligible $o$. If $o$ is eligible, the outcast property gives $H((S\cup\{o\})\setminus\{o\})=H$, so $\marg oH$ is the term $D_o$ of $D(S\cup\{o\})$. A uniform $(j+1)$-subset of $O$ together with a uniform element $o$ of it is the same as a uniform $j$-subset $I_j$ together with a uniform $o\in O\setminus I_j$. Since all terms of $D$ are nonnegative, it follows that $D_{j+1}(t)\ge\eta_j\,\E L(R_t\cup I_j)$, and it suffices to bound $L$ from below.

If $o\in O\setminus I_j$ is not eligible, then $H(S\cup\{o\})=H$, and if $|H|=k$, exchanging $o$ for any $x\in H$ shows $\marg oH\le\marg o{H\setminus\{x\}}\le f(H)-f(H\setminus\{x\})$. The same bound holds for elements $o\in S\setminus H$. If $|H|<k$, all elements of $O\setminus I_j$ are eligible and $S\setminus H=\emptyset$.

For the first inequality, the terminal marginals $f(H)-f(H\setminus\{x\})$ sum to at most $v$, so each element of $O$ that is not eligible and not in $H$ has marginal at most $v/k$ at $H$. There are at most $k$ such elements, and $\sum_{o\in O}\marg oH\ge1-v$, so $L\ge1-2v$. For the second, let $s=v-h\ge0$. If $o$ is not eligible, then $I_j\cup\{o\}$ is feasible in $S\cup\{o\}$, so $\marg o{I_j}\le s$. If $o$ is eligible, then $\marg oH\ge\marg o{I_j\cup H}\ge\marg o{I_j}-\marg H{I_j}\ge\marg o{I_j}-\frac{k}{k-j}s$. If $C\le k-j$ elements are eligible, then $(k-j-C)s+\frac{Ck}{k-j}s\le ks$, so $L\ge\sum_{o\in O\setminus I_j}\marg o{I_j}-ks\ge1-h-ks=1+(k-1)h-kv$. For the third, if $|H|=k$ then $H\setminus I_j\subseteq R$ has at least $k-j$ elements, whose terminal marginals sum to at most $f(H\setminus I_j)\le g(R)$; so each non-eligible $o$ has $\marg oH\le g(R)/(k-j)$, and all of them together at most $g(R)$. Also $\sum_{o\in O\setminus I_j}\marg oH\ge f(O\cup H)-f(H\cup I_j)\ge1-v-\marg{I_j}H$. With \eqref{mon:eq:S8}, $L\ge1-\frac k{k-j}v+\frac j{k-j}h-g(R)$. In all three cases, take expectations, using $\E h\ge j/k$ and $\E g(R_t)=g_0(t)$.
\end{proof}

\paragraph{The linear relaxation.}
Fix $k\ge2$, $p$, and an integer $m$, and divide $[p,1]$ into cells $[a_i,a_{i+1}]$ of length $\delta=(1-p)/m$, with $a_i=p+i\delta$. For an instance, put $v_{ij}=g_j(a_i)$ for $0\le i\le m$, and $z_{ij}=\delta^{-1}\int_{a_i}^{a_{i+1}}g_j$ and $w_{ij}=\delta^{-1}\int_{a_i}^{a_{i+1}}D_j$ for $0\le i<m$. By \cref{mon:lem:profiles} these numbers satisfy the following linear constraints: the bounds of (i); $v_{ij}\le z_{ij}\le v_{i+1,j}$, $v_{ij}\le v_{i,j+1}$, and $z_{ij}\le z_{i,j+1}$; the averages over each cell of (iii) and (iv), which are linear in $z$ and $w$; and the averages of (ii), which, since $\int_a^btg'\,dt=bg(b)-ag(a)-\int_a^bg$, read
\[
 \frac{a_{i+1}v_{i+1,j}-a_iv_{ij}}{\delta}+(j-1-\alpha)z_{ij}-jz_{i,j-1}-qw_{ij}\le0 .
\]
For the objective, \eqref{mon:eq:Gprime} gives $G(t)\ge tG(a_{i+1})/a_{i+1}$ on a cell, and $G(a_{i+1})=\sum_j\beta_j(a_{i+1})v_{i+1,j}$. With the substitution $t=pe^u$,
\[
 \int_a^b tK_G(t)\,dt=\frac p{k-1}\Bigl[\Bigl(\frac pa\Bigr)^{k-1}Q\bigl(x_a\bigr)-\Bigl(\frac pb\Bigr)^{k-1}Q\bigl(x_b\bigr)\Bigr],\qquad Q(x)=\sum_{r=0}^{k-1}(k-r)\frac{x^r}{r!},\quad x_t=(k-1)\ln\frac tp .
\]
For the $D$ part, $K_D\beta_j$ is log-concave as a function of $u$: the factor $t^{j-k-1}(1-t)^{k-j}$ has logarithm $\mathrm{const}-(k+1-j)u+(k-j)\ln(1-pe^u)$, and $P_k-1$ is log-concave in $u$. Indeed, as a function of $x=(k-1)u$, $P_k'/P_k=1-\frac{x^{k-1}/(k-1)!}{P_k}$ is nonincreasing, so $(P_k')^2\ge P_kP_k''\ge(P_k-1)P_k''$, as $P_k''\ge0$. So $K_D\beta_j$ attains its minimum on a cell at an endpoint, and $\int K_D\beta_jD_j\ge\delta\min\{K_D\beta_j(a_i),K_D\beta_j(a_{i+1})\}\,w_{ij}$, as $D_j\ge0$. Hence the ratio of $\textsc{ExactPrefix}(k,p)$ is at least the minimum of
\[
 \sum_{i<m}\sum_{j\le k}\Bigl(\frac{\beta_j(a_{i+1})}{a_{i+1}}\int_{a_i}^{a_{i+1}}tK_G\,dt\Bigr)v_{i+1,j}+\sum_{i<m}\sum_{j\le k}\delta\min_{t\in\{a_i,a_{i+1}\}}K_D(t)\beta_j(t)\,w_{ij}
\]
over this polytope. All variables are nonnegative, so the objective coefficients may be replaced by smaller ones. If the program is $\min\{c^{\top}x: Ax\le b,\ \ell\le x\le\bar u\}$, then every $y\ge0$ gives the lower bound $-y^{\top}b+\sum_s\min\{r_s\ell_s,r_s\bar u_s\}$, where $r=c+A^{\top}y$.

\begin{proposition}\label{mon:prop:lp}
Let $p_k=19/50$ for $k\in\{2,3\}$, $p_k=2/5$ for $k\in\{4,5\}$, and $p_k=21/50$ for $6\le k\le21$. For each $2\le k\le21$, with $m=100$ cells, there is a vector $y\ge0$ with entries in $10^{-18}\mathbb Z$ whose lower bound exceeds the value of $\rho_k$ in \cref{mon:thm}. Hence $\textsc{ExactPrefix}(k,p_k)$ has ratio greater than that value.
\end{proposition}

\begin{proof}
A computer check rebuilds the program in exact rational arithmetic, takes precomputed rational multipliers, and evaluates the bound exactly. The objective coefficients are replaced by rational lower bounds, obtained from enclosures of $\ln(a_i/p)$ by the series $\ln z=2\sum_{r\ge0}w^{2r+1}/(2r+1)$, $w=(z-1)/(z+1)$, with an explicit remainder bound, using that $P_k$ and $Q$ have nonnegative coefficients; all roundings are directed so as to weaken the bound. The multipliers were found with a floating-point LP solver, but the validity of the bound does not depend on it. The certified values are $0.42642$ for $k=2$, decrease with $k$, and equal $0.369684$ at $k=21$.
\end{proof}

\subsection{The insured price policy}\label{mon:sec:price}

Let $k\ge2$, $0<p<1$, $0<T\le1$, and integers $1\le m<k$ and $0\le s<k$; put $c=k-m$ and $b=k-s$. Let $h(t)=e^{t-T}$, and for $t\in[0,T]$ and integers $j\ge0$ let
\[
 \varphi(t,j)=\E\pos{j+\Pi_{b(T-t)}-c},\qquad \Delta(t,j)=\varphi(t,j+1)-\varphi(t,j)=\Pr\bigl[j+\Pi_{b(T-t)}\ge c\bigr].
\]

\begin{framed}
\noindent\textsc{PoissonPrice}$(p,T,m,s)$. Reject the elements with timestamp below $p$ (the sample $S_0$), and let $M=\max_{i\in S_0}(f(\{i\})-f_0)$, or $M=0$ if $S_0=\emptyset$. Rescale the remaining time to $[0,1]$. At rescaled time $t$, with $N_{t-}$ the number of elements accepted so far, the price is
\[
 \theta_t=M\,\Delta(t,N_{t-})/h(t)=M\,e^{T-t}\,\Pr\bigl[N_{t-}+\Pi_{b(T-t)}\ge c\bigr].
\]
An arrival at time $t\le T$ is accepted if and only if fewer than $k$ elements have been accepted and its marginal value with respect to the accepted set exceeds $\theta_t$. Later arrivals are rejected.

\noindent\textsc{Record}$(p)$. Order the elements by singleton value, breaking ties by independent uniform marks. Reject the elements with timestamp below $p$, accept the first later arrival that has higher priority than every earlier arrival, and stop.

\noindent\textsc{InsuredPrice}$(p,T,m,s,\bar C)$. With probability $(1-p)/(1-p+\bar C)$ run \textsc{PoissonPrice}$(p,T,m,s)$, and otherwise run \textsc{Record}$(p)$.
\end{framed}

The price policy uses at most $n+1$ queries: $f_0$, the singletons of the sample, and one value $f(A\cup\{i\})$ per later arrival. The record rule uses at most $n$ queries. The state of the price policy is the accepted set, since $M$ is frozen and the price depends only on $N=|A|$ and $t$; a rejected element changes nothing.

\paragraph{Release times and bursts.}
For integers $b',d\ge1$, the function $u\mapsto e^u\Pr[\Pi_{b'u}\ge d]$ is continuous and strictly increasing from $0$ to $\infty$; let $u_d(b')$ be the point where it equals $1$. It is nondecreasing in $d$ and nonincreasing in $b'$. Call an element $i\notin S_0$ \emph{normal} if $f(\{i\})-f_0\le M$, and \emph{exceptional} otherwise. A normal element has marginal at most $M$, so it is accepted at time $t$ only if $\theta_t<M$; then $M>0$, and since $\Delta(t,j)=1$ for $j\ge c$, the state $N_{t-}=c-d$ must have $d\in\{1,\dots,c\}$ and $e^{T-t}\Pr[\Pi_{b(T-t)}\ge d]<1$, that is,
\begin{equation}\label{mon:eq:release}
 t>T-u_d(b).
\end{equation}
Let $Y'$ be geometric with $\Pr[Y'=\ell]=p(1-p)^\ell$ for $\ell\ge0$, and $V_1,V_2,\dots$ independent uniforms on $[0,1]$, independent of $Y'$. For $t\in[0,T]$ and $\upsilon\in(0,\infty]^c$ let $\mathfrak B_k(t;\upsilon)$ be the event that, among $V_1,\dots,V_{Y'}$, for some $d\le c$ at least $m+d-1$ lie in $((T-\upsilon_d)_+,t]$, or at least $k$ lie in $(0,t]$. Let
\[
 \pi_k(s,p)=1-p-\frac1k\E\pos{s-\mathrm{Bin}(k,p)},\qquad a=\pi_k(s,p)(1-e^{-T}),\qquad C=\E\pos{\Pi_{bT}-c},
\]
\[
 q(t)=\Pr\bigl[\mathfrak B_k(t;(u_d(b))_{d\le c})\bigr],\qquad \Lambda^*=\int_0^Te^{t-T}q(t)\,dt .
\]

\begin{theorem}\label{mon:thm:insured}
For every $\bar C\ge C$, every $n\ge k$, and every nonnegative monotone submodular $f$,
\[
 \E f(A)\ \ge\ f_0+\frac{1-p}{1-p+\bar C}\,(a-\Lambda^*)\,(\OPT-f_0)
\]
for \textsc{InsuredPrice}$(p,T,m,s,\bar C)$. The bound remains valid if each $u_d(b)$ in the definition of $q$ is replaced by a larger number.
\end{theorem}

\begin{proof}
Assume $f_0=0$. We first show
\begin{equation}\label{mon:eq:affine}
 \E f(A_{\mathrm{price}})\ge(a-\Lambda^*)\OPT-C\,\E M .
\end{equation}
Condition on $S_0$, let $R=E\setminus S_0$, and write $\E_0,\Pr_0$. The rescaled timestamps of $R$ are independent uniforms. Let $\sigma$ be the time at which the $k$th element is accepted ($\sigma=\infty$ if this does not happen), and $F(t)=\E_0f(A_t)$.

\emph{Deletion.} Fix $i\in R$ and the timestamps of the other elements, and run the policy on $E\setminus\{i\}$; mark its quantities with a superscript $-$. Let $I_i^-(t)$ indicate that $i$ would be accepted at time $t$ against this deleted run (this includes $t\le T$ and $|A_{t-}^-|<k$), and $w_i^-(t)=\marg i{A^-_{t-}}$. Since a rejected element changes nothing, the two runs agree before $i$ arrives, and afterwards if $i$ is rejected. So $i$ is accepted at time $u$ with density $I_i^-(u)$, and then gains $w_i^-(u)$ and pays $\theta_u^-$. Define the score $z_i(t)=\ind[t<\sigma]\,\pos{\marg i{A_t}-\theta_{t+}}$, where $\theta_{t+}=M\Delta(t,N_t)/h(t)$. If $i\in A_t$ then $z_i(t)=0$; otherwise the two runs agree up to $t$. So $z_i(t)\le z_i^-(t)$ on every path, and $I_i^-(t)(w_i^-(t)-\theta_t^-)=z_i^-(t)$ for all but finitely many $t$. Let $Z(t)=\sum_{i\in R}\E_0[I_i^-(t)\theta_t^-]$. Then $F(t)=\sum_{i\in R}\E_0\int_0^tI_i^-w_i^-\,du$ is Lipschitz, and for every $B\subseteq R$ and almost every $t$ (at which also $\theta_{t+}=\theta_t$ almost surely),
\[
 F'(t)-Z(t)=\sum_{i\in R}\E_0z_i^-(t)\ge\sum_{i\in B}\E_0z_i(t).
\]

\emph{Drift.} Let $|B|\le b$. If $t<\sigma$, submodularity and monotonicity give $\sum_{i\in B}z_i(t)\ge f(A_t\cup B)-f(A_t)-b\theta_{t+}\ge f(B)-f(A_t)-b\theta_{t+}$. If $t\ge\sigma$, all scores vanish and $0\ge f(B)-f(A_t)-f(B)$. Hence
\[
 F'(t)+F(t)\ge f(B)\bigl(1-\Pr_0[\sigma\le t]\bigr)+Z(t)-b\,\E_0\bigl[\theta_t\ind[t<\sigma]\bigr]\qquad\text{for a.e. }t\in[0,T].
\]

\emph{Potential.} Differentiating the Poisson expectation in its mean, $\partial_t\varphi(t,j)=-b\Delta(t,j)$. Along a path, $\varphi(t,N_t)$ decreases at rate $b\Delta(t,N_t)=bh(t)\theta_t/M$ and jumps by $\Delta(t,N_{t-})=h(t)\theta_t/M$ at each acceptance, normal or exceptional. Since $\varphi\ge0$ and $\varphi(0,0)=C$,
\[
 \int_0^{\sigma\wedge T}h\theta\,dN-b\int_0^{\sigma\wedge T}h\theta\,dt=M\bigl(\varphi(\sigma\wedge T,N_{\sigma\wedge T})-C\bigr)\ge-MC
\]
on every path. The expectations of the two integrals are $\int_0^ThZ\,dt$ and $\int_0^Th\,\E_0[\theta_t\ind[t<\sigma]]\,dt$. Multiplying the drift inequality by $h$, integrating over $[0,T]$, and using $(hF)'=h(F'+F)$ and $F(0)=0$, we get, for every $B\subseteq R$ with $|B|\le b$,
\begin{equation}\label{mon:eq:cond}
 \E_0f(A_T)\ge f(B)(1-e^{-T})-f(B)\int_0^Te^{t-T}\Pr_0[\sigma\le t]\,dt-MC .
\end{equation}

\emph{Thinned benchmark.} Let $B'=O\setminus S_0$ if $|O\setminus S_0|\le b$, and otherwise a uniformly random $b$-subset of $O\setminus S_0$. With $X=|O\cap S_0|\sim\mathrm{Bin}(k,p)$ we have $|B'|=k-\max(s,X)$, and by symmetry each element of $O$ lies in $B'$ with probability $\pi=\E|B'|/k=\pi_k(s,p)$. \Cref{lem:random-set} applied to $Y\mapsto f(O\setminus Y)$ and $Y=O\setminus B'$ gives $\E f(B')\ge\pi_k(s,p)\OPT$. Average \eqref{mon:eq:cond} over $B'$ given $S_0$, bound $f(B')\le\OPT$ in the negative term, and take expectations over $S_0$; \eqref{mon:eq:affine} follows once we show $\Pr[\sigma\le t]\le q(t)$.

\emph{The cap.} Suppose $\sigma\le t$. If no normal element was accepted, at least $k$ exceptional elements arrived in $(0,t]$. Otherwise let $r\le t$ be the time of the last normal acceptance and $N_{r-}=c-d$. By \eqref{mon:eq:release}, $r>T-u_d(b)$, and the remaining $m+d-1$ acceptances are exceptional elements arriving in $(r,t]$. So one of $c+1$ events occurs, each of the form ``at least $y$ exceptional elements arrive in a fixed interval''. Their union is increasing in the set of arrival times. The number $X'$ of exceptional elements is stochastically dominated by $Y'$: every exceptional element has a larger singleton value than every sampled element, so if $X'\ge j$, the $j$ elements that come first in a fixed order by nonincreasing singleton value are all outside $S_0$, which has probability $(1-p)^j$. Given $S_0$, the rescaled times of the exceptional elements are independent uniforms. Hence $\Pr[\sigma\le t]\le q(t)$. Enlarging $u_d(b)$ enlarges the windows, so the last statement of the theorem follows.

\emph{Insurance.} Number the elements in priority order, with values $v_1\ge\dots\ge v_n\ge0$, $v_i=f(\{i\})$, and $v_{n+1}=0$. Then $M$ is the value of the first sampled element in this order, so $\E M=\sum_j(v_j-v_{j+1})(1-(1-p)^j)$. Let $P_i$ be the probability that \textsc{Record}$(p)$ accepts $i$. Given that $i$ arrives at $t>p$ and all $i-1$ elements of higher priority arrive later, which has probability $(1-t)^{i-1}$, element $i$ is accepted if no element arrived before $t$, or if the best element that arrived before $t$ arrived before $p$; this has probability at least $p/t$. So $\sum_{i\le j}P_i\ge\int_p^1\frac p{t^2}(1-(1-t)^j)\,dt\ge(1-p)(1-(1-p)^j)$, and summation by parts gives $\E f(A_{\mathrm{record}})=\sum_j(v_j-v_{j+1})\sum_{i\le j}P_i\ge(1-p)\E M$.

\emph{Mixture.} With $\alpha=(1-p)/(1-p+\bar C)$, \eqref{mon:eq:affine} and the insurance bound give $\E f(A)\ge\alpha(a-\Lambda^*)\OPT+\bigl((1-\alpha)(1-p)-\alpha C\bigr)\E M$, and the last bracket equals $\alpha(\bar C-C)\ge0$.
\end{proof}

The event $\mathfrak B_k$ has an exact finite expression. After the reflection $V\mapsto t-V$ of $[0,t]$, its windows become initial segments $[0,\omega]$, and its complement says that the counts $\nu(\omega)=\#\{j\le Y':V_j\le\omega\}$ stay below given caps at finitely many points. These probabilities are given by the following product formula, which one checks by summing the multinomial distribution of the counts against the geometric law of $Y'$ with the negative binomial series.

\begin{lemma}\label{mon:lem:prod}
For $0=\omega_0<\omega_1<\dots<\omega_r\le1$ and integers $0=n_0\le n_1\le\dots\le n_r$,
\[
 \Pr[\nu(\omega_i)=n_i,\ 1\le i\le r]=\prod_{i=1}^r\binom{n_i}{n_i-n_{i-1}}x_i^{n_i-n_{i-1}}(1-x_i)^{n_{i-1}+1},\qquad x_i=\frac{(1-p)(\omega_i-\omega_{i-1})}{p+(1-p)\omega_i}.
\]
\end{lemma}

\begin{proof}
Let $\Omega_i=p+(1-p)\omega_i$, $\delta_i=\omega_i-\omega_{i-1}$, and $N=n_r$. Summing the multinomial probabilities over $Y'=\ell\ge N$ and using $\sum_{l\ge0}\binom{N+l}{l}z^l=(1-z)^{-N-1}$ with $z=(1-p)(1-\omega_r)$, the left side equals $\frac{N!}{\prod_i(n_i-n_{i-1})!}\prod_i((1-p)\delta_i)^{n_i-n_{i-1}}\cdot p\,\Omega_r^{-N-1}$. On the right, $x_i=(1-p)\delta_i/\Omega_i$ and $1-x_i=\Omega_{i-1}/\Omega_i$, the binomial coefficients multiply to the same multinomial coefficient, and the powers of $\Omega$ telescope to $\Omega_0/\Omega_r^{N+1}=p\,\Omega_r^{-N-1}$.
\end{proof}

\begin{lemma}\label{mon:lem:mono}
Let $0<T<1$ and integers $1\le m<K'$, $0\le s<K'$. For $L\ge1$ let $\Sigma=\sum_{l=1}^L\prod_{i=0}^{l-1}\tau_i$, where $\tau_i=T$ for $i\le m-1-s$ and $\tau_i=(K'-s)T/(K'-m+1+i)$ for $i\ge\max(0,m-s)$. If $\eta\ge1/\Sigma$ satisfies $T\eta\le\ln(1+\eta)$, then $C(k)=\E\pos{\Pi_{(k-s)T}-(k-m)}$ is nonincreasing on the integers $k\ge K'$.
\end{lemma}

\begin{proof}
Write $\mu=(k-s)T$, $c=k-m$, $j=c+1$, and $\mathcal H(\mu')=\Pr[\Pi_{\mu'}\ge j]$. Since $\partial_\mu\E\pos{\Pi_\mu-c}=\Pr[\Pi_\mu\ge c]$,
\[
 C(k+1)-C(k)=\int_\mu^{\mu+T}\mathcal H(\mu')\,d\mu'-\mathcal H(\mu).
\]
Now $(\ln\mathcal H)'=\Pr[\Pi_{\mu'}=j-1]/\Pr[\Pi_{\mu'}\ge j]=:\eta(\mu')$, and $1/\eta(\mu')=\sum_{l\ge1}\mu'^l/(j(j+1)\cdots(j+l-1))$ is increasing, so $\mathcal H(\mu')\le\mathcal H(\mu)e^{\eta(\mu)(\mu'-\mu)}$ and the difference is at most $\mathcal H(\mu)\bigl((e^{\eta(\mu)T}-1)/\eta(\mu)-1\bigr)$. This is nonpositive when $T\eta(\mu)\le\ln(1+\eta(\mu))$, and since $x\mapsto\ln(1+x)-Tx$ is concave and vanishes at $0$, it suffices that $\eta(\mu)\le\eta$. The $l$th term of the series for $1/\eta(\mu)$ is $\prod_{i<l}\frac{(k-s)T}{k-m+1+i}$, and each factor is at least $\tau_i$: for $i\le m-1-s$ because $k-m+1+i\le k-s$, and otherwise because the factor increases with $k$. So $1/\eta(\mu)\ge\Sigma\ge1/\eta$.
\end{proof}

\begin{corollary}\label{mon:cor:rows}
For each row $(K,g,p,T,m,s)$ in a finite list of certified rows (one row for every $22\le K\le60$ and $29$ rows for $65\le K\le1000$; for example, the rows with $K=22$, $98$, and $1000$ have $g=0.3706$, $0.5000$, and $0.5872$, with $(p,T,m,s)$ equal to $(0.1365,0.6974,2,3)$, $(0.0482,0.7982,2,5)$, and $(0.0073,0.9060,2,7)$), there is a rational $\bar C$ such that, for every $k\ge K$, \textsc{InsuredPrice}$(p,T,m,s,\bar C)$ run with budget $k$ has ratio greater than $g$.
\end{corollary}

\begin{proof}
Fix a row and $k\ge K$. We bound the quantities of \cref{mon:thm:insured} at budget $k$ uniformly in $k$.
(i) $\pi_k(s,p)\ge\pi_K(s,p)$, since $\mathrm{Bin}(k+1,p)$ is $\mathrm{Bin}(k,p)$ plus an independent Bernoulli variable and the factor $1/k$ decreases.
(ii) The check finds $K'\ge K$ satisfying the hypothesis of \cref{mon:lem:mono}, so $C(k)\le\bar C:=\max_{K\le\kappa\le K'}C(\kappa)$, rounded up. (In all rows $K'=K$.)
(iii) The check finds numbers $\bar u_1\le\dots\le\bar u_{K-m}\le T$ such that for each $d$ either $\bar u_d=T$ or $e^{\bar u}\Pr[\Pi_{(K-s)\bar u}\ge d]\ge1$ for some $\bar u\le\bar u_d$; in the second case $u_d(k-s)\le u_d(K-s)\le\bar u_d$. In both cases the $d$th window at budget $k$ is contained in $((T-\bar u_d)_+,t]$. For $d>K-m$, and for the final event, at least $K$ points lie in $(0,t]$. Hence $q_k(t)\le\bar q(t):=\Pr[\mathfrak B_K(t;\bar u)]$. As $e^{t-T}\bar q(t)$ is nondecreasing in $t$, a right-endpoint Riemann sum $\bar\Lambda$ bounds $\Lambda^*$, and each $\bar q(t_i)$ is computed from \cref{mon:lem:prod} by a recursion over at most $K-m+1$ points.
By \cref{mon:thm:insured} with the enlarged release times, it remains to check $\frac{1-p}{1-p+\bar C}\bigl(\pi_K(s,p)(1-e^{-T})-\bar\Lambda\bigr)>g$. A computer check verifies all of this in integer fixed-point arithmetic with denominator $10^{50}$ and directed rounding, with rigorous bounds for exponentials, logarithms, and Poisson probabilities; floating point is used only to propose the numbers $\bar u_d$, which are then verified exactly.
\end{proof}

\begin{proof}[Proof of \cref{mon:thm}]
For $k=1$, use $\textsc{ExactPrefix}(1,1/e)$ and \cref{mon:lem:master}. For $2\le k\le21$, use \cref{mon:prop:lp}. For $22\le k<10^6$, use the row of \cref{mon:cor:rows} with the largest $K\le k$; the values in the table are those of rows. For $k\ge10^6$, \cref{cor:table} gives at least $0.6092$, which exceeds $0.5872$. Finally, all values for $2\le k\le21$ exceed $0.3696$, every row has $g\ge0.3706$, and $0.3696>1/e$.
\end{proof}

\subsection{Large budgets}\label{sec:deterministic}

For $k\ge10^6$ we use a variant of \AP{} without the attenuation test, whose smaller reserve gives explicit guarantees already for $k\ge8\cdot10^4$ (\cref{cor:table}), which the constants in \cref{sec:algorithm} do not. Here the prices use a reflected queue rather than an exact potential.

For $k\ge2$ set $b=\ceil{\sqrt k}$, $p=1/b$, $m=\ceil{2\ln k}$, $\lambda=\ln(k+1)/b$, $T=1-((m+1)(b+1)+m)/k$, and $h(t)=e^{t-T}$. Assume $T>0$. Use the sample, the time rescaling, and the horizon of \AP. Maintain $Q\ge0$, starting at $0$, which drains at rate $k$ while positive. The price is $\theta_t=\Phi_M'(Q)/h(t)$, where
\[
 \Phi_M(q)=\frac Mk\Bigl(\frac{e^{\lambda q}-1}{\lambda}-q\Bigr),\qquad\Phi_M'(q)=\frac Mk(e^{\lambda q}-1).
\]
On arrival of $e$ with marginal $w$: if $f(\{e\})>M$, set $M\gets f(\{e\})$, $Q\gets 0$, and accept $e$ if and only if $w>0$; otherwise accept $e$ if and only if $w>\theta_t$, and set $Q\gets Q+1$ on acceptance. Stop at $k$ acceptances.

\begin{theorem}\label{thm:deterministic-variant}
With $D_\lambda=(e^\lambda-1-\lambda)/\lambda$ and $\epsilon_k=(b+2)b^34^{-m}/3$,
\[
 \frac{\E[f(A)]}{\OPT}\ge\frac{(1-1/b)(1-e^{-T})-D_\lambda}{1+D_\lambda}-\epsilon_k.
\]
\end{theorem}

\begin{proof}
Analyze the uncapped execution, conditional on the sample. Define scores, the deleted run, and net gain as before, with $\tau\equiv0$. The conclusion of \cref{lem:deletion}(c) becomes $z_i(t)\le z_i^-(t)$: either $i$ was accepted and $z_i(t)=0$, or the runs agree, or $i$ was a rejected record of nonpositive marginal and both scores vanish. Also $\E[\mathrm{net}_i\mid\omega_i,t_i=t]\ge z_i^-(t)\ge\E[z_i(t)\mid\omega_i]$, so \cref{lem:net} holds. As in \cref{lem:value}, using $f(A_t\cup O)\ge f(O)$ directly,
\[
 \E_0[f(A_T)]\ge(1-e^{-T})f(O)+\E_0\Bigl[\int_0^Th\theta\,dN-k\int_0^Th\theta\,dt\Bigr].
\]

\emph{Prices.} Between arrivals, $\Phi_M(Q)$ falls at rate $k\Phi_M'(Q)=kh\theta$, also at $Q=0$. A record resets it to zero. A normal acceptance increases it by exactly $\Phi_M'(Q)+D_\lambda(\Phi_M'(Q)+M/k)$, and $\Phi_M'(Q)=h\theta<w$. Since $\Phi\ge0$, telescoping gives $\int h\theta\,dN-k\int h\theta\,dt\ge-D_\lambda(f(A_T)+\OPT\cdot B/k)$, where $B$ is the uncapped number of acceptances.

\emph{Counts.} If $Q\ge b$, then $\theta_t\ge M$ and no normal element is accepted. So $Q<b+1$ and between records there are at most $k\ell+b+1$ normal acceptances in time $\ell$. With $J$ the number of strict singleton records after the sample and before $T$, it follows that $B\le kT+(J+1)(b+1)+J=k+(b+2)(J-m)$.

\emph{Records.} Let $X$ be the number of tie-broken records after the sample. Given that $S\sim\mathrm{Bin}(n,p)$ elements are sampled, $\E[4^X\mid S]=\prod_{x=S+1}^n(1+3/x)=\frac{(n+1)(n+2)(n+3)}{(S+1)(S+2)(S+3)}$. The identity $\E[1/((S+1)(S+2)(S+3))]\le 1/(p^3(n+1)(n+2)(n+3))$ then gives $\E[4^X]\le b^3$, so $\Pr[J\ge j]\le b^34^{-j}$ and $\E[(J-m)_+]\le b^34^{-m}/3$. The element ranked $x$ makes a record at time $u$ only if the $x-1$ better-ranked elements arrive later, so the density of records at $u$ is at most $\sum_x(1-u)^{x-1}\le1/u$. Thus $\E[J]\le\ln b\le m$ and $\E[B]\le k$.

Combining these bounds yields $(1+D_\lambda)\E[f(A_T)]\ge(1-e^{-T})\E[f(O)]-D_\lambda\OPT$. Since $f$ is monotone, \cref{lem:random-set} gives $\E[f(O^*\setminus S_0)]\ge(1-p)\OPT$. Finally $\E[f(A_T)]-\E[f(A)]\le\OPT\,\E[(B-k)_+]\le \epsilon_k\OPT$.
\end{proof}

\begin{corollary}\label{cor:table}
For every $n\ge k$ and monotone objective, the algorithm above has the following finite guarantees.
\begin{center}
\begin{tabular}{@{}ll@{}}
\toprule
Budget & Ratio strictly greater than \\
\midrule
$k\ge8\cdot10^4$ & $0.5644$ \\
$k\ge10^6$ & $0.6092$ \\
$k\ge10^9$ & $0.6310$ \\
$k\ge10^{12}$ & $0.63207$ \\
\bottomrule
\end{tabular}
\end{center}
\end{corollary}

\begin{proof}
For $k\ge400$ put $u(k)=(2\ln k+2)/\sqrt k+(6\ln k+5)/k$ and $\bar\lambda(k)=\ln(k+1)/\sqrt k$. Then $1-T\le u$, $\lambda\le\bar\lambda$, and $p\le k^{-1/2}$, using $b\le\sqrt k+1$ and $m\le2\ln k+1$. Using $m\ge 2\ln k$ and $4\ln 2>11/4$ one checks that $\epsilon_k<\frac12k^{-3/4}$. Both $u$ and $\bar\lambda$ are decreasing in this range, and $(e^x-1-x)/x$ is increasing in $x$. The bound of \cref{thm:deterministic-variant} is increasing in $T$ and decreasing in $D_\lambda$ and in $p$, so it is at least
\[
 \mathcal{L}(k)=\frac{(1-k^{-1/2})(1-e^{-(1-u(k))})-\overline D(k)}{1+\overline D(k)}-\frac12k^{-3/4},
 \qquad\overline D(k)=\frac{e^{\bar\lambda(k)}-1-\bar\lambda(k)}{\bar\lambda(k)},
\]
which is increasing in $k$ whenever $u(k)<1$. A computer check verifies in exact rational arithmetic that $u(8\cdot10^4)<1$ and that $\mathcal L$ exceeds each constant at the corresponding endpoint.
\end{proof}

\section{A better bound from fans}\label{bh:sec}

\Cref{thm:transfer} charges every rejection of a carrier of an invisible bad atom at full price, and it ignores the rank that a rejected item takes away from its own gadget. For one large fan per level both effects can be computed, and they give a better constant. For $t\in[0,1]$ let
\[
 \varrho(t)=\frac t{1-t+t^2},\quad \psi(t)=\int_t^1\varrho,\quad \chi=1-\varrho^2,\quad \omega=\chi-\psi^2,\quad \psi_2(t)=t\,\psi(t)^2 .
\]
Heuristically, $\varrho(t)$ is the probability that an endpoint of a bad edge of an infinite fan, with independent uniform arrival times, is joined to the hub by edges arriving before time~$t$, so that a bad edge arriving at time $t$ increases the rank with probability $\chi(t)$; the rigorous finite version is the quantity $\varrho_L$ below. Let $\kappa=\int_0^1\varrho^2$. With $q=1-t+t^2$ we have $\varrho^2=t^2/q^2$, and the symmetry $t\mapsto1-t$ and $t^2+(1-t)^2=2q-1$ give $\kappa=\int_0^1q^{-1}\,dt-\frac12\int_0^1q^{-2}\,dt=\frac{2\sqrt3\pi}9-\frac12\bigl(\frac{4\sqrt3\pi}{27}+\frac23\bigr)=\frac{4\sqrt3\pi}{27}-\frac13=0.47279\ldots$. Since $1-t+t^2\ge\frac34$, we have $\varrho(t)\le\frac43t$ and $\psi\le\psi(0)\le\frac23$. Let
\[
 V(x)=\kappa+\int_0^x\bigl(\omega(t)-2t^2\psi(t)^2\bigr)\,dt,\qquad x_0=\tfrac1{24},\quad c_1=V'(x_0),\quad c_0=V(x_0)-c_1x_0,\quad \delta=\tfrac{37}{600},
\]
\[
 C_{\mathrm{fan}}=\frac{1-\delta}2+\delta c_0+\frac{\delta^2c_1^2}{2(1-\delta)}=0.4991365873\ldots .
\]

\begin{theorem}\label{bh:thm}
For every $\varepsilon>0$ there is a finite family $\mathcal F$ as in \cref{thm:transfer} such that, for every algorithm as there and every $k\ge n_0$, some $f\in\mathcal F$ satisfies $\E[f(A)]\le(C_{\mathrm{fan}}+\varepsilon)\OPT$, and every member of $\mathcal F$ has largest singleton value at most $\varepsilon\OPT$. In particular, the best ratio $\rho_k$ with budget $k$ satisfies $\limsup_{k\to\infty}\rho_k\le C_{\mathrm{fan}}<0.4991366$.
\end{theorem}

The construction is that of \cref{thm:transfer} with $\delta=37/600$, except that the copies of $\mathsf M$ at a level $j>N$ are replaced by a single fan. Let $R_j=\delta n_jQ$, and let $\mathsf F_j$ be the \emph{fan} of rank $R_j$: the graphic matroid on the vertices $\{0,1,\dots,R_j\}$ with spokes $g_i=0i$ and bad edges $b_i=(i-1)i$, where $b_1$ and $g_1$ are parallel. Its value is $u_j\varphi_{\mathsf F_j}$. Its $R_j$ spokes are the remaining $\delta Q$ good atoms of the $n_j$ items of level $j$, assigned by a uniformly random bijection, so that each item \emph{owns} $\delta Q$ spokes. A fixed set $\mathcal B_{j,d}$ of $\pi_dR_j$ bad edges is sent to level $j-d$, for each $1\le d\le N$; these sets partition the bad edges. The counts $s_\ell$, the injections, and the sizes $n_\ell$ are as before, with $r=1$ in the definition of $n_\ell$, the bijections are independent of each other and of the injections, and $\mathcal F$ now consists of the objectives for all choices of injections and bijections. Since the bad path is a spanning tree and the spokes of an item are independent, \cref{lem:transfer-valid} holds verbatim. We let $\mathcal T_j$ be the set of bad edges of $\mathsf F_j$ that increase the rank of the set of atoms of $\mathsf F_j$ that arrived strictly before them. Each bad edge is the only atom of $\mathsf F_j$ carried by its carrier, so the relaxation $\bar f\ge f$, the paragraph on invisible roles, and the bound \eqref{eq:pairbound} on the pair part carry over unchanged. Only the matroid bound \eqref{eq:matbound} is replaced.

\paragraph{One fan.}
Fix a level $j>N$ and write $R=R_j$, $n=n_j$, $q=\delta Q$, so $R=nq$. Let $o(i)$ be the owner of $g_i$, let $u_i$ be its arrival time, and let $z_i$ be the arrival time of the carrier of $b_i$. Given the injections, the carriers are distinct items of lower levels, so the $z_i$ are independent uniform variables, independent of the bijection and of the arrival times $(t_a)$ of the level-$j$ items. Let $\epsilon_j=\sup_t|F_j(t)-t|$, where $F_j(t)$ is the fraction of level-$j$ items that arrive by time $t$. Let $G_{<t}$ be the graph on $\{0,\dots,R\}$ formed by the spokes and bad edges that arrive before time $t$. A bad edge $b_i$ lies in $\mathcal T_j$ exactly when its endpoints are not connected in $G_{<z_i}$.

For an integer $L\ge1$, a vertex $p$ and a time $t$, let $\mathsf L_p(t)$ be the event that some $k$ with $p-L<k\le p$ has $u_k<t$ and $z_{k+1},\dots,z_p<t$, and let $\mathsf R_p(t)$ be the event that some $k$ with $p\le k<p+L$ has $u_k<t$ and $z_{p+1},\dots,z_k<t$. Each of them implies that $p$ is joined to $0$ in $G_{<t}$. If all the times involved are independent and uniform, then, scanning from $p$, each has probability
\[
 \varrho_L(t)=t\sum_{m<L}\bigl(t(1-t)\bigr)^m,\qquad 0\le\varrho(t)-\varrho_L(t)=\varrho(t)\bigl(t(1-t)\bigr)^L\le4^{-L}.
\]
If the owner times are replaced by the empirical law of $(t_a)$, the probability changes by at most $L\epsilon_j$, since each of the at most $L$ spokes involved is early with probability within $\epsilon_j$ of $t$.

\begin{lemma}\label{bh:bd}
Let $X=h(\tau,z)$, where $\tau$ is a uniformly random bijection between two sets of size $R$, and $z=(z_1,\dots,z_R)$ are independent and independent of $\tau$. If $h$ changes by at most $c$ when two values of $\tau$ are swapped or one $z_i$ is changed, then $\Pr[X\le\E X-\Delta]\le e^{-\Delta^2/(4Rc^2)}$ and $\Pr[X\ge\E X+\Delta]\le e^{-\Delta^2/(4Rc^2)}$ for every $\Delta>0$.
\end{lemma}

\begin{proof}
Reveal $\tau$ one value at a time and then $z_1,\dots,z_R$. At a step of the first kind, the conditional laws of $\tau$ given two different values at the current point are exchanged by a swap, so the corresponding conditional expectations of $X$ differ by at most $c$, and so does each of them from their average. Thus the Doob martingale has $2R$ increments bounded by $c$, and the Azuma--Hoeffding inequality applies.
\end{proof}

For $Z\subseteq[n]$, let $\lambda_Z$ be the measure on $[0,1]$ with mass $1/n$ at $t_a$ for each $a\in Z$, let $G_{\bar Z}$ be the set of spokes owned outside $Z$, and let
\[
 \Gamma(\lambda)=\int\psi^2\,d\lambda+\iint\psi_2(u\vee u')\,d\lambda(u)\,d\lambda(u').
\]

\begin{lemma}[Rank credit]\label{bh:credit}
Let $L\ge1$ and $\zeta\in(0,1)$. Conditionally on $(t_a)$, with probability at least $1-\zeta$, for every $Z\subseteq[n]$,
\[
 R-r_{\mathsf F_j}(G_{\bar Z}\cup\mathcal T_j)\ge R\bigl(\Gamma(\lambda_Z)-\theta\bigr),\quad
 \theta=4L\epsilon_j+4^{1-L}+\frac{(2L+2)^2}R+(8L+8)\sqrt{\frac{n\ln2+\ln(1/\zeta)}R}.
\]
\end{lemma}

\begin{proof}
For $L<i<R-L$ consider the events
\begin{align*}
 E^1_i&:\ u_i<\min(z_i,z_{i+1}),\ \ \mathsf L_{i-1}(z_i),\ \ \mathsf R_{i+1}(z_{i+1});\\
 E^2_i&:\ z_{i+1}<\max(u_i,u_{i+1})<\min(z_i,z_{i+2}),\ \ \mathsf L_{i-1}(z_i),\ \ \mathsf R_{i+2}(z_{i+2}).
\end{align*}
On $E^1_i$, both $b_i$ and $b_{i+1}$ join two vertices that are already joined to $0$, so neither lies in $\mathcal T_j$, and $\{i\}$ is a component of the forest $\mathcal T_j$. On $E^2_i$, when $b_{i+1}$ arrives, the only other edges at $i$ and $i+1$ that have arrived are spokes, and at most one of them; so $i$ and $i+1$ are not connected and $b_{i+1}\in\mathcal T_j$. When $b_i$ and $b_{i+2}$ arrive, both spokes have arrived, so $b_i,b_{i+2}\notin\mathcal T_j$, and $\{i,i+1\}$ is a component of $\mathcal T_j$. In particular $E^1_i$ and $E^2_i$ are disjoint. Let $N(Z)$ be the number of indices $i$ such that $E^1_i$ holds and $o(i)\in Z$, or $E^2_i$ holds and $o(i),o(i+1)\in Z$. Each such component avoids $0$ and meets no edge of $G_{\bar Z}$, so it is a component of $G_{\bar Z}\cup\mathcal T_j$ by itself, and $R-r_{\mathsf F_j}(G_{\bar Z}\cup\mathcal T_j)\ge N(Z)$.

The events for index $i$ involve the spokes at the $2L+2$ positions $i-L,\dots,i+L+1$. Their owners are a uniformly random sequence of distinct tokens, $q$ per owner; drawing the tokens independently instead changes the probability of the events by at most $\binom{2L+2}2/R$, and makes the owners independent and uniform on $[n]$. Then the spoke times are independent with the empirical law of $(t_a)$, the spoke at $i$ has time in $du$ with owner in $Z$ with probability $\lambda_Z(du)$, and the other spokes and the $z$'s are independent of it. Integrating over $z_i,z_{i+1}$, and in the second case over $z_{i+1}<\max(u_i,u_{i+1})$, the probability of the counted event is at least
\[
 \int\bigl(\psi_L(u)-L\epsilon_j\bigr)_+^2\,d\lambda_Z(u)+\iint(u\vee u')\bigl(\psi_L(u\vee u')-L\epsilon_j\bigr)_+^2\,d\lambda_Z(u)\,d\lambda_Z(u')-\binom{2L+2}2\frac1R,
\]
where $\psi_L(u)=\int_u^1\varrho_L\ge\psi(u)-4^{-L}$. Since $\psi\le1$ and $\lambda_Z$ has mass at most $1$, this is at least $\Gamma(\lambda_Z)-4L\epsilon_j-4^{1-L}-\binom{2L+2}2/R$. Summing over $i$ and accounting for the $2L+1$ excluded indices, and using $\Gamma\le1$, gives $\E N(Z)\ge R(\Gamma(\lambda_Z)-4L\epsilon_j-4^{1-L})-(2L+2)^2$.

A spoke position enters the events of at most $2L+2$ indices, and an edge time of at most $2L+1$, so \cref{bh:bd} applies to $N(Z)$ with $c=4L+4$. The probability that $N(Z)<\E N(Z)-\Delta$ for some of the $2^n$ sets $Z$ is at most $2^ne^{-\Delta^2/(4Rc^2)}\le\zeta$ for $\Delta=2cR\sqrt{(n\ln2+\ln(1/\zeta))/R}$.
\end{proof}

\begin{lemma}[Inventory]\label{bh:inventory}
Let $\mathcal B$ be a fixed set of $s$ bad edges of $\mathsf F_j$, and for $t\in[0,1]$ let $I(t)$ be the number of $b_i\in\mathcal B$ with $z_i\ge t$ whose endpoints are not connected in $G_{<t}$. Let $L,p\ge1$ be integers and $\zeta\in(0,1)$. Conditionally on $(t_a)$, with probability at least $1-\zeta$, for all $t\in[0,1]$,
\[
 I(t)\le s\bigl((1-t)\chi(t)+\theta'\bigr),\qquad
 \theta'=2L\epsilon_j+2\cdot4^{-L}+\frac{2L^2}R+\frac5p+\frac{2L+8L\sqrt{R\ln(p/\zeta)}}{s}.
\]
\end{lemma}

\begin{proof}
Both conditions in the definition of $I(t)$ become weaker as $t$ decreases, so $I$ is nonincreasing. Let $I^+(t)$ be the number of $b_i\in\mathcal B$ with $z_i\ge t$ for which $\mathsf L_{i-1}(t)\cap\mathsf R_i(t)$ fails, where indices with $i\le L$ or $i>R-L+1$ always count. Then $I\le I^+$. As in the proof of \cref{bh:credit}, the events for $b_i$ involve $2L$ spoke positions, and an interior $b_i$ counts with probability at most $(1-t)\bigl(1-(\varrho_L(t)-L\epsilon_j)_+^2\bigr)+\binom{2L}2/R\le(1-t)\chi(t)+2\cdot4^{-L}+2L\epsilon_j+2L^2/R$. \Cref{bh:bd} applies to $I^+(t)$ with $c=4L$, and a union bound over $t\in\{0,\frac1p,\dots,\frac{p-1}p\}$ shows that, with probability at least $1-\zeta$, $I^+(t)\le s((1-t)\chi(t)+\theta'-5/p)$ at all these points. For $t$ between two consecutive points $t'\le t$, use $I(t)\le I(t')\le I^+(t')$ and $|((1-t)\chi)'|=|\chi+2(1-t)\varrho\varrho'|\le1+2\cdot\frac{16}9<5$, as $0\le\varrho'=(1-t^2)/(1-t+t^2)^2\le\frac{16}9$.
\end{proof}

\begin{lemma}[Baseline]\label{bh:baseline}
$\E[R-|\mathcal T_j|]\le R(\kappa+o(1))$.
\end{lemma}

\begin{proof}
Let $L<i\le R-L$. The edge $b_i$ is not in $\mathcal T_j$ exactly when $i-1$ and $i$ are both joined to $0$ in $G_{<z_i}$, which does not contain $b_i$. If $i-1$ is joined to $0$ but $\mathsf L_{i-1}(z_i)$ fails, then the edges $b_{i-L},\dots,b_{i-1}$ arrive before $z_i$ and the spokes at $i-L,\dots,i-1$ after it; similarly on the right. The $2L$ spokes involved have distinct owners with probability at least $1-\binom{2L}2/n$, and then, unconditionally, all the times involved are independent and uniform. Hence $\Pr[b_i\notin\mathcal T_j]\le\int_0^1\varrho^2+2\cdot4^{-L}+2L^2/n$. There are at most $2L$ other indices, $\int_0^1\varrho^2=\kappa$, and $n\to\infty$; then let $L\to\infty$.
\end{proof}

Let $\lambda_\ell$ be the measure $\lambda_Z$ for the set $Z$ of rejected items of level $\ell$, so that $\lambda_\ell([0,1])=1-P_\ell$.

\begin{lemma}[Rejected invisible roles]\label{bh:hazard}
Let $j>N$, $1\le d\le N$, $\ell=j-d$, and let $X$ be the number of edges of $\mathcal B_{j,d}\cap\mathcal T_j$ whose carriers are rejected. Then for every algorithm $\E X\le\pi_dR_j\bigl(\E\int\chi\,d\lambda_\ell+o(1)\bigr)$.
\end{lemma}

\begin{proof}
Let $s=\pi_dR_j$. Fix auxiliary integers $L,p\ge1$ and reals $\zeta,\upsilon>0$ and $\gamma\in(0,\frac12)$, let $\theta'_\upsilon$ be $\theta'$ of \cref{bh:inventory} for $\mathcal B=\mathcal B_{j,d}$ with $\epsilon_j$ replaced by $\upsilon$, and let $\mathcal G$ be the event that the bound of \cref{bh:inventory} holds with $\theta'_\upsilon$, that $\epsilon_j\le\upsilon$, and that $\sup_t|F_\ell(t)-t|\le\eta'$, with $\eta'$ as in the smoothing step. Then $\Pr[\mathcal G^c]\le\zeta+o(1)$, because unconditionally the times $z_i$ are independent and uniform and independent of everything else at level $j$.

Condition as in the proof of \cref{lem:invisible}, and use its notation $\nu=n_\ell$, items $1,\dots,\nu$ in arrival order at times $t_1<\dots<t_\nu$, and indicators $Z_i,Z^0_i$. Let $\iota_i$ indicate that item $i$ carries an edge of $\mathcal B_{j,d}\cap\mathcal T_j$. All atoms of $\mathsf F_j$ arriving before $t_i$ are carried by items whose atoms have been revealed, so $G_{<t_i}$ and $I(t_i)$ are determined before the decision on item $i$; and since item $i$ carries each unrevealed atom with probability $1/(\nu-i+1)$, and an atom of $\mathcal B_{j,d}$ carried by item $i$ lies in $\mathcal T_j$ exactly when it is counted in $I(t_i)$, the conditional probability of $\iota_i=1$ is $I(t_i)/(\nu-i+1)$. Call $i$ \emph{bad} if $t_i\le1-\gamma$ and $I(t_i)/(\nu-i+1)>\frac s\nu(\chi(t_i)+\varkappa)$, where $\varkappa=(\theta'_\upsilon+\eta')/(\gamma-\eta')$, and let $i^*$ be the first bad index. Badness is determined before the decision on item $i$. On $\mathcal G$ no index is bad, since $\nu-i+1=\nu(1-F_\ell(t_i^-))\ge\nu(1-t_i-\eta')$. As $Z_i=Z^0_i$ when $\iota_i=1$,
\[
 X\le\sum_{i<i^*,\ t_i\le1-\gamma}\iota_iZ^0_i+\#\{b\in\mathcal B_{j,d}:\text{its carrier arrives after }1-\gamma\}+s\,\ind[\mathcal G^c].
\]
The expectation of the first sum is at most $\frac s\nu\E\sum_i(\chi(t_i)+\varkappa)Z^0_i\le s\bigl(\E\int\chi\,d\lambda_\ell+s_\ell/n_\ell+\varkappa\bigr)$, because $Z^0_i\ne Z_i$ for at most $s_\ell$ items. The second term has expectation $s\gamma$. Hence $\E X/s\le\E\int\chi\,d\lambda_\ell+s_\ell/n_\ell+\varkappa+\gamma+\zeta+o(1)$. As $Y\to\infty$, $s/R_j=\pi_d$ is fixed and $R_j\to\infty$, so $\varkappa\to(2L\upsilon+2\cdot4^{-L}+5/p)/\gamma$. Let $\upsilon\to0$, then $L,p\to\infty$, and then $\gamma,\zeta\to0$.
\end{proof}

\paragraph{The fan part.}
Let $Z_j$ be the set of rejected items of level $j$, and let $\mathcal T_A\subseteq\mathcal T_j$ be the edges with accepted carriers. Since $f\le\bar f$ and every accepted spoke is in $G_{\bar Z_j}$, the fan of level $j$ contributes to $\bar f(A)$ at most
\[
 u_j\bigl(r(G_{\bar Z_j}\cup\mathcal T_A)-|\mathcal T_A|\bigr)\le u_j\bigl(R_j-|\mathcal T_j|\bigr)+u_j|\mathcal T_j\setminus\mathcal T_A|-u_j\bigl(R_j-r(G_{\bar Z_j}\cup\mathcal T_j)\bigr).
\]
We have $u_jR_j=\delta$ and $|\mathcal T_j\setminus\mathcal T_A|=\sum_dX_{j,d}$. Apply \cref{bh:baseline}, \cref{bh:hazard}, and \cref{bh:credit} with $Z=Z_j$, which may depend on the whole run; in the last one, $n_j\ln2/R_j=\ln2/(\delta Q)\to0$, $\E\epsilon_j\to0$, and on the failure event we use $\Gamma\le1$. Letting $Y\to\infty$ and then $L\to\infty$ and $\zeta\to0$, the fan of level $j$ contributes at most $\delta\bigl(\kappa+\sum_d\pi_d\E\int\chi\,d\lambda_{j-d}-\E\Gamma(\lambda_j)\bigr)+o(1)$. Each level $\ell$ receives weight $\pi_d$ from at most one level $j=\ell+d$ for each $d$, $\chi\ge0$, and $0\le\Gamma\le1$, so the expected fan part of $\bar f(A)$ is at most
\begin{equation}\label{bh:fanbound}
 \delta\,\E\sum_{\ell=1}^m\Psi(\lambda_\ell)+\delta N+o(m),\qquad
 \Psi(\lambda)=\kappa+\int\omega\,d\lambda-\iint\psi_2(u\vee u')\,d\lambda(u)\,d\lambda(u').
\end{equation}

\begin{lemma}\label{bh:scalar}
On $[0,1]$: (a) $\varrho$ increases from $0$ to $1$, $\varrho'\ge(1-t)(1+t)$, and $(1-t)\varrho\le\psi\le1-t$; (b) $\omega'=-2\varrho(\varrho'-\psi)\le0$, $\omega(1)=0$, and $\omega\le1$; (c) $g(t,y)=-\omega'(t)y+\psi_2'(t)y^2$ is nondecreasing in $y\in[0,\frac12]$ and $|\partial_yg|\le6$; (d) $V$ is concave on $[0,\frac12]$; (e) $\ell(x)=c_0+c_1x$ satisfies $\ell\ge V$ on $[0,\frac12]$ and $\ell(x)\ge\kappa+\int_0^x\omega$ on $[\frac12,1]$.
\end{lemma}

\begin{proof}
(a) $\varrho'=(1-t^2)/(1-t+t^2)^2$ and $1-t+t^2\le1$; the bounds on $\psi$ follow from $\varrho\le1$ and monotonicity. (b) Use $\psi'=-\varrho$ and $\varrho'\ge1-t\ge\psi$. (c) $\partial_yg=-\omega'+2\psi_2'y$ with $\psi_2'=\psi^2-2t\psi\varrho$. If $\psi_2'\ge0$ this is at least $-\omega'\ge0$; otherwise it is at least $-\omega'+\psi_2'=2\varrho(\varrho'-(1+t)\psi)+\psi^2\ge0$, since $\varrho'\ge(1-t)(1+t)\ge(1+t)\psi$. Also $|\omega'|\le2\varrho'\le\frac{32}9$ and $|\psi_2'|\le2$. (d) $V''=\omega'-4t\psi(\psi-t\varrho)\le0$ for $t\le\frac12$, as $\psi\ge(1-t)\varrho\ge t\varrho$. (e) The first claim holds because $\ell$ is tangent to $V$ at $x_0$. For the second, $\frac d{dx}(\ell-\kappa-\int_0^x\omega)=c_1-\omega(x)\ge c_1-\omega(\frac12)>0$ for $x\ge\frac12$, and the difference is positive at $x=\frac12$; both margins exceed $0.04$ by the certificate below.
\end{proof}

\begin{lemma}\label{bh:profile}
Let $\lambda$ be a measure on $[0,1]$ of mass $x\le1$ with $\lambda[0,t]\le t+e$ for all $t$. Then $\Psi(\lambda)\le c_0+c_1x+6e$.
\end{lemma}

\begin{proof}
Let $D(t)=\lambda[0,t]$. By Fubini and $\omega(1)=\psi_2(1)=0$, $\int\omega\,d\lambda=-\int_0^1D\omega'\,dt$, and, since $u\vee u'$ has distribution function $D^2$ under $\lambda\otimes\lambda$, $\iint\psi_2(u\vee u')=-\int_0^1D^2\psi_2'\,dt$. So $\Psi(\lambda)=\kappa+\int_0^1g(t,D(t))\,dt$. If $x\le\frac12$, let $y_t=\min(t,x)$; then $D(t)\le x\le\frac12$, and either $D(t)\le y_t$ or $y_t<D(t)\le y_t+e$, so $g(t,D(t))\le g(t,y_t)+6e$ by \cref{bh:scalar}(c). Integrating by parts, $\int_0^1g(t,y_t)\,dt=\int_0^x\omega-2\int_0^xt^2\psi^2=V(x)-\kappa$, and \cref{bh:scalar}(e) concludes. If $x>\frac12$, drop the last term of $\Psi$, which is nonnegative, and use $-\omega'\ge0$ and $D(t)\le\min(t,x)+e$ to get $\Psi(\lambda)\le\kappa+\int_0^x\omega+e\le c_0+c_1x+e$.
\end{proof}

\begin{proof}[Proof of \cref{bh:thm}]
On the event of the smoothing step, each $\lambda_\ell$ satisfies the hypothesis of \cref{bh:profile} with $e=\eta'$, because every rejected item has arrived; off it, $\Psi\le\kappa+1$. Since the bound of \cref{bh:profile} is affine in $x$, \eqref{bh:fanbound} is at most $\delta m\,\E[c_0+c_1\bar x]+\delta N+o(m)$, where $\bar x=1-\bar P$. Combining it with \eqref{eq:pairbound}, the smoothing step, and \cref{lem:control} exactly as in the proof of \cref{thm:transfer},
\[
 \frac{\E[f(A)]}m\le\frac{(1+\delta)N}m+\E\Bigl[(1-\delta)\frac{1-\bar x^2}2+\delta(c_0+c_1\bar x)\Bigr]+\frac9{8\mathcal H_N}+o(1)
 \le C_{\mathrm{fan}}+\frac{(1+\delta)N}m+\frac9{8\mathcal H_N}+o(1),
\]
since the bracket is maximized at $\bar x=\delta c_1/(1-\delta)$. Conclude as in the proof of \cref{thm:transfer}, using \cref{lem:transfer-valid}, and choosing $N$, then $m$, then $Y$ large.
\end{proof}

\begin{remark}\label{bh:remark}
A computer check encloses $\pi$ and $\sqrt3$ rationally, expands $\varrho(t)=t(1+t)/(1+t^3)$ in its power series, whose coefficients lie in $\{-1,0,1\}$, and computes $\kappa$, $c_0$, $c_1$, and $C_{\mathrm{fan}}$ in exact rational arithmetic with rigorous truncation bounds. It gives $c_1=0.63238805347\ldots$, $c_0=0.47285764611\ldots$, $C_{\mathrm{fan}}=0.49913658734\ldots$, and the margins $c_1-\omega(\frac12)=0.2758\ldots$ and $\ell(\frac12)-\kappa-\int_0^{1/2}\omega=0.0404\ldots$ used in \cref{bh:scalar}(e). Without the two-vertex term, that is, bounding $\omega\le\omega(0)=1-\pi^2/27$ and optimizing $\delta$, the same argument gives the closed form $\kappa-a^2+a\sqrt{a^2+1-2\kappa}=0.4991383\ldots$ with $a=1-\pi^2/27$. Optimizing $\delta$ and $x_0$ changes $C_{\mathrm{fan}}$ only in the eleventh decimal.
\end{remark}

\subsection{Grouped spokes}\label{bu3:sec}

In the fans of \cref{bh:thm} every spoke has its own owner. We now let $h$ consecutive spokes share an owner. A rejected owner then removes almost a whole block of rank. Moreover, an internal bad edge of a block can be retained only before the block's owner arrives, and this cancels most of the charge for rejected invisible roles. Half of the value is put into such fans. Let $h=12$, and for $s,t,x\in[0,1]$ let
\[
 v(s)=\frac{s}{1-(1-s)s^h},\qquad
 \beta_h=\frac{h-1}{2h}+\frac1h\int_0^1\bigl[1+(h-1)(1-s)s^h\bigr]v(s)^2\,ds,
\]
\[
 \vartheta(t)=\frac{t^{h+3}}{h+3}+\int_t^1(1-s)s^{h+1}\,ds,\qquad
 c_{\mathrm b}=\frac{1-2\vartheta(\frac12)}h=\frac{567979}{6881280},\qquad
 \sigma(x)=\frac{2x^3}3-\frac{x^4}4,
\]
\[
 \chi_h(t)=\frac{h-(h-1)t-t^2}h,\qquad
 \gamma_h(t)=\frac{h-1-(h-1)t-t^2+2\vartheta(t)}h .
\]
Let $L=41$, $\tau_0=\frac12$, $\tau_i=\frac{49+i}{100}$ for $1\le i\le L$, $\nu_i=\frac2h(\vartheta(\tau_i)-\vartheta(\tau_{i-1}))$, $\alpha_i=(1-\tau_i)^2-(1-\tau_{i+1})^2$ for $i<L$, and $\alpha_L=(1-\tau_L)^2$. For $x\in[0,1]$ and $y\in[0,1]^L$ let
\[
 \Phi(x,y)=\frac{1-x^2}4+\frac{\beta_h}2+\frac12\Bigl(c_{\mathrm b}x-\sum_{i=1}^L\nu_i(x-y_i)\Bigr)-\frac{\sigma(x)}{2h}-\frac{h-1}{4h}\sum_{i=1}^L\alpha_iy_i^2,
\]
and $C_{\mathrm{blk}}=\max\{\Phi(x,y):0\le y_1\le\dots\le y_L\le x\le1\}$.

\begin{theorem}\label{bu3:thm}
\Cref{bh:thm} holds with $C_{\mathrm{fan}}$ replaced by $C_{\mathrm{blk}}$, and $C_{\mathrm{blk}}<0.4970528$. In particular, $\limsup_{k\to\infty}\rho_k\le C_{\mathrm{blk}}<0.4970528$.
\end{theorem}

\paragraph{The construction.}
It is that of \cref{bh:thm}, with the following changes. We take $\delta=\frac12$; nothing before the last step of the proof of \cref{thm:transfer} uses $\delta<\frac12$. We let $Q$ be the least multiple of $hQ_0$ with $Q\ge(\log Y)^2$, and we put $q=Q/(2h)$. Thus each item owns $hq$ pair components, $q\to\infty$, and $q^2/Y\to0$ as $Y\to\infty$. At a level $j>N$ we have $R_j=hM_j$ with $M_j=n_jq$. We split the vertices $1,\dots,R_j$ of $\mathsf F_j$ into $M_j$ consecutive \emph{blocks} of $h$ vertices. Block $p$ has the vertices $(p-1)h+1,\dots,ph$. Its \emph{boundary edge} is $b_{(p-1)h+1}$, and for $p=1$ this is the edge $01$. Its \emph{internal edges} are $b_{(p-1)h+2},\dots,b_{ph}$. The two boundary edges \emph{at} block $p$ are its own and that of block $p+1$; the latter is absent for $p=M_j$. A uniformly random bijection assigns $q$ tokens per item to the $M_j$ blocks, and an item owns all spokes of its $q$ blocks. The set $\mathcal B_{j,d}$ consists of all bad edges of $\pi_dM_j$ consecutive blocks, and these intervals of blocks partition $[M_j]$. Since $Q_0\pi_d=2N!\,\mathrm{den}(\mathcal H_N)/d$, all these numbers are integers. As in \cref{bh:sec}, \cref{lem:transfer-valid}, the relaxation $\bar f\ge f$, the paragraph on invisible roles and \eqref{eq:pairbound} carry over.

Fix $j>N$ and write $n=n_j$, $M=M_j$, $R=R_j$. Let $t_a$ be the time of item $a$ of level $j$, let $U_p$ be the time of the owner of block $p$, and let $z_i$ be the carrier time of $b_i$. As in \cref{bh:sec}, the $z_i$ are independent and uniform, and they are independent of the bijection and of $(t_a)$; $\mathcal T_j$, $\epsilon_j$ and $\lambda_\ell$ are as there. Let $Z$ be the set of rejected items of level $j$, let $\mathcal T_A\subseteq\mathcal T_j$ be the set of edges with accepted carriers, and let $x_j=\lambda_j([0,1])$ and $\Lambda_\ell(t)=\lambda_\ell([0,t])$.

\begin{lemma}[Local facts]\label{bu3:local}
Let $p$ be a block with owner time $t$, and let $e$ be a boundary edge at $p$, between $p$ and $p'=p+o$ with $o=\pm1$, with carrier time $V$.
\begin{enumerate}[nosep,label=(\alph*)]
\item Every internal edge of $p$ in $\mathcal T_j$ has carrier time less than $t$.
\item If $e\in\mathcal T_j$, then $V<t$ or $U_{p'}>V$. Moreover $e\notin\mathcal T_j$ on each of the following events.
$W_1$: $V<t$, $U_{p-1},U_{p+1}<V$, and the $h$ bad edges on the path from the endpoint of $e$ in $p$ through $p$ to block $p-o$ arrive before $V$.
$W_2$: $t<V<U_{p'}$, $U_{p+2o}<V$, and the $h$ bad edges on the path from the endpoint of $e$ in $p'$ through $p'$ to block $p+2o$ arrive before $V$.
\item Let $e$ be a boundary edge between blocks with owner times $t,u$. Then $e\in\mathcal T_j$ if $t>V$ and the internal edge of the left block at $e$ arrives after $V$, or if $t<V<u$ and the internal edge of the right block at $e$ arrives after $V$.
\item If both boundary edges at $p$ arrive after $t$, then every internal edge of $p$ that arrives before $t$ lies in $\mathcal T_j$. Moreover, for every set $S$ of atoms of $\mathsf F_j$ that arrive before $t$, deleting the internal edges of $p$ from $S$ does not change $\varphi_{\mathsf F_j}(S)$.
\end{enumerate}
If the owner times involved are distinct independent uniform variables, independent of the carrier times, then, conditionally on $t$ (and on $u$ in (c)), the event in the first part of (b) has probability $\frac{1+t^2}2$ and $W_1\cup W_2$ has probability $\vartheta(t)$. The event in (c) has probability $w-\frac{w^2}2$ with $w=\max(t,u)$, and the hypothesis of (d) has probability $(1-t)^2$.
\end{lemma}

\begin{proof}
(a) After time $t$, both endpoints are joined to $0$ by spokes. (b) If $V>t$ and $U_{p'}<V$, the same holds for $e$. On $W_1$, the endpoint of $e$ in $p'$ is joined to $0$ by its spoke, and the other endpoint is joined to $0$ through block $p$ and the spoke of a vertex of block $p-o$. On $W_2$, the endpoint in $p$ is joined to $0$ by its spoke, and the other endpoint through $p'$ and block $p+2o$. Neither path uses $e$. (c) The endpoint in question has no spoke and no other bad edge before $V$, so it is isolated in $G_{<V}$. (d) Before $t$, no spoke of $p$ and no boundary edge at $p$ has arrived. So the internal edges of $p$ in $S$ form a forest on the vertices of $p$ that meets no other edge of $S$, and each of them adds one to the rank. For the probabilities, integrate over $V=s$: the integrands are $\ind[s<t]+\ind[s>t](1-s)$; $\ind[s<t]s^{h+2}+\ind[s>t](1-s)s^{h+1}$; $(1-s)\ind[s<\max(t,u)]$. The events $W_1$ and $W_2$ are disjoint.
\end{proof}

\begin{lemma}[Forest count]\label{bu3:forest}
Let $V_Z$ be the set of vertices of blocks owned by $Z$. Let $I_Z$ and $I_Z^-$ be the numbers of internal edges of these blocks in $\mathcal T_j$ and in $\mathcal T_j\setminus\mathcal T_A$, respectively. Let $B_Z$ be the number of pairs of a block owned by $Z$ and a boundary edge at it lying in $\mathcal T_j$, and let $E_Z$ be the number of boundary edges in $\mathcal T_j$ between two blocks owned by $Z$. If $G_{\bar Z}$ is the set of spokes owned outside $Z$, then
\[
 r(G_{\bar Z}\cup\mathcal T_A)-|\mathcal T_A|\le R-|\mathcal T_j|+|\mathcal T_j\setminus\mathcal T_A|-\bigl(|V_Z|-I_Z-B_Z\bigr)-E_Z-I_Z^- .
\]
\end{lemma}

\begin{proof}
Let $\Xi$ be the number of components of the forest $(\{0,\dots,R\},\mathcal T_A)$ contained in $V_Z$. They meet neither $0$ nor a spoke of $G_{\bar Z}$, so $r(G_{\bar Z}\cup\mathcal T_A)\le R-\Xi$, while $|\mathcal T_A|=|\mathcal T_j|-|\mathcal T_j\setminus\mathcal T_A|$. At most $I_Z-I_Z^-+E_Z$ edges of $\mathcal T_A$ lie inside $V_Z$. At most $B_Z-2E_Z$ edges of $\mathcal T_A$ leave $V_Z$, each joining at most one component to the outside. Hence $\Xi\ge|V_Z|-(I_Z-I_Z^-+E_Z)-(B_Z-2E_Z)$.
\end{proof}

\begin{lemma}[Baseline]\label{bu3:baseline}
$\E[R-|\mathcal T_j|]\le R(\beta_h+o(1))$.
\end{lemma}

\begin{proof}
A bad edge $b_i$ arriving at time $s$ is outside $\mathcal T_j$ exactly when both its endpoints are joined to $0$ in $G_{<s}$. In a doubly infinite chain of blocks with independent uniform times, an endpoint is joined to $0$ on its side with probability $v(s)$. Indeed, its block is rooted with probability $s$, and otherwise its $h$ bad edges towards the next block must all have arrived. A boundary edge is therefore outside $\mathcal T_j$ with probability $v(s)^2$. An internal edge is outside $\mathcal T_j$ with probability $s+(1-s)s^hv(s)^2$, where $s^h$ accounts for the other $h$ edges of the path between the neighbouring blocks. Averaging over the $h$ positions gives $\beta_h$. In the fan, fix $A\ge1$ and an edge whose block is at distance more than $A$ from both ends. The $2A+1$ blocks around it have distinct owners with probability at least $1-(2A+1)^2/n$. Exploring at most $A$ unrooted blocks on each side changes the probability by at most $2(1-s)^As^{hA}\le2(h+1)^{-A}$. Let $Y\to\infty$ and then $A\to\infty$, as in \cref{bh:baseline}.
\end{proof}

\begin{lemma}[Rejected blocks]\label{bu3:credit}
$\E\bigl[|V_Z|-I_Z-B_Z\bigr]\ge R\bigl(\E\int\gamma_h\,d\lambda_j-o(1)\bigr)$.
\end{lemma}

\begin{proof}
Fix an item $a$. Let $C_a$ be the number of internal edges of its $q$ blocks that arrive before $t_a$. Given $t_a$, $C_a$ is binomial with mean $(h-1)qt_a$, so $\E|C_a-(h-1)qt_a|\le\sqrt{hq}$. Let $\mathcal B_a$ be the event that a block of $a$ is within distance $2$ of an end, or two blocks of $a$ are within distance $4$, or two of the at most $4q$ blocks at distance $1$ or $2$ from blocks of $a$ have the same owner, or one of them is owned by $a$. Then $\Pr[\mathcal B_a]\le6/n+8q/n+16q^2/n$. On $\mathcal B_a^c$, let $D_a$ be the sum over the $2q$ pairs $(p,e)$ of a block $p$ of $a$ and a boundary edge $e$ at $p$ of $\ind[V<t_a\text{ or }U_{p'}>V]-\ind[W_1\cup W_2]$, in the notation of \cref{bu3:local}(b). On $\mathcal B_a$, let $D_a=2q$. In both cases $D_a$ is at least the number of such pairs with $e\in\mathcal T_j$. Given the bijection on $\mathcal B_a^c$ and given $t_a$, the $q$ blocks contribute independent terms in $[0,2]$, with mean $1+t_a^2-2\vartheta(t_a)$ each by \cref{bu3:local}. Hence $\E|D_a-q(1+t_a^2-2\vartheta(t_a))|\le\sqrt q+2q\Pr[\mathcal B_a]$. Now $|V_Z|-I_Z-B_Z\ge\sum_{a\in Z}(hq-C_a-D_a)$ by \cref{bu3:local}(a). Replace $C_a$ and $D_a$ by their centres, at a total cost of at most $\sum_a(|C_a-\dots|+|D_a-\dots|)$ whatever $Z$ is. The centres give $hq\sum_{a\in Z}\gamma_h(t_a)=R\int\gamma_h\,d\lambda_j$. The total cost is $O(n\sqrt{hq}+q^3)=o(R)$.
\end{proof}

\begin{lemma}[Shared boundaries]\label{bu3:shared}
$\E E_Z\ge\frac Rh\bigl(\E\sigma(x_j)-o(1)\bigr)$.
\end{lemma}

\begin{proof}
For $S\subseteq[n]$, let $J(S)$ be the number of boundary edges between two blocks owned by $S$ at which the event of \cref{bu3:local}(c) holds; then $E_Z\ge J(Z)$. Condition on $(t_a)$. At each of the $M-1$ boundaries between blocks, the two tokens form a uniformly random ordered pair of distinct tokens. Let $x=|S|/n$ and let $g(s)$ be the fraction of items in $S$ with time at most $s$. Then $g(s)\le\min\{x,s+\epsilon_j\}$, and \cref{bu3:local}(c) gives
\[
 \E[J(S)\mid(t_a)]\ge M\int_0^1(1-s)\Bigl[x^2-g(s)^2-\frac{x-g(s)}M\Bigr]ds\ge M\Bigl(\int_0^x(1-s)(x^2-s^2)\,ds-2\epsilon_j-\frac1M\Bigr),
\]
and the last integral is $\sigma(x)$. Apply \cref{bh:bd} with $R$ replaced by $M$, the bijection between blocks and tokens, and $z_p$ the vector of carrier times of the bad edges of block $p$. A swap of two tokens changes $J(S)$ by at most $4$; changing $z_p$ changes it by at most $2$, since $h\ge3$. So for fixed $\varepsilon>0$, $\Pr[J(S)\le\E[J(S)\mid(t_a)]-\varepsilon M]\le e^{-\varepsilon^2M/64}$. A union bound over the $2^n$ sets $S$ gives failure probability at most $2^ne^{-\varepsilon^2nq/64}\to0$, because $q\to\infty$. This holds for every $S$ at once, in particular for $S=Z$, which may depend on the whole run. Since $J\ge0$, $\sigma\le1$, $\E\epsilon_j\to0$ and $\varepsilon$ is arbitrary, the lemma follows.
\end{proof}

\begin{lemma}[Rejected owners of isolated blocks]\label{bu3:joint}
\[
 \E I_Z^-\ge R\Bigl(\frac{h-1}h\sum_{d=1}^N\pi_d\,\E\int(1-t)^2\Lambda_{j-d}(t^-)\,d\lambda_j(t)-o(1)\Bigr).
\]
\end{lemma}

\begin{proof}
Fix an item $a$ and $d$, and let $\ell=j-d$. Let $\mathcal P$ be the set of blocks of $a$ in the class-$d$ interval, other than the first and last block of $\mathsf F_j$. Let $\mathcal P'\subseteq\mathcal P$ be the set of those whose two boundary edges arrive after $t_a$, and let $K_a=(h-1)|\mathcal P'|$ be the number of their internal edges, all of which lie in $\mathcal B_{j,d}$. Given $t_a$, $|\mathcal P|$ has mean in $[q\pi_d-2/n,q\pi_d]$ and variance at most $q\pi_d$. Given the bijection, $|\mathcal P'|$ is a sum of $|\mathcal P|$ indicators of mean $(1-t_a)^2$, dependent only for adjacent blocks. Hence $\E|K_a-(h-1)q\pi_d(1-t_a)^2|\le h(3\sqrt{q\pi_d}+2/n)$.

Condition on all arrival times, the bijections, the algorithm's coins, the injections into levels other than $\ell$, and the injection into $\ell$ of all atoms other than these $K_a$ edges. This is legitimate, because $\mathcal P'$ is determined by the times and the placement of the boundary edges, which is revealed first. The $K_a$ edges are then injected uniformly into the remaining items of level $\ell$. By \cref{bu3:local}(d), each value $f(S)$ with $S$ a set of items arriving before $t_a$ is the same for every placement. So the set of items of level $\ell$ rejected before $t_a$ is fixed, and it has $n_\ell\Lambda_\ell(t_a^-)$ elements. Let $D$ be the number of the $K_a$ edges that land on this set. Then $\E|D-K_a\Lambda_\ell(t_a^-)|\le\frac12\sqrt{K_a}+2K_a s_\ell/n_\ell$. By \cref{bu3:local}(d), each of these $D$ edges lies in $\mathcal T_j\setminus\mathcal T_A$, so if $a\in Z$ it is counted in $I_Z^-$. Therefore $I_Z^-\ge\sum_{a\in Z}\sum_dD_{a,d}$. Multiplying the two absolute deviations by $\ind[a\in Z]$, which may depend on the placement, does not increase them. We get
\[
 \E I_Z^-\ge\sum_a\sum_d\Bigl((h-1)q\pi_d\,\E\bigl[\ind[a\in Z](1-t_a)^2\Lambda_{j-d}(t_a^-)\bigr]-O\bigl(h\sqrt{q}+hq\,s_{j-d}/n_{j-d}\bigr)\Bigr),
\]
which is the claim, since $R=hnq$, $N$ is fixed and $q\to\infty$.
\end{proof}

\begin{lemma}[Rejected invisible roles]\label{bu3:hazard}
Let $1\le d\le N$, $\ell=j-d$, and let $X$ be the number of edges of $\mathcal B_{j,d}\cap\mathcal T_j$ whose carriers are rejected. Then $\E X\le\pi_dR\bigl(\E\int\chi_h\,d\lambda_\ell+o(1)\bigr)$.
\end{lemma}

\begin{proof}
For $t\in[0,1]$, call an edge of $\mathcal B_{j,d}$ \emph{eligible at $t$} if it is internal to a block whose owner time exceeds $t$, or it is a boundary edge at a block whose owner time exceeds $t$. By \cref{bu3:local}(a),(b), every edge of $\mathcal T_j$ is eligible at its carrier's arrival time. Let $\mathcal E(t)$ be the set of eligible edges and $w=\pi_dR$. Follow the proof of \cref{lem:invisible} with the $w$ fixed atoms replaced by the set $\mathcal E(t_i)$, which the conditioning there determines. The same argument gives $\E X\le\E\frac1\nu\sum_i|\mathcal E(t_i)|Z_i^0+2\sqrt w$. Given the bijection, a fraction $1/h$ of $\mathcal B_{j,d}$ are boundary edges, eligible with probability at most $1-t^2$, and the rest are eligible with probability $1-t$, so $\E|\mathcal E(t)|\le w\chi_h(t)$. Changing whether $t_a>t$ for one item changes $|\mathcal E(t)|$ by at most $c_a\le(h+1)q$, and $\sum_ac_a\le2w$. So $\mathrm{Var}|\mathcal E(t)|\le\frac14\sum_ac_a^2\le\frac12(h+1)qw$, and $\E(|\mathcal E(t)|/w-\chi_h(t))_+\le\sqrt{(h+1)/(2h\pi_dn)}$. This also holds at the times $t_i$, which are independent of level $j$. As in \cref{lem:invisible}, $\sum_i\chi_h(t_i)Z_i^0\le\nu\int\chi_h\,d\lambda_\ell+s_\ell$, since $0\le\chi_h\le1$.
\end{proof}

\begin{lemma}\label{bu3:harm}
On the event of the smoothing step, let $x_j$ ($j\in[m]$) be nondecreasing right-continuous paths with $x_j(0)=0$ and $x_j(v)-x_j(u)\le F_j(v)-F_j(u)$ for $u\le v$, and let $\bar x=\frac1m\sum_jx_j(1)$. Then
\[
 \sum_{j>N}\sum_{d=1}^N\pi_d\int x_{j-d}(t^-)\,dx_j(t)\ge\frac m2\bar x^2-\frac{9m}{8\mathcal H_N}-N-4m\eta'.
\]
\end{lemma}

\begin{proof}
Add the rows $j\le N$, each at most $1$, and replace $x(t^-)$ by $x(t)$, as in the smoothing step. Let $\tilde x_j(t)=\inf_{s\le t}\{x_j(s)+t-s\}$; as there, $\tilde x_j$ is $1$-Lipschitz and $\Delta_j=x_j-\tilde x_j\in[0,2\eta']$. Integrating by parts against the continuous nondecreasing path $\Pi\tilde x$,
\[
 \int(\Pi x)_j\,dx_j-\int(\Pi\tilde x)_j\,d\tilde x_j=\int(\Pi\Delta)_j\,dx_j+(\Pi\tilde x(1))_j\Delta_j(1)-\int\Delta_j\,d(\Pi\tilde x)_j\ge-2\eta'.
\]
By \cref{lem:control}, whose statement is equivalent to $\int_0^1p^{\mathsf T}\Pi a\,dt\ge\frac m2\bar P^2-\frac{9m}{8\mathcal H_N}$, we get $\sum_j\int(\Pi\tilde x)_j\,d\tilde x_j\ge\frac m2(\bar x-2\eta')^2-\frac{9m}{8\mathcal H_N}$.
\end{proof}

\begin{proof}[Proof of \cref{bu3:thm}]
As in the paragraph on the fan part in \cref{bh:sec}, the fan of level $j$ contributes at most $u_j(r(G_{\bar Z}\cup\mathcal T_A)-|\mathcal T_A|)$ to $\bar f(A)$, and $u_jR=\frac12$. By \cref{bu3:forest,bu3:baseline,bu3:credit,bu3:shared,bu3:joint,bu3:hazard}, its expectation is at most
\[
 \frac12\Bigl[\beta_h+\sum_d\pi_d\E\!\int\!\chi_h\,d\lambda_{j-d}-\E\!\int\!\gamma_h\,d\lambda_j-\frac{\E\sigma(x_j)}h-\frac{h-1}h\sum_d\pi_d\E\!\int(1-t)^2\Lambda_{j-d}(t^-)\,d\lambda_j(t)\Bigr]+o(1).
\]
Sum over $j>N$. Each level $\ell$ receives weight $\pi_d$ from at most one $j=\ell+d$, $\chi_h\ge0$, $\gamma_h\le1$ and $\sigma\le1$. So we may sum the second, third and fourth terms over all $\ell\in[m]$ at a cost of at most $2N$. Since $\vartheta'(t)=t^{h+1}(2t-1)$, $\vartheta$ has its minimum at $\frac12$ and increases on $[\frac12,1]$. Hence $\chi_h-\gamma_h=\frac{1-2\vartheta}h\le c_{\mathrm b}-\sum_i\nu_i\ind[t>\tau_i]$, and $\int(\chi_h-\gamma_h)\,d\lambda_\ell\le c_{\mathrm b}x_\ell-\sum_i\nu_i(x_\ell-\Lambda_\ell(\tau_i))$. The function $\sigma$ is convex on $[0,1]$, so $\sum_\ell\sigma(x_\ell)\ge m\sigma(\bar x)$, where $\bar x=\frac1m\sum_\ell x_\ell=1-\bar P$. Since $(1-t)^2\ge\sum_i\alpha_i\ind[t\le\tau_i]$, the last term is at least $\sum_i\alpha_i$ times the left side of \cref{bu3:harm} for the paths $\Lambda_j(\min\{t,\tau_i\})$. These paths qualify, because rejected items have arrived. With $\bar y_i=\frac1m\sum_j\Lambda_j(\tau_i)$, \cref{bu3:harm} and $\sum_i\alpha_i=\frac14$ bound this term below by $m\sum_i\alpha_i\bar y_i^2/2-O(m/\mathcal H_N+N+m\eta')$ on the event of the smoothing step. Off that event we use $0\ge m\bar y_i^2/2-m/2$. The pair part is bounded by \eqref{eq:pairbound} with $1-\delta=\frac12$, the smoothing step and \cref{lem:control}, by $\frac m2\E[\bar P-\frac12\bar P^2]+\frac{9m}{16\mathcal H_N}+o(m)=m\E\frac{1-\bar x^2}4+\frac{9m}{16\mathcal H_N}+o(m)$. Altogether, since $0\le\bar y_1\le\dots\le\bar y_L\le\bar x\le1$,
\[
 \frac{\E[f(A)]}m\le\E\,\Phi(\bar x,\bar y)+O\Bigl(\frac Nm+\frac1{\mathcal H_N}\Bigr)+o(1)\le C_{\mathrm{blk}}+O\Bigl(\frac Nm+\frac1{\mathcal H_N}\Bigr)+o(1).
\]
Conclude as in the proof of \cref{thm:transfer}, choosing $N$, then $m$, then $Y$ large. The numerical bound is \cref{bu3:remark}.
\end{proof}

\begin{remark}\label{bu3:remark}
Let $r=\frac2{25}$, $a_*=r\sigma'(r)-\sigma(r)$, and $c_*=c_{\mathrm b}-\sum_i\nu_i-\sigma'(r)/h$. Since $\sigma(x)\ge\sigma'(r)x-a_*$ on $[0,1]$, completing the square in $x$ and in each $y_i$ separately gives
\[
 C_{\mathrm{blk}}\le\frac14+\frac{\beta_h}2+\frac{a_*}{2h}+\frac{c_*^2}4+\sum_{i=1}^L\frac{h\,\nu_i^2}{4(h-1)\alpha_i}.
\]
To bound $\beta_h$, write $z=(1-s)s^h\in[0,\frac1{h+1}]$, expand $v^2=s^2\sum_{b\ge0}(b+1)z^b$, and use $\int_0^1s^2z^b\,ds=\frac{(hb+2)!\,b!}{((h+1)b+3)!}$. Bound the terms with $b\ge9$ by $s^2(1+(h-1)z)\le2$ and $z\le\frac1{h+1}$. All other quantities are polynomials in rationals; for example, $\vartheta(t)=\frac1{210}-\frac{t^{14}}{14}+\frac{2t^{15}}{15}$. A computer check verifies, in exact rational arithmetic, that $\beta_{12}=0.49082217997\ldots$ (to eleven digits), that $c_{\mathrm b}=\frac{567979}{6881280}$, and that the right side above is less than $0.4970528$. Weaker credits give larger constants. Without the last lemma (that is, dropping the $\nu_i$ and $\alpha_i$), the bound is $0.4970995$. Without \cref{bu3:shared} as well, it is $\frac14+\frac{\beta_h}2+\frac{c_{\mathrm b}^2}4=0.4971143$. Without the events $W_1,W_2$, so with $c_{\mathrm b}$ replaced by $\frac1h$, it is $0.4971472$.
\end{remark}

\end{document}